\documentclass[aps,prd,twocolumn,superscriptaddress,nofootinbib]{revtex4-2}
\usepackage{blindtext}

\usepackage{mathtools,mathalfa,amssymb,amsmath,amsfonts,amsthm,bm}
\usepackage[T1]{fontenc}
\usepackage{tipa}
\usepackage{wasysym}
\usepackage{appendix}
\usepackage{calrsfs}
\usepackage{tikz}
\usetikzlibrary {decorations.pathmorphing}

\usepackage{color}
\usepackage[mathscr]{euscript}

\renewcommand{\d}{\mathrm{d}}
\newcommand{\re}{\mathrm{Re}}
\newcommand{\im}{\mathrm{Im}}

\newcounter{definition}
\definecolor{wuppergreen}{RGB}{137, 186, 23}

\newcommand{\pullback}[1]{\hbox{\lower0.5ex\hbox{${}_{\leftarrow}$}}\kern-1.9ex{#1}}
\newcommand{\pullbacklong}[1]{\hbox{\lower0.85ex\hbox{${}_{\longleftarrow}$}}\kern-3.0ex{#1}}
\newcommand{\pullbackllong}[1]{\hbox{\lower0.85ex\hbox{${}_{\longleftarrow\!\!-\!\!-\!\!-\!\!-}$}}\kern-6.4ex{#1}}

\begin{document}
\title{
Black Hole Tomography: Unveiling Black Hole Ringdown via Gravitational Wave Observations 
}
\author{Ariadna Ribes Metidieri}
\email{ariadna.ribesmetidieri@ru.nl}
\affiliation{Institute for Mathematics, Astrophysics and Particle Physics, Radboud University, Heyendaalseweg 135, 6525 AJ Nijmegen, The Netherlands}

\author{B\'eatrice Bonga}
\email{bbonga@science.ru.nl}
\affiliation{Institute for Mathematics, Astrophysics and Particle Physics, Radboud University, Heyendaalseweg 135, 6525 AJ Nijmegen, The Netherlands}

\author{Badri Krishnan} \email{badri.krishnan@ru.nl}
\affiliation{Institute for Mathematics, Astrophysics and Particle
  Physics, Radboud University, Heyendaalseweg 135, 6525 AJ Nijmegen,
  The Netherlands}
\affiliation{Max-Planck-Institut f{\"u}r Gravitationsphysik (Albert
  Einstein Institute), Callinstra{\ss}e 38, 30167 Hannover, Germany}
\affiliation{Leibniz Universit{\"a}t Hannover, 30167 Hannover,
  Germany}

\begin{abstract}

  During the post-merger regime of a binary black hole merger, the
  gravitational wave signal consists of a superposition of
  quasi-normal modes (QNMs) of the remnant black hole.  
   It has been observed empirically,
  primarily through numerical simulations and heuristic arguments,
  that the infalling radiation at the horizon is also composed of a
  superposition of QNMs.  In this paper we provide an analytic
  explanation for this observation in the perturbative regime.
  Our analysis is based on a characteristic initial value
  formulation where data is prescribed on the horizon (modeled as a
  perturbed isolated horizon), and on a transversal null-hypersurface
  which registers the outgoing radiation.  This allows us to reformulate the
  traditional QNM problem in a fully 4-dimensional setting.
  Using a mode-decomposition, we
  demonstrate that the radiation modes crossing $\mathcal{H}$ are
  highly correlated with the outgoing modes crossing $\mathcal{I}$, and provide explicit expressions linking $\widetilde{\Psi}_0$ at the horizon with $\widetilde{\Psi}_4$ at null infinity. 
  \end{abstract}

\maketitle

\section{Introduction}
Binary black hole mergers are one of the most important natural
phenomena where non-perturbative and dynamical aspects of general
relativity play a key role.  The process by which two black hole
horizons merge and form a remnant black hole has several interesting
aspects which so far have been primarily understood through numerical
simulations.  Similarly, the calculations of the emitted gravitational
wave signal and developments of waveform models for the merger regime
typically requires numerical relativity results.  Apart from the
merger itself, in the inspiral and post-merger regimes, analytical
methods and black hole perturbation theory in particular, remain very
useful. In this paper, we shall model the perturbative post-merger
regime as a perturbed isolated horizon.  This is part of a general
quasi-local framework for studying black hole physics which has found
many applications in quantum and classical gravity, including
numerical relativity (see
e.g. \cite{Ashtekar:2004cn,Booth:2005qc,Gourgoulhon:2005ng,Hayward:2000ca}).

The eventual goal here is one of the most important themes in
gravitational wave astronomy, namely to obtain information about the
merging compact objects (and other gravitational wave sources) from
gravitational wave observations.  Achieving this goal requires a
detailed understanding of the problem using analytical or numerical
solutions of the Einstein field equations. When we infer the masses
and spins of the black holes from gravitational wave events, we
need a gravitational wave signal model that incorporates the
necessary effects with the appropriate degree of accuracy.  More
detailed information can be obtained in some
circumstances.  Black hole perturbation theory turns out to be useful
for this purpose in several cases.  Here we mention two situations of
particular relevance for this paper.  For both of these examples, the
horizon is very close to being \emph{isolated} in the sense that its
area is almost constant and it is useful to model the horizon as
a perturbed isolated horizon:
\begin{itemize}
\item Consider first extreme mass ratio inspirals (EMRIs), in which a small
  compact object, like a stellar-mass black hole or neutron star,
  spirals into a supermassive black hole. These systems are one of the
  targets of the space-based gravitational-wave detector LISA.  Given that the small
  object undergoes a large number of cycles in the field of the
  massive object, it might be possible to ``map'' the spacetime around
  the black hole and to infer its multipole moments with high accuracy
  \cite{Ryan:1995wh}.  Given that the multipole moments of a black
  hole are related to the geometry of the horizon (see
  e.g. \cite{Ashtekar:2004gp}), we would in effect be measuring the
  horizon geometry using observations of these extreme-mass-ratio
  systems.  To linear order in perturbation theory, the mass and
  angular momentum of the black hole do not change; these change only at
  second order in the amplitude of the radiation falling into the
  black hole. However, as we shall see, the higher multipole moments
  depend \emph{linearly} in the down-horizon wave
  amplitude. Therefore, given a sufficiently loud GW signal, the
  spacetime mapping project could be used to measure time-dependent
  black hole multipole moments.

\item The second example is the ringdown regime.  After the
  binary black hole merger, the gravitational wave signal is a
  superposition of damped sinusoids referred to as quasi-normal modes
  (QNMs), with frequencies and damping times determined by the remnant
  black hole parameters.  Sufficiently late in the post-merger regime,
  typically about $\sim 6-10 GM/c^3$ after the peak of the waveform
  \cite{Kamaretsos:2011um}, perturbation theory becomes applicable.
  QNMs are special perturbations of a Kerr black hole that satisfy
  purely dissipative boundary conditions
  \cite{vishveshwara:1970zz,chandrasekhar}. Given the dissipative
  nature of QNMs, they are prescribed by a complex frequency
  $\omega_{\ell mm}$ (or alternatively a frequency and damping time).
  The complex frequencies are labeled by three integers $(l,m,n)$:
  $\ell \geq 2$ and $|m|\leq \ell$ are the usual angular 
  harmonic
  numbers, while $n\geq 0$ is the overtone index
  associated with the radial part of the solution. The fundamental
  mode refers to $n=0$ while $n\geq 1$ are the overtones. For a Kerr
  black hole, the frequencies $\omega_{\ell mn}(M,a)$ have been
  calculated and tabulated \cite{Leaver:1985ax,berti_ringdown}; see
  also \cite{Berti:2009kk,Kokkotas:1999bd} for reviews.
  
  QNMs are seen both numerically in simulated binary black hole
  spacetimes (see e.g. \cite{Giesler:2019uxc,Baibhav:2023clw}) and
  observationally in data from the LIGO and Virgo observatories (see
  e.g. \cite{LIGOScientific:2016lio,LIGOScientific:2019fpa,LIGOScientific:2020tif,Isi:2019aib,Capano:2021etf}).
  If the notion of correlations mentioned above is viable, then these
  must be evident in the infalling radiation at the remnant black
  horizon as well.  This has been empirically shown to be the case in
  numerical simulations \cite{Mourier:2020mwa,Khera:2023oyf}, but an
  analytical proof is still missing.  This paper aims to fill this gap
  in the literature.  We shall show that, under certain reasonable
  assumptions within linear perturbation theory and in the situation
  when there is no external incoming radiation from past null
  infinity, the QNMs do indeed appear in the infalling radiation and in fact they do so with the same well-known frequencies and
  damping times mentioned above.\footnote{To some extent, these results could have been anticipated from the Teukolsky-Starobinsky identities and/or the metric reconstruction procedure. Here, this link is made concrete and explicit.}  
\end{itemize}

Our work is based on the characteristic initial value formalism (see
e.g. \cite{1990RSPSA.427..221R,Chrusciel:2012ap,Friedrich:1983vi,Friedrich:2000qv,Friedrich:1983vi}).
In a companion paper \cite{metidieri2024tidal} (henceforth referred to
as ``Paper I''), the application of this formalism in the context of
tidally perturbed isolated horizons has been explained in great detail
(see also
\cite{Lewandowski:1999zs,Lewandowski:2018khe,Lewandowski:2000nh,Dobkowski-Rylko:2018ahh,Lewandowski:2014nta,Flandera:2024awl,Scholtz:2017ttf,Flandera:2016qwg}).
In Paper I, the horizon
was required to be exactly isolated so that the infalling horizon
fluxes vanish identically. 
Here we drop this
restriction.
This allows us to pursue the idea of accessing the
horizon geometry with gravitational wave observations by a method
sometimes referred to as ``black hole tomography''
\cite{Ashtekar:2021kqj}.  The underlying idea is that, when there is no
incoming external radiation from past null infinity, both the
infalling and outgoing radiation must be generated by the spacetime
dynamics in the vicinity of the black hole.  This indicates the
possibility of correlations between the gravitational wave fluxes
observed by gravitational wave detectors, and the flux of infalling
radiation across the black hole horizon
\cite{Rezzolla:2010df,Jaramillo:2011rf,Jaramillo:2011re,Alic:2012df,Jaramillo:2012rr}.
There is now strong numerical evidence that such correlations do exist
and can be used as a probe of the horizon dynamics.

Let us illustrate this for the black hole ringdown example.  
Observations of the QNMs by current and future gravitational-wave observatories provide a possible avenue of
observationally testing the nature of the remnant object.  The newly
formed remnant horizon is initially distorted in the sense that
horizon geometry is markedly different from the Kerr
geometry.  These differences can be quantified by appropriate horizon
multipole moments.  The horizon loses these distortions as it
approaches its final Kerr state by \emph{absorbing} just the right
amount of gravitational radiation.  This situation is depicted in
Fig.~\ref{fig:horizon-late-time}.  
There are two aspects of this ringdown process that we wish to highlight:
\begin{itemize}
\item First, at sufficiently late times, the horizon geometry can be
  considered to be a perturation of a Kerr horizon.  This perturbation
  consists of two distinct, though related, aspects: i) the higher
  horizon multipole moments are perturbed away from the appropriate
  Kerr values; ii)
  there is a small amount of infalling radiation.  
\item Secondly, the properties of the infalling radiation are mirrored by the properties of the outgoing radiation that
  reaches future null infinity $\mathcal{I}^+$ and can potentially be detected by a gravitational-wave observatory.  This is an empirical
  observation seen in various numerical simulations.  A general proof
  of this empirical observation would establish the existence of
  correlations between the observed radiations and horizon dynamics;
  we could then infer properties of the horizon (which is otherwise
  inaccessible to us classically) based on gravitational-wave
  observations.  It is evident that the details of these correlations must depend on the dynamical equations.  Thus,
  when we infer properties of the horizon based on observations, we
  are assuming the validity of the Einstein equations and our
  conclusions about the infalling radiation would differ in other
  theories of gravity.
\end{itemize}

Given this context, we can now state the main results of this
paper. We reformulate the QNM problem using the Newman-Penrose formalism tailored to isolated horizons of perturbed slowly spinning black holes. This formalism can be viewed as Schwarzschild perturbation theory using Newman-Penrose language with ``atypical'' gauge conditions. However, the appeal to the isolated horizon framework (which may seem cumbersome for the uninitiated) is exactly what motivated these peculiar, yet powerful, gauge conditions.  This reformulation also sheds light on the minimal conditions required for QNM solutions to appear: the
Schwarzschild QNM frequencies are obtained when we impose that the solutions to the perturbed Weyl scalars are analytic and that the spacetime is stable towards the future. This naturally selects solutions with no incoming radiation from past null infinity; that there is no radiation from the horizon is already embedded in our construction from the start. 

Second, we show that
the outgoing field $\Psi_4$ on $\mathcal{N}$ (which can be taken to coincide with $\mathcal{I}^+$) is determined by $\Psi_0$
on the horizon, and for the QNM case, the same modes appear in
both places. We also provide an explicit formula linking the amplitudes of $\Psi_0$ at the horizon and those of $\Psi_4$ at $\mathcal{N}$ . This can be considered as analytical evidence for black hole tomography in this perturbative context.  We also clarify the link with the Teukolsky-Starobinsky identities which also relate $\Psi_0$ and $\Psi_4$. 
Finally, we also provide explicit expressions for $\widetilde{\Psi}_2$, which encodes geometric information about the horizon.

The plan for the rest of this paper is as follows.
Sec.~\ref{sec:perturbed-ih} reviews the standard formulations of the QNM problem and provides the necessary background for
understanding perturbed isolated horizons. In Sec.~\ref{sec:perturbed-intrinsic-geometry}, we provide the explicit expressions describing the perturbed isolated horizon by a small flux of infalling gravitational radiation.
In Sec.~\ref{sec:psi0-psi4}, we solve these expressions explicitly on the horizon and we relate our results to the Teukolsky-Starobinksy identities.
Sec.~\ref{sec:QNM-frequencies} specializes to the QNM problem and in Sec.~\ref{sec:tomography} we show the explicit link between the horizon amplitudes of $\Psi_0$ with the amplitudes of $\Psi_4$ at null infinity.  Finally,
Sec.~\ref{sec:multipole-moments} describes the Weyl scalar $\Psi_2$. We conclude in Sec.~\ref{sec:conclusions}. In the appendices, we have collected supplementary material.

\section{Background}
\label{sec:perturbed-ih}

In this section, we collect the necessary background material which
will be used in this paper. We begin with the standard formalism
for defining and calculating QNMs within black hole perturbation
theory. This is followed by setting up our calculational framework
for perturbations of isolated horizons, which will later lead to a
reformulation of QNMs.

\subsection{Formulations of the quasi-normal mode problem}
\label{subsec:standard-qnm}
QNMs arise in various physical situations such as
optical cavities, acoustics and bound quantum states (see
e.g. \cite{dyatlov2019mathematical}).  These can be related to a
scattering problem from a potential satisfying purely dissipative
boundary conditions. We consider the
one-dimensional wave equation with a potential $V(x)$
\begin{equation}
  \label{eq:wave-eq}
  \left(\partial_t^2 -  \partial_x^2 + V(x)\right) \psi(t,x) = 0\,.
\end{equation}
QNMs correspond to solutions of the above
equation which are purely outgoing at
$x\rightarrow \pm\infty$, i.e.
\begin{subequations}\label{eq:qnm-boundary}
    \begin{eqnarray}  
    \psi &\sim& e^{-i\omega(t-x)} \qquad x\rightarrow \infty\,, \\
  \psi &\sim& e^{-i\omega(t+x)} \qquad x\rightarrow -\infty \,.
\end{eqnarray}
\end{subequations}
These boundary conditions imply that the system is dissipative with no
in-coming energy from any part of the boundary.  Such solutions turn
out to exist only for a discrete set of complex frequencies $\omega_j$
(labeled here by an integer $j$), which are independent of the initial
data. Modes with $\im[\omega_j]<0$ correspond to decaying
solutions.  These frequencies are referred to as either QNM or
resonant frequencies.  Alternatively, resonant frequencies are also
defined as poles of the Green's function (first suggested in
\cite{vishveshwara1970stability}). The importance of the QNMs lies in
the fact that they determine the behavior of the solution of
Eq.~(\ref{eq:wave-eq}) at late times once all other waves have
dissipated away from the system.  Rigorous results are known for
potentials of compact support (see
e.g. \cite{dyatlov2019mathematical}). In these cases, it can be shown
that if $\psi(t,x)$ is a solution of Eq.~(\ref{eq:wave-eq}) with
sufficiently localized initial data then at late times
\begin{equation}
  \label{eq:resonantexpansion}
  \psi(t,x) = \sum_{\im[\omega_j] > -\Omega} e^{-i\omega_j t}\psi_j(x) + \mathcal{O}(e^{-t\Omega})\,.
\end{equation}
Here $\Omega$ is a positive real number, and the sum is over the least
damped QNMs.  Thus, at late times the solution can be written as an
expansion over the QNMs (i.e. as damped sinusoids as far as their time
evolution is concerned), and the expansion is dominated by the longest
lived QNMs.

Following \cite{chandrasekhar,Chandrasekhar:1985kt} the above
formalism is directly applicable to the QNMs of a black hole.
Consider a Schwarzschild black hole of mass $M$ with the usual metric
outside the event horizon
\begin{equation}
  ds^2 = -\left(1-\frac{2M}{\mathfrak{r}}\right)dt^2 + \left(1-\frac{2M}{\mathfrak{r}}\right)^{-1}d\mathfrak{r}^2 + \mathfrak{r}^2d\Omega^2\,,
\end{equation}
where $d\Omega^2 = d\theta^2 + \sin^2\theta d\phi^2$ and $\{t\,,\mathfrak{r}\,, \theta\,, \phi\}$ are the usual Schwarzschild coordinates. 
Following an
angular mode decomposition of the perturbations to this metric, we
obtain wave-equations of the form of Eq.~\eqref{eq:wave-eq} for the
radial functions. The coordinate $x$ in this case is the ``tortoise
coordinate'' $dr_\star = (1-2M/\mathfrak{r})^{-1}d\mathfrak{r}$. The potentials are the well-known Regge-Wheeler and Zerilli potentials for the axial and polar
perturbations, respectively.  Applying the boundary conditions of
Eq.~\eqref{eq:qnm-boundary} leads to the QNM frequencies
$\omega_{\ell m n}(M)$ for a Schwarzschild black hole.  A similar but
somewhat more complicated analysis leads to the QNM frequencies
$\omega_{\ell mn}(M,a)$ of a Kerr black hole of mass $M$ and angular
momentum $J$, with $a = J/M$.

\begin{figure}
    \centering    \includegraphics[width=1\columnwidth]{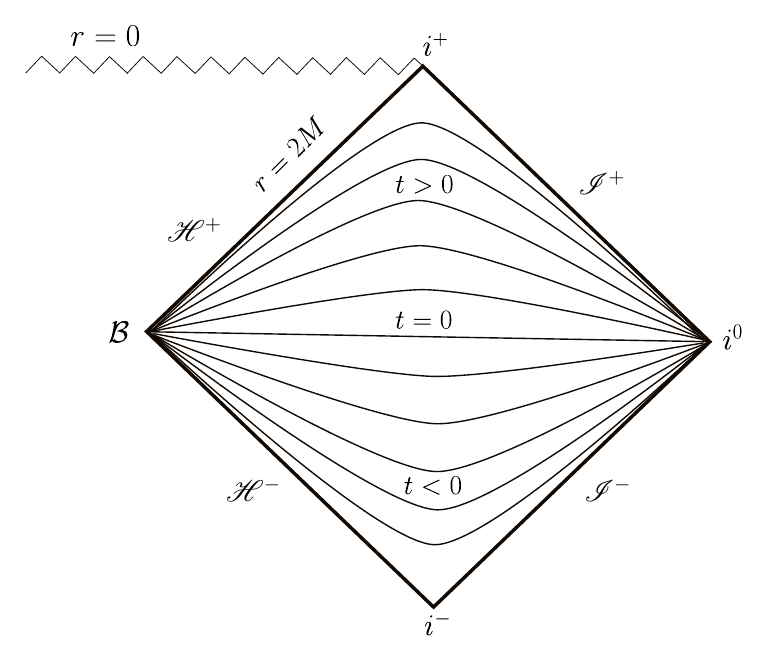}
    \includegraphics[width=1\columnwidth]{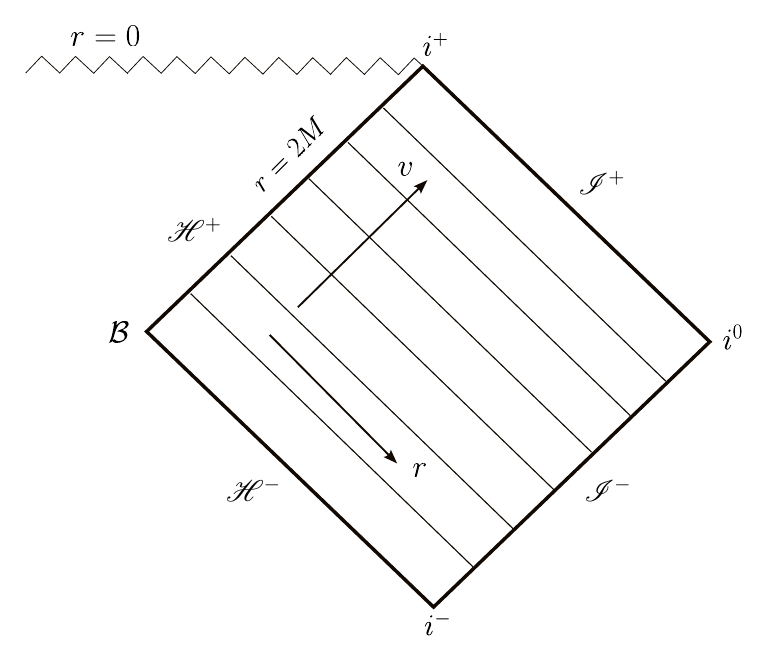}
    \caption{This figure shows a portion of the maximally extended
      Schwarzschild spacetime with two different slicings. The top panel shows the
      constant time slices of Schwarzschild in the traditional
      formulation of black hole perturbation theory.  The future
      horizon is $\mathcal{H}^+$, the past horizon (the white hole) is
      $\mathcal{H}^-$, while past and future null infinity are
      $\mathcal{I}^-$ and $\mathcal{I}^+$, respectively. The
      singularity is at $r=0$.  All the constant time slices end up at
      the bifurcate cross-section $\mathcal{B}$ of the horizon where
      $r_\star\rightarrow -\infty$. The lower panel shows the same
      portion of spacetime but now in the ingoing
      Eddington-Finkelstein coordinates $(v,r)$.  The constant $v$
      surfaces shown in the conformal diagram are null surfaces and $r$ is a radial coordinate along these null surfaces. }
    \label{fig:schw-slices}
\end{figure}

While the above formalism is widely used, it is not entirely
satisfactory when we wish to study the horizon; the tortoise
coordinate $r_\star$ is not horizon penetrating since
$r_\star\rightarrow -\infty$ when $\mathfrak{r}\rightarrow 2M$.  From a 4-dimensional perspective, the constant time slices for the
Regge-Wheeler and Zerilli equations correspond to the spacelike
hypersurfaces shown in Fig.~\ref{fig:schw-slices}.  It is clear that
these slices are not suitable for studying fluxes across the future
horizon $\mathcal{H}^+$ or future null infinity $\mathcal{I}^+$, nor
correlations between these fluxes.  

There have been several efforts to generalize the above discussion in horizon penetrating coordinates (see e.g. \cite{Sarbach:2001qq,Preston:2006ze} for the use of such coordinates in the more general context of black hole perturbation theory).  One of
the most developed formalisms is the use of hyperboloidal slices which
intersect both $\mathcal{H}^+$ and $\mathcal{I}^+$; see
e.g. \cite{Zenginoglu:2007jw,PanossoMacedo:2024nkw,Zenginoglu:2024bzs,PanossoMacedo:2018hab,PanossoMacedo:2023qzp}.
These studies have been used primarily for numerical studies though it is certainly plausible that these could be used for analytical studies as well.

Here we shall pursue a different approach and prescribe data on null
surfaces rather than spacelike surfaces. Since one of our goals
is to study possible correlations between fluxes
across $\mathcal{H}^+$ and $\mathcal{I}^+$, we would like to prescribe
data directly on these surfaces and use these data to construct the
near-horizon spacetime.  Since both $\mathcal{H}^+$ and
$\mathcal{I}^+$ are null surfaces, we are naturally led to the
characteristic initial value formulation. Recent work by Mongwane \textit{et al}
\cite{Mongwane:2024vao} provides an important step in this direction.
Working with the Bondi-Sachs \cite{Bondi:1962px} or Newman-Unti
\cite{Newman:1962cia} form of the metric using the coordinates
$(u,\mathfrak{r},\theta,\phi)$ where $u$ is the outgoing null coordinate
$u = t-r_\star$, Mongwane \textit{et al} show that the Schwarzschild
QNM frequencies are recovered.  In their analysis, the boundary
condition at the horizon appears in the behavior of certain metric
coefficients as $\mathfrak{r}\rightarrow 2M$.  We extend this analysis in three directions:
\begin{enumerate}
\item The inner boundary is taken to be a non-extremal isolated
  horizon. This allows us to extend the analysis to black holes with  more general multipole moment structure, and allows us to study  the infalling radiative flux across the horizon.  
\item We shall see that the QNMs arising in our analysis nicely extend to past null infinity and it is clear that they satisfy the no-incoming radiation condition. 
  \item  We use the Newman-Penrose formalism instead of metric formalism, which should allow for an easier generalization to the Kerr spacetime.
\end{enumerate}
We will use the analog of the Bondi-Sachs coordinates near
$\mathcal{I}^+$, but now adapted to the future horizon
$\mathcal{H}^+$. The ingoing Eddington-Finkelstein coordinates
$(v,\mathfrak{r},\theta,\phi)$ for the Schwarzschild coordinates
provide the prototypical example.  Here $v=t+r_\star$ and the metric
in these coordinates is
\begin{equation}
  ds^2 = -\left(1-\frac{2M}{\mathfrak{r}}\right)dv^2 + 2dv\,d\mathfrak{r} + \mathfrak{r}^2d\Omega^2\,.
\end{equation}
We shall see that the above construction will yield a more transparent
formulation of the QNM problem.  In other respects however, the results
are equivalent to the standard formulation.  In particular, we shall obtain, as in the standard formulation, solutions which are divergent
at spatial infinity. 

\subsection{Perturbed Isolated Horizons}
\label{subsec:perturbed-ih}

\begin{figure}
    \centering
 \begin{tikzpicture}[decoration=snake]
    \node (center) at (0,0) {};
    \shadedraw[top color=wuppergreen, bottom color=gray, opacity=0.2] (-2.5,1.5)--(0,4)--(2.5,1.5) to[bend left] (0,1.5) to[bend right] (-2.5,1.5);
    \draw[very thick] (-4,0)--(0,4);
    \draw[very thick] (0,4)--(4,0);
    \draw[very thick, dashed] (4,0)--(0,-4);
    \draw[->,decorate,thick,rotate=45] (1.5,0)--(2.5,0);
    \draw[->,decorate,thick,gray,rotate=-45] (-1.5,0)--(-2.5,0);
    \draw[->,decorate,thick,gray,rotate=-45] (-1.5,1)--(-2.5,1);
    \draw[->,decorate,thick,gray,rotate=-45] (-1.5,-1)--(-2.5,-1);
    \draw[->,decorate,thick,rotate=45] (1.5,1.)--(2.5,1.);
    \draw[->,decorate,thick,rotate=45] (1.5,-1.)--(2.5,-1.);
    \node (delta) at (-2.5,2.5) {\huge $\Delta$};
    \node (delta) at (2.5,2.5) {\huge $\mathcal{N}$};
    \node (delta) at (2.5,-2.5) {\huge $\mathcal{I}^-$};
    \node (psi0) at (-2.3,0.1) {\Large $\widetilde\Psi_0$};
    \node (psi4) at (2.3,0.1) {\Large $\widetilde\Psi_4$};
\end{tikzpicture}
\caption{Here $\Delta$ is a null surface (a perturbed isolated
  horizon) and the null surface $\mathcal{N}$ transverse to it can be
  seen as an approximate version of future null infinity.  In
  particular, the space of all allowed gravitational fields living in
  the shaded region is such that $\Delta$ is a perturbed isolated
  horizon. The Weyl tensor component $\Psi_4$ is responsible for the
  flux crossing $\mathcal{N}$ while the Weyl tensor component $\Psi_0$
  is responsible for fluxes across $\Delta$. }
    \label{fig:horizon-late-time}
\end{figure}
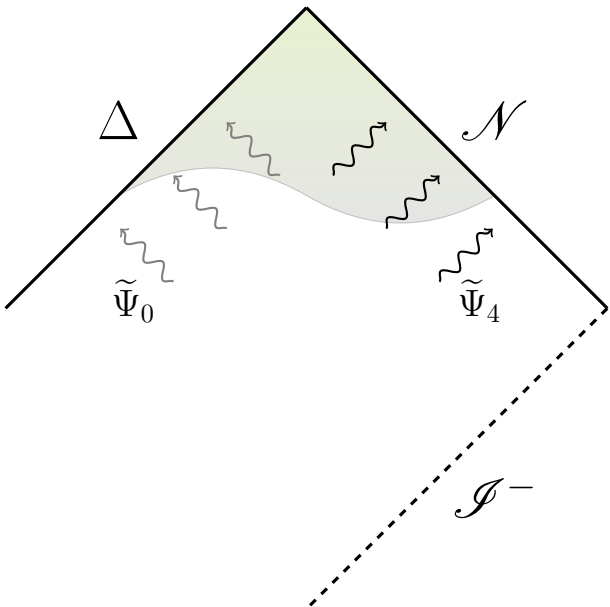

The notion of isolated horizons, meant to model a black hole in
equilibrium in an otherwise dynamical spacetime, is discussed in paper
I. These horizons do not have any gravitational wave fluxes crossing
the horizon.  Here we relax this condition, include dynamical
perturbations and allow for small amounts of infalling radiation.
Before discussing these perturbed isolated horizons, let us first
consider isolated horizons.  It is not our purpose to repeat the
various definitions here; readers can consult paper I for that
purpose.  Here we mostly aim to set up notation.  Readers familiar
with black hole perturbation theory can think of the Schwarzschild
horizon, but modified with small amounts of shear for the null
generator of the horizon.  The gauge conditions we employ here will
generally be tied to properties of the horizon, and this differs from those used typically, e.g. \cite{Chandrasekhar:1985kt,Martel:2005ir}. Readers
familiar with the Newman-Penrose formalism will also find this
discussion familiar since we will use it extensively. A brief summary of the Newman-Penrose formalism is provided in
Appendix~\ref{subsec:np-nutshell}.

There are several relevant definitions here, with increasingly
stringent conditions: Non-Expanding Horizon (NEH), weakly isolated
horizon (WIH) and isolated horizon (IH).  The basic object is a null
surface $\Delta$ with topology $S^2\times\mathbb{R}$.  The
``$S^2$ part'' of the surface are Riemannian manifolds and
correspond to cross-sections of the horizon, while the ``$\mathbb{R}$
part'' are null curves.  The degenerate metric on $\Delta$ is denoted
by $q_{ab}$, and $\ell^a$ denotes a future-directed null-normal
(unique up to rescalings by positive functions).  There is a
derivative operator $D_a$ on $\Delta$ compatible with $q_{ab}$, and it
is the pull-back of the spacetime derivative operator
$D_a = \pullbacklong{\nabla_a}$; an arrow on a covariant index,
e.g. $\pullbacklong{X_a}$, denotes the pullback to $\Delta$ and can
then only be contracted with vector fields tangent to $\Delta$.  The
1-form $\omega_a$ is a part of $D_a$ and is defined as 
\begin{equation}
  D_a\ell^b = \omega_a\ell^b\,.
\end{equation}
No conditions are imposed on $D_a$ (or $\omega_a$) on a NEH, on a WIH $\omega_a$ is time independent, while on a IH $D_a$ is time
independent. The surface gravity associated with $\ell^a$ is
\begin{equation}
  \kappa_{(\ell)} = \omega_a\ell^a\,.
\end{equation}
The 1-form $\omega_a$ projected to a cross-section determines the angular momentum and higher
spin moments, while the curvature of the induced metric on the cross-section determines
the mass multipole moments.  Moreover, 
$\omega_a$ is entirely determined by its curl $D_{[a}\omega_{b]}$ and
the divergence $D_a\omega^a$.  The curl
contains all the information about the angular momentum and higher spin multipole moments,
while the divergence is a ``gauge'' quantity and specifying it determines the foliation.

Pick a cross-section $S_0$, and introduce angular coordinates
$(\theta,\phi)$ or holomorphic coordinates $(z,\bar{z})$ on it.  Use $\ell^a$ to transport these coordinates everywhere on
$\Delta$.  Now construct the past-directed null geodesics starting
with $-n^a$ at $\Delta$, and let $r$ be the affine parameter such that $n^a\nabla_a r = -1$. Finally, use parallel transport along $-n^a$ to obtain a coordinate system $(v,r,\theta, \phi)$ or $(v,r,z,\bar{z})$ and a null tetrad in a neighborhood of $\Delta$.

The phase space here is the spacetime $(M,g_{ab})$ of all solutions of
the Einstein field equations such that $\Delta$ is an isolated
horizon.  This is tailored to the situation shown in
Fig.~\ref{fig:horizon-late-time} for a black hole approaching
equilibrium.  This could be the remnant black hole formed in a binary
black hole merger, or from gravitational collapse.  It could however
also be modified for a black hole which was isolated in the past, such as
in the inspiral phase of a binary black hole merger (in this case the
black holes would be Kerr black holes in the distant past).  For such
situations, we would consider \emph{past} directed null normals which
are complete in the distant past.

For an isolated horizon, both the expansion $\theta_{(\ell)}$ and
shear $\sigma^{(\ell)}_{ab}$ vanish identically, and for a perturbed
isolated horizons they are both small.  Following
\cite{Ashtekar:2021kqj,Hawking:1972hy}, it turns out that we can still
take surfaces with vanishing expansion but with non-vanishing shear.
{{To see this, start with the Raychaudhuri equation in vacuum for the
affinely parameterized null-generators of $\Delta$ (now including also
terms with non-vanishing expansion):
\begin{equation}
  \mathcal{L}_\ell\theta_{(\ell)} = \kappa_{(\ell)}\theta_{(\ell)} -\frac{1}{2}\theta_{(\ell)}^2 - |\sigma_{(\ell)}|^2 \,.
\end{equation}
For a perturbed isolated horizon, $\theta_{(\ell)}$ is small.  If
$\theta_{(\ell)}$ were to vanish exactly, then the Raychaudhuri
equation tells us that the shear should also vanish.  This implies
that $\sigma^{(\ell)}_{ab}$ is also of the first order of smallness on
a perturbed isolated horizon.  $\kappa_{(\ell)}$ is the acceleration
of $\ell^a$ and it is the surface gravity.  However, we are also free
to re-parameterize the geodesic to choose $\kappa_{(\ell)}$ as
convenient.  For the moment, following \cite{Hayward_1993}, it will be
convenient to choose $\kappa_{(\ell)} = \theta_{(\ell)}/2$ so that
\begin{equation}
  \mathcal{L}_\ell\theta_{(\ell)} = -|\sigma_{(\ell)}|^2 \,.
\end{equation}
In this form, it is evident that within first-order perturbation theory, we
must have
\begin{equation}
  \mathcal{L}_\ell\theta_{(\ell)} = 0\,.
\end{equation}
Assuming further that the expansion vanishes asymptotically at late
times, it must vanish everywhere on $\Delta$ to first order. However,
$\sigma^{(\ell)}_{ab}$ can now be non-vanishing. In fact,
non-vanishing shear represents the infalling flux of gravitational
waves. }} In paper I we considered perturbations where the shear
vanishes and so the horizon is still, exactly, an isolated horizon.
The horizon multipole moments were allowed to vary. Here we allow
for a non-vanishing shear as well.

\section{Perturbations of the intrinsic horizon geometry}
\label{sec:perturbed-intrinsic-geometry}

Before we can formulate the QNM problem, the first necessary
ingredient is to specify the boundary data on the inner boundary,
i.e. the horizon. To this end, we need to specify a perturbed isolated horizon as discussed earlier.  In particular, we need to specify the free data and to determine the rest of the horizon geometry from this free data.  The hurried reader can skip the details of this construction and look at the results in Eqs.~\eqref{eq:eqs-horizon} and \eqref{eq:eqs-constraint}. 

In
the following, we shall distinguish between two kinds of perturbations of a stationary black hole: tidal vs. radiative.  Tidal perturbations
as discussed in paper I, are time independent, and modeled as a
variation of the horizon multipole moments. These variations are
time independent and meant to model a well-separated binary system
where the perturbation changes sufficiently slowly.  Within this
approximation, the horizon area, angular momentum and higher
multipoles all remain time independent. This is however an
approximation and a more accurate model includes the phenomenon of
tidal heating, i.e. incorporating infalling fluxes of gravitational
radiation across the horizon.

Let us now consider the necessary modifications to the horizon
structure when we include infalling radiation.  The tidal distortions
represent, within phase space, a variation within the subspace of
spacetimes admitting an isolated horizon; here we will go away from
this subspace.  There are two important fields that we need to
consider at the horizon:
The shear of $\ell^a$ is now allowed to be non-vanishing, and we shall keep terms
linear in this quantity. A second, and related quantity is the Weyl
tensor component $\Psi_0$ which again, vanishes identically on a
isolated horizon and here will be taken to be of the first order of
smallness.

Notationally, the perturbations to the spin coefficients, Weyl
scalars, and tetrad components sourced by this radiative perturbation
will be denoted with the symbol $\widetilde{\quad}$. The background
quantities will be denoted without any symbol unless otherwise
explicitly specified. From this point onward, the analysis will be
perturbative to first order in the radiative perturbation. So,
whether a quantity without the $\widetilde{\quad}$ symbol denotes a
general quantity or its background value should be clear from the
context. We will develop the following analysis for perturbations of a
slowly rotating isolated horizon, as the background quantities
specified in App.~\ref{subsec:gauge-choices} have been set to zero to
simplify the analysis. However, notice that the same formalism applies
as long as the background is taken to be an isolated horizon in the
sense of Paper I, and in particular, the same discussion for the
equations at the horizon goes through with very little modifications
for a general isolated horizon (i.e., the Kerr isolated horizon).  As
in paper I, we shall follow the Newman-Penrose formalism 
and use the
symbol ``$\triangleq$'' to denote equalities which are satisfied only at the horizon.   

Let us briefly recall the construction of paper I.  There, we started with
prescribing a perturbation of the Weyl tensor component $\Psi_2$ on a
single cross-section $S_0$, which corresponds to a tidal
perturbation of the horizon multipole moments.  The field equations on an isolated horizon then guaranteed that $\Psi_2$ is time independent on the horizon: $D\Psi_2=0$. Introducing now a non-vanishing $\Psi_0$ at the horizon, it turns out that
$\Psi_2$
may no longer be time independent at the horizon, and this dynamical
evolution will be forced by $\Psi_0$. 

Using the gauge discussed in Paper I, and summarized in
App.~\ref{sec:field-equations}, the perturbations to the tetrad
components simplify greatly at the horizon.  In particular since
$\ell^a$ continues to have vanishing expansion at the horizon, we can
define the null coordinate $v$ as before following
$\ell^a\nabla v = 1$. On the other hand, as we have just argued, the
angular dyad $(m^a,\bar{m}^a)$ will be perturbed as the geometry of
the cross-sections is modified by the infalling radiation. Thus:
\begin{equation}
  \widetilde{\ell}^a\triangleq 0\,,\quad\widetilde{m}^a\partial_a \triangleq\widetilde{\xi}^z\partial_z+\widetilde{\xi}^{\bar{z}}\partial_{{\bar{z}}}\,.
\end{equation}
The radial coordinate $r$ is the affine parameter along $-n^a$ and
this continues to be true even in our present case; thus the
radial coordinate $r$ and $n^a$ ``adapt'' to the infalling
radiation so that we can take $n^a$ to be unperturbed everywhere:
\begin{equation}
    \widetilde{n}^a=0\,.
\end{equation} 
This coordinate is normalized such that it takes the value $r=0$ at
the horizon. Notice that this radial coordinate is related to the radial Schwarzschild coordinate $\mathfrak{r}$ through $\mathfrak{r}=r+c$. The parameter $c$ defined in Paper I, which appears in
the background quantities (see Appendix~\ref{sec:field-equations})
represents the size of the unperturbed horizon, i.e.,
\begin{equation}
  c=1/(2\kappa_{(\ell)}) = R_0=2M
\end{equation}
for a Schwarzschild background. The quantity $c$ is related to the area radius of the horizon. In the simple case of a Schwarzschild black hole, this quantity coincides with the horizon radius. However, for a rotating black hole, the area radius and the horizon position differ~\footnote{For Kerr with mass  $M$ and angular momentum $M \, a$, the horizon is located at $r_+=M + \sqrt{M^2-a^2}$ while the area radius is $c=\sqrt{2M r_+}$.}.  Hence, we use this quantity here instead of the usual $2M$ factor to ease the generalization to more general backgrounds.

Away from the horizon, we can no longer take $\ell^a$ to be
unperturbed.  More generally, away from the horizon, the
remaining tetrad elements are perturbed so that
\begin{align}
    \widetilde{\ell}^a \partial_a= \widetilde{U} \partial_r +\widetilde{X}^z\partial_z+\widetilde{{X}}^{\bar{z}} \partial_{\bar{z}}\,,\\
     \widetilde{m}^a \partial_a= \widetilde{\Omega} \partial_r +\widetilde{\xi}^z\partial_z+\widetilde{\bar{\xi}}^{\bar{z}} \partial_{\bar{z}}\,.
\end{align}
Given our gauge conditions, it follows that the perturbation to
$\Psi_1$ needs to be of the same order of smallness as
$\widetilde{\Psi}_0$ so the field equations at the horizon are
self-consistent. Gathering the
perturbative versions of the angular field equations (see
Eq.~\eqref{eqs:angular} in Appendix~\ref{sec:field-equations}) with
$\widetilde{\Psi}_1\sim \mathcal{O}(\widetilde{\Psi}_0)$, we obtain the
following set of equations which need to hold at the horizon
\begin{subequations}\label{eq:angular-perturbed-2}
\begin{align}\label{eq:angular-perturbed-2:1}
     \bar{\eth}\widetilde{\sigma} \triangleq &\widetilde{\Psi}_1\\\label{eq:angular-perturbed-2:2}
     \eth \widetilde{\lambda} -\bar{\eth}\widetilde{\mu} \triangleq & \mu \widetilde{\pi} -\widetilde{\Psi}_3\\\label{eq:angular-perturbed-2:3}
     -2\re \widetilde{\Psi}_2 \triangleq & \eth\widetilde{a} + \bar{\eth} \widetilde{\bar{a}}+\widetilde{\eth}a + \widetilde{\bar{\eth}}\bar{a}\\\label{eq:angular-perturbed-2:4}
    -2i \im \widetilde{\Psi}_2  \triangleq &\eth \widetilde{\pi} -\bar{\eth}\widetilde{\bar{\pi}}\,.
\end{align}
\end{subequations} 
Repeating the same procedure for the Weyl scalar's evolution equations~\eqref{eqs:evolution-bianchi} 
 yields
\begin{subequations}\label{eq:time-evolution-Weyl-perturbed}
    \begin{align}\label{eq:time-evolution-Weyl-perturbed:1}
        D\widetilde{\Psi}_1 -\kappa_{(l)} \widetilde{\Psi}_1\triangleq & \bar{\eth} \widetilde{\Psi}_0\\\label{eq:time-evolution-Weyl-perturbed:2}
        D\widetilde{\Psi}_2 \triangleq &\bar{\eth} \widetilde{\Psi}_1\\\label{eq:time-evolution-Weyl-perturbed:3}
        D\widetilde{\Psi}_3 +\kappa_{(l)} \widetilde{\Psi}_3 \triangleq &\bar{\eth} \widetilde{\Psi}_2 +3\widetilde{\pi}\Psi_2\\\label{eq:time-evolution-Weyl-perturbed:4}
        D\widetilde{\Psi}_4 +2\kappa_{(l)} \widetilde{\Psi}_4 \triangleq & \bar{\eth} \widetilde{\Psi}_3 -3\widetilde{\lambda} \Psi_2\,.
    \end{align}
\end{subequations} Notice that the last two equations can be written as second-order differential equations so that it is clear that $\widetilde{\Psi}_3$ and $\widetilde{\Psi}_4$ are sourced by $\widetilde{\Psi}_1$ and $\widetilde{\Psi}_2$ respectively.~\footnote{Start by taking the $D$ derivative of Eqs.~\eqref{eq:time-evolution-Weyl-perturbed:3} and~\eqref{eq:time-evolution-Weyl-perturbed:3}. We can use the fact that the $D$ and the $\eth$ derivatives commute for the unperturbed quantities (see Eq.~\eqref{eqs:commutation}) to rewrite the terms $D\bar{\eth} \tilde{\Psi}_2=\bar{\eth} D\tilde{\Psi}_2$ and $D\bar{\eth} \tilde{\Psi}_3=\bar{\eth}D \tilde{\Psi}_3$. Using Eqs.~\eqref{eq:time-evolution-Weyl-perturbed:2}, Eq.~\eqref{eq:time-evolution-spin-perturbed:2} and the gauge $\tilde{\epsilon}\triangleq 0$, we obtain Eq.~\eqref{eq:second-order-psi3-psi4:1}. Eq.~\eqref{eq:second-order-psi3-psi4:2} follows from using Eqs.~\eqref{eq:time-evolution-Weyl-perturbed:2} and~\eqref{eq:time-evolution-spin-perturbed:4}.  Expanding all the terms and collecting the terms proportional to $\bar{\eth} \tilde{\Psi}_3 -3\tilde{\lambda} \Psi_2$, we can the left-hand side of Eq.~\eqref{eq:time-evolution-Weyl-perturbed:4} to simplify the expression and obtain Eq.~\eqref{eq:second-order-psi3-psi4:2}.}
\begin{subequations}\label{eq:second-order-psi3-psi4}
    \begin{align}\label{eq:second-order-psi3-psi4:1}
        D^2\widetilde{\Psi}_3+\kappa_{(l)}D \widetilde{\Psi}_3 \triangleq &\bar{\eth}^2 \widetilde{\Psi}_1 +3\Psi_2\widetilde{\bar{\Psi}}_1\\\label{eq:second-order-psi3-psi4:2}
        D^2 \widetilde{\Psi}_4 +3\kappa_{(l)} D\widetilde{\Psi}_4 +2\kappa_{(l)}^2\widetilde{\Psi}_4\triangleq & \bar{\eth}^2 \widetilde{\Psi}_2 -3\mu \Psi_2 \widetilde{\bar{\sigma}}
    \end{align}
\end{subequations}
Finally, we can again rewrite the perturbative evolution equations for the spin coefficients at the horizon. The first equation in~\eqref{eqs:timeevolution} yields
\begin{equation}
    \bar{\eth} \widetilde{\kappa}\triangleq 0\,.
\end{equation} This expression implies that the $\widetilde{\kappa}$ is only a function of the time coordinate $v$. Nevertheless, in our construction of the perturbed horizon, we would like the perturbed null generators at the horizon $\ell$ to still be geodesic. Hence, we choose
\begin{equation}
    \widetilde{\kappa} \triangleq 0\,.
\end{equation} The remaining equations in~\eqref{eqs:timeevolution} yield
\begin{subequations}\label{eq:time-evolution-spin-perturbed}
    \begin{align}\label{eq:time-evolution-spin-perturbed:1}
        D\widetilde{\sigma} -\kappa_{(l)} \widetilde{\sigma} \triangleq& \widetilde{\Psi}_0\\\label{eq:time-evolution-spin-perturbed:2}
          D \widetilde{\pi}  \triangleq & \bar{\delta}(\widetilde{\epsilon} + \widetilde{\bar{\epsilon}}) +\widetilde{\bar{\Psi}}_1
     \\\label{eq:time-evolution-spin-perturbed:3}
    D\widetilde a -\bar{\delta} (\widetilde{\epsilon}-\widetilde{\bar{\epsilon}}) \triangleq &- a (\widetilde{\epsilon}-\widetilde{\bar{\epsilon}})-\bar{a} \widetilde{\bar{\sigma}}-\widetilde{\bar{\Psi}}_1\\\label{eq:time-evolution-spin-perturbed:4}
     D\widetilde{\mu} +\kappa_{(\ell)} \widetilde{\mu}\triangleq & -\mu (\widetilde{\epsilon}+\widetilde{\bar{\epsilon}})+\eth \widetilde{\pi} + \widetilde{\Psi}_2\\
    D\widetilde{\lambda} +\kappa_{(\ell)} \widetilde{\lambda} \triangleq & \bar{\eth} \widetilde{\pi} + \mu\widetilde{\bar{\sigma}}\,. 
    \end{align}
\end{subequations}
Notice that there is no differential equation specifying the time-evolution equation of $\widetilde{\Psi}_0$ in Eq.~\eqref{eq:time-evolution-Weyl-perturbed}. From this absence, one might conclude that the time dependence of $\widetilde{\Psi}_0$ can be freely specified at the horizon. However, as we will discuss in detail in Sec.~\ref{sec:psi0-psi4}, this apparent freedom is restricted by the field equations  through the Teukolsky equation for $\widetilde{\Psi}_0$.\footnote{The radial dependence of $\Psi_4$ can be ``freely'' specified in the sense that there is no independent radial field equation for the perturbation to $\Psi_4$. However, as discussed in Paper I, the radial dependence of $\Psi_4$ is restricted by the Teukolsky equation.}

For simplicity, we choose $\widetilde{\epsilon}$ to be at least second order in the perturbation. By doing so, we are assuming the change in surface gravity to be negligible. Notice that this is a gauge choice, which is useful because it considerably simplifies the system of differential equations at the horizon. However, this choice is not unique. Another useful option, discussed in App.~\ref{sec:alternative-gauge} is to fix $\widetilde{\mu}\triangleq0$, which forces $\widetilde{\epsilon}\neq0$ at the horizon. Regardless of the gauge choice the analysis carried out in this paper still applies.

 Setting $\widetilde{\epsilon}\triangleq0$, the second and third differential equations in~\eqref{eq:time-evolution-spin-perturbed} simplify to
\begin{equation}\label{eq:evolution-pi-a}
    D\widetilde{\pi} \triangleq \widetilde{\bar{\Psi}}_1\,,\quad D\widetilde{a} \triangleq -\bar{a} \widetilde{\bar{\sigma}}-\widetilde{\bar{\Psi}}_1 \, ,
\end{equation} which relates the change in the cross-section's connection and rotation one-form with the perturbation of $\Psi_1$. 

It will also be useful to extract an angular equation for $\widetilde{a}$ by taking the $\eth$ operator of Eq.~\eqref{eq:evolution-pi-a}
\begin{equation}\label{eq:Detha}
    \eth D\widetilde{a}\triangleq D\eth \widetilde{a} \triangleq-\widetilde{\bar{\sigma}}\eth \bar{a} -\bar{a}\eth \widetilde{{\bar{\sigma}}}-\eth \widetilde{\bar{\Psi}}_1\,,
\end{equation} where we have used that the unperturbed $D$ and $\eth$ operators at the horizon commute (see Eq.~\eqref{eqs:commutation}). Using the Schwarzschild value for the unperturbed connection and tetrad in Eqs.~\eqref{eq:Schwarzschild-nonvanishing-data} and~\eqref{eq:tetrad-unperturbed}, we see that $\eth \bar{a} \triangleq \delta\bar{a} +\bar{a}^2\triangleq \bar{a}^2$. Further, using the complex conjugate of Eq.~\eqref{eq:angular-perturbed-2:1} and Eq.~\eqref{eq:time-evolution-Weyl-perturbed:2} we can rewrite Eq.~\eqref{eq:Detha} as
\begin{equation}
    D\eth \widetilde{a} \triangleq -\bar{a} (\bar{a} \widetilde{\bar{\sigma}} +\widetilde{\bar{\Psi}}_1 ) -D\widetilde{\bar{\Psi}}_2\,,
\end{equation} where we can identify the first term on the right-hand side with the right-hand side of Eq.~\eqref{eq:evolution-pi-a}. Hence, using again Eq.~\eqref{eq:evolution-pi-a} we obtain
\begin{equation}\label{eq:Detha-simplified}
    D\eth \widetilde{a} \triangleq \bar{a}D \widetilde{a} -D\widetilde{\bar{\Psi}}_2\Rightarrow D(\eth \widetilde{a} - \bar{a}\widetilde{a} +\widetilde{\bar{\Psi}}_2)\triangleq 0 \, ,
\end{equation} where we have used $D\bar{a} \triangleq0$. Eq.~\eqref{eq:Detha-simplified} shows that the combination  $\eth \widetilde{a} - \bar{a}\widetilde{a} +\widetilde{\bar{\Psi}}_2$ is time-independent, i.e., 
\begin{equation}
    \eth \widetilde{a} - \bar{a}\widetilde{a} +\widetilde{\bar{\Psi}}_2 \triangleq g(z,\bar{z})
\end{equation} with $g(z,\bar{z})$ a smooth function of the angular coordinates. The freedom in the connection of the cross-sections of the horizon was extensively discussed in Paper I, so we shall be brief. All of the functions $g(z,\bar{z})$ that leave the curvature of the cross-section invariant form an equivalence class. Different representatives of the equivalence class yield a different set of perturbed angular coordinates on the sphere. For simplicity, we choose the representative with $g(z,\bar{z}) \triangleq0$  so that 
\begin{equation}
    \eth\widetilde{a} -\bar{a} \widetilde{a} \triangleq -\widetilde{\bar{\Psi}}_2\,.
\end{equation}
Combining this expression with the expression for the real part of $\widetilde{\Psi}_2$ in Eq.~\eqref{eq:angular-perturbed-2} we obtain
\begin{equation}
    \widetilde{\eth} a + \widetilde{\bar{\eth}}\bar{a} \triangleq -\bar{a}\widetilde{a}-a\widetilde{\bar{a}} \Rightarrow \widetilde{\delta}a+\widetilde{\bar{\delta}} \bar{a}\triangleq 0\,.
\end{equation}This implies that $\re [\widetilde{\delta} a ]\triangleq 0$, but it can also mean that $\widetilde{\xi}^z \triangleq 0$. To see this, we will use the evolution equations for the tetrad at the horizon, i.e., 
\begin{equation}
    D \Omega-\delta U= \kappa +\rho \Omega+\sigma\bar{\Omega}\,,\quad D\xi^i-\delta X^i=(\bar{\rho}+\epsilon-\bar{\epsilon}) \xi^i+\sigma \bar{\xi}^i\,.
\end{equation} Perturbing these equations to first order and evaluating them at the horizon, we obtain
\begin{equation}\label{eq:diff-eq-tetrad-horizon}
    D\widetilde{\Omega} -\delta\widetilde{U} \triangleq 0\,,\quad D\widetilde{\xi}^i-\delta \widetilde{X}^i \triangleq\widetilde{\sigma} \bar{\xi}^i+(\widetilde{\epsilon}-\widetilde{\bar{\epsilon}}) \xi^{i}\,,
\end{equation} where we have used that $\Omega=0$, $ U\triangleq 0$ and $D\xi^i\triangleq X^ i \triangleq 0$ for the unperturbed tetrad components (see Eq.~\eqref{eq:tetrad-unperturbed}).  Furthermore, to preserve the orthogonality of the perturbed tetrad vectors, we need to set at the horizon
\begin{equation}\label{eq:Omega-U-X-tilde}
    \widetilde{\Omega} \triangleq \widetilde{U} \triangleq \widetilde{X}^i\triangleq0\,,
\end{equation} which is a solution of Eq.~\eqref{eq:diff-eq-tetrad-horizon}. Therefore, to fully determine the change of the tetrad vectors at the horizon we just need to solve  
\begin{equation}
    D\widetilde{\xi}^z\triangleq (\widetilde{\epsilon}-\widetilde{\bar{\epsilon}})\xi^z\,,\quad   D\widetilde{\xi}^{\bar{z}}\triangleq\widetilde{\sigma}  \bar{\xi}^{\bar{z}}\,,
\end{equation} which we obtained by simplifying the second expression in Eq.~\eqref{eq:diff-eq-tetrad-horizon} with Eq.~\eqref{eq:Omega-U-X-tilde} and recalling that  $\xi^{\bar{z}}=\bar{\xi}^z=0$ for the background quantities. 
We see that the choice $\widetilde{\epsilon} \triangleq0$ implies that 
\begin{equation}
   D \widetilde{\xi}^z \triangleq0\,.
\end{equation} In other words, the perturbation $\widetilde{\xi}^z$ is time-independent as pointed out in Paper I for time-independent perturbations of the Schwarzschild geometry, and we can set them to zero so the tetrad in the horizon's cross-section can be written as 
\begin{equation}
    m^A\partial_A \triangleq \xi^z\partial_z + \widetilde{\xi}^{\bar{z}} \partial_{\bar{z}}\,.
\end{equation} From this expression it is straightforward to check that the area of the cross-section, denoted by $S$, changes only to second order in the perturbation as we assumed. Using  
\begin{equation}
    g^{z\bar{z}}_{{S}}=g^{\bar{z}z}_{{S}} = \xi^z \widetilde{\bar{\xi}^z} \widetilde{\xi}^{\bar{z} }\,,\quad g^{zz}_S = 2\xi^z \widetilde{\bar{\xi}^z}
\end{equation} and inverting the metric of the cross-section, it follows that
\begin{equation}
    A=\int \d z\d\bar{z} \sqrt{|g_{{S}}|} = \int \d z\d\bar{z} \left[\left(\frac{1}{\xi^z}\right)^2 +\mathrm{O}[ \widetilde{\bar{\xi}^z} \widetilde{\xi}^{\bar{z} }] \right]
 \end{equation} the perturbation to the area's cross-section is at least second order in the perturbation. 
In summary, the time-dependent perturbation due to $\widetilde{\Psi}_0$ and $\widetilde{\Psi}_1$ decouples from the time-independent perturbations due to the tidal distortion. This means that we can combine these two effects by simply adding the perturbations. Notice that the ``transverse'' tetrad component encodes the time-dependent perturbations and it is proportional to the shear. 

Contrary to the perturbed horizon due to static tidal perturbations that we considered before in paper I, the foliation of the horizon is not freely specifiable anymore. To see this, consider the second equation in~\eqref{eq:time-evolution-Weyl-perturbed}, which specifies the time evolution of $\widetilde{\Psi}_2$. This equation needs to be consistent with Eq.~\eqref{eq:evolution-pi-a} since the time-evolution of the spin coefficient $\widetilde{\pi}$ is also sourced by $\widetilde{\Psi}_1$. This means that $\widetilde{\pi}$ and $\widetilde{\Psi}_2$ need to be related. In particular, the equations 
\begin{equation}
    D\widetilde{\Psi}_2\triangleq \bar{\eth} \widetilde{\Psi}_1\,,\quad D\widetilde{\pi} \triangleq \widetilde{\bar{\Psi}}_1
\end{equation} need to be consistent with one another. To check this, we take the complex conjugate of the second equation and compute its $\bar{\eth}$ derivative. Using that $D \bar{\eth} \triangleq\bar{\eth} D$, we obtain
\begin{equation} \label{eq:foliation}   D\bar{\eth}\widetilde{\bar{\bar{\pi}}} \triangleq D\widetilde{\Psi}_2 \Rightarrow \eth \widetilde{\pi} \triangleq \widetilde{\bar{\Psi}}_2 + f(z,\bar{z})  \, .
\end{equation} Recall that the quantity $\eth\pi$ specifies the foliation of the horizon by cross-sections, as reviewed in Sec.~\ref{subsec:perturbed-ih}. In paper I this quantity was pure gauge, as we could choose the foliation as we pleased. However, Eq.~\eqref{eq:foliation} implies that the value of $\eth\widetilde{\pi}$ is not purely gauge anymore. As stressed by the function $f(z,\bar{z})$, if both $\widetilde{\pi}$ and $\widetilde{\Psi}_2$ are time independent, then any time-independent function $f(z,\bar{z})$ can be used to define $\eth \widetilde{\pi}$, which is consistent with what we found in paper I. However, we see that when introducing a time-dependent perturbation, 
the time-dependent part of $\eth \pi$ is fully determined by $\Psi_2$.  Here, we will choose the time-independent function to vanish $f(z,\bar{z})=0$, to separate the part of the perturbation sourced by the time-dependent part of $\widetilde{\Psi}_0$. This equation naturally links the linear momentum of the horizon (related to the divergence of $\omega$) with the force exerted on the hole due to $\widetilde{\Psi}_0$ from a purely tidal perturbation encoded in $\widehat{\Psi}_2$, where we used the notation of Paper I in which $\widehat{\quad }$ denotes a static perturbation due to tidal disruption.   Therefore,
\begin{equation}\label{eq:cuts}
    \eth\widetilde{\pi}\triangleq \widetilde{\bar{\Psi}}_2\,.
\end{equation} Notice the different sign of Eq.~\eqref{eq:cuts} with the gauge choice in Eq.~(123) of Paper I. 
This sign difference illustrates that while the force due to the tidal perturber is attractive, $\widetilde{\Psi}_0$ is ``kicking'' the horizon in the opposite direction. Notice that the time-evolution equation for $\widetilde{\mu}$ simplifies if we combine it with this ``gauge choice''~\eqref{eq:cuts}
\begin{equation}\label{eq:evolution-mu-2}
    D\widetilde{\mu} +\kappa_{(l)} \widetilde{\mu} \triangleq \widetilde{\bar{\Psi}}_2+\widetilde{\Psi}_2=2 \re \widetilde{\Psi}_2\,.
\end{equation} This equation implies that the expansion of the null vector $n$ cannot be chosen to vanish anymore. Rather, its evolution is sourced by the real part of $\widetilde{\Psi}_2$. 

Gathering Eqs.~\eqref{eq:angular-perturbed-2}-\eqref{eq:evolution-mu-2} we obtain a system of 15 differential equations to prescribe 10 quantities (the spin coefficients, Weyl scalars and tetrad functions  $\widetilde{\sigma}$, $\widetilde{\pi}$, $\widetilde{\mu}$, $\widetilde{\lambda}$, $\widetilde{a}$, $\widetilde{\Psi}_1$, $\widetilde{\Psi}_2$, $\widetilde{\Psi}_3$, $\widetilde{\Psi}_4$, and $\widetilde{\xi}^{\bar{z}}$) as a function of the perturbation to $\widetilde{\Psi}_0$. This implies that we can separate this system in a group of 10 differential equations that will be used to solve the initial data at the horizon using $\widetilde{\Psi}_0$ as the source 
\begin{subequations}\label{eq:eqs-horizon}
    \begin{align}\label{eq:eqs-horizon:1}
      D\widetilde{\sigma} -\kappa_{(l)} \widetilde{\sigma} \triangleq& \widetilde{\Psi}_0\,,\\
      D\widetilde{\Psi}_1-\kappa_{(l)} \widetilde{\Psi}_1 \triangleq & \bar{\eth}\widetilde{\Psi}_0\,,\\
      D\widetilde{\Psi}_2 \triangleq &\bar{\eth}\widetilde{\Psi}_1\,,\\
      D\widetilde{\pi} \triangleq & \widetilde{\bar{\Psi}}_1\,,\\
      D\widetilde{\mu} +\kappa_{(l)} \widetilde{\mu} \triangleq& \widetilde{\Psi}_2+\widetilde{\bar{\Psi}}_2\,,\\
      D\widetilde{\lambda} +\kappa_{(l)} \widetilde{\lambda} \triangleq & \bar{\eth} \widetilde{\pi} +\mu \widetilde{\bar{\sigma}}\,,\\
      D\widetilde{\Psi}_3+\kappa_{(l)} \widetilde{\Psi}_3 \triangleq &  \bar{\eth}\widetilde{\Psi}_2+3\widetilde{\pi}\Psi_2 \,,\\   
      D\widetilde{\Psi}_4+2\kappa_{(l)}\widetilde{\Psi}_4 \triangleq & \bar{\eth} \widetilde{\Psi}_3-3\widetilde{\lambda} \Psi_2 \,,\\
      D\widetilde{a}\triangleq & -\bar{a}\widetilde{\bar{\sigma}}-\widetilde{\bar{\Psi}}_1\,,\\
      D\widetilde{\xi}^{\bar{z}}& \triangleq\widetilde{\sigma}  \bar{\xi}^{\bar{z}}\,.
    \end{align}
\end{subequations}
As shown in the Appendix, the above equations can be combined to yield the following constraint equations
\begin{subequations}\label{eq:eqs-constraint}
  \begin{align}
    \eth \widetilde{\lambda}-\bar{\eth}\widetilde{\mu}\triangleq& \mu\widetilde{\pi}-\widetilde{\Psi}_3\,,\\
    -2\re \widetilde{\Psi}_2 \triangleq& \eth\widetilde{a}+\bar{\eth}\widetilde{\bar{a}} -a\widetilde{\bar{a}}-\bar{a}\widetilde{a}\,,\\
    -2i\im \Psi_2\triangleq&\eth\widetilde{\pi}-\bar{\eth}\widetilde{\bar{\pi}}\,,\\
    \bar{\eth}\widetilde{\sigma}\triangleq & \widetilde{\Psi}_1\,,\\
    \eth\widetilde{\pi} \triangleq& \widetilde{\bar{\Psi}}_2\,.
  \end{align}
\end{subequations}
These equations contain only angular derivatives, i.e. no radial or
time derivatives.  In the next section we shall solve
Eqs.~\eqref{eq:eqs-horizon} (and hence also
Eqs.~\eqref{eq:eqs-constraint}) explicitly.

\section{The relationship between $\widetilde{\Psi}_0$ and $\widetilde{\Psi}_4$}
\label{sec:psi0-psi4}

The initial data at the horizon is determined by solving Eqs.~\eqref{eq:eqs-horizon} for the spin coefficients and Weyl scalars at the horizon as a function of the perturbation to $\Psi_0$. 
The next step in obtaining the metric of a ringing-down black hole would be to integrate the radial equations with this initial data. This will be discussed in Sec.~\ref{sec:multipole-moments}. However, since one of the goals of this work is to provide an explicit construction of the geometry of the perturbed, ringing-down horizon (and its neighborhood), we provide an explicit solution of Eqs.~\eqref{eq:eqs-horizon} in this section, which will be used in the following discussion.   

The radial behavior of $\Psi_4$ is related to its angular and temporal
behavior through the Teukolsky equation. This particular feature of
$\Psi_4$ (which other spin coefficients and Weyl scalars do not share,
e.g. $\Psi_2$) can be understood as a consequence of $\Psi_4$ not
having an independent radial equation.  Similarly, $\Psi_0$ does not
have an independent time-evolution equation. As discussed above, this
could give the impression that one can prescribe the time dependence
of $\widetilde{\Psi}_0$ (or alternatively of $\widetilde{\sigma}$)
``freely'' at the horizon. This is, however, not true. The Weyl scalar
$\Psi_0$ also satisfies a Teukolsky equation, which links its radial,
angular, and temporal behavior. A remarkable feature of the Teukolsky
equation for linear perturbations around a Schwarzschild background is
that it is
separable~\cite{Teukolsky:1972my,Chandrasekhar:1985kt}. This feature
is also present in the coordinate system that we use in our construction. Hence, the solutions of the Teukolsky equations can be
expressed as products of purely radial, purely angular, and purely
temporal functions. This means in particular, that the form of
$\widetilde{\Psi}_0$ and $\widetilde{\Psi}_4$ at the horizon will
inherit a similar structure.

For completeness, we first show that the Teukolsky equations for $\widetilde{\Psi}_0$ and $\widetilde{\Psi}_4$ are separable. The Teukolsky equation for $\widetilde{\Psi}_4$ was already discussed in~\cite{metidieri2024tidal}, so we shall be brief. The Teukolsky equation for $\widetilde{\Psi}_0$ can be easily obtained by using the transformation
\begin{equation}
    \ell \leftrightarrow n\,,\quad m\leftrightarrow \bar{m}
\end{equation} on the Teukolsky equation for $\widetilde{\Psi}_4$. Hence, we obtain the Teukolsky equations for $\widetilde{\Psi}_0$ and $\widetilde{\Psi}_4$ 
\begin{subequations}
\begin{align}\label{eq:Teukolsky-4&0}
     [  D \Delta-\eth\bar{\eth}+\mu D -(5\rho+2\epsilon) \Delta -4\mu(\rho+\epsilon) -2\Psi_2] \widetilde{\Psi}_0=&0\\
  [  \Delta D -\bar{\eth}\eth+5\mu D +(4\epsilon-\rho) \Delta +4\mu(5\epsilon-\rho) -6\Psi_2] \widetilde{\Psi}_4=&0\,.
\end{align}     
\end{subequations}
Using that  $(r+c)^2 \eth\bar{\eth}$ is independent of the $r$ coordinate, we can separate the angular part of Eqs.~\eqref{eq:Teukolsky-4&0}.
Using that $\Psi_0$ and $\Psi_4$ have spin-weight 2 and -2, we write the ansatz
\begin{align}\label{eq:psi0-psi4-ansatz}
     \widetilde{\Psi}_0 &=\sum_{l,m} \psi^{(0)}_{lm}(v,r) \,{}_{2}Y_{lm}\,,\\
     \widetilde{\Psi}_4 &= \sum_{l,m} \psi^{(4)}_{lm}(v,r)\, {}_{-2} Y_{lm}
\end{align}
which for every $l,m$ yields the following two differential equations for the radial and temporal parts of $\widetilde{\Psi}_0$ and $\widetilde{\Psi}_4$
\begin{widetext}
\begin{subequations}
 \begin{align}\label{eq:diff-eqs-psi0-psi4-r&v}
   & \mathcal{O}_{T_0}\widetilde{\Psi}_0=\sum_{l,m} [ D \Delta+\mu D -(5\rho+2\epsilon) \Delta -4\mu(\rho+\epsilon) -2\Psi_2 +\frac{(l+2)(l-1)}{2(r+c)^2}] \psi_{lm}^{(0)}(v,r) \, {}_2 Y_{lm} = 0\\
    & \mathcal{O}_{T_4}\widetilde{\Psi}_4 =\sum_{l,m}  [\Delta D +5\mu D +(4\epsilon-\rho) \Delta +4\mu(5\epsilon-\rho) -6\Psi_2+\frac{(l+2)(l-1)}{2(r+c)^2}]\psi_{lm}^{(4)}(v,r) \, {}_{-2} Y_{lm} = 0
\end{align}   
\end{subequations}

\end{widetext}
To proceed further, we use the inverse Fourier transform for $\psi_{lm}^{(0)}$/
$\psi_{lm}^{(4)}$
\begin{equation}
    \psi_{lm}^{(0,4)}(v,r) = \frac{1}{2\pi} \int_{-\infty}^\infty \d\omega e^{-i \omega v} X_{lm}^{(0,4)}(r,\omega) \, .
\end{equation}
It is important to distinguish this from a resonant expansion as in
Eq.~\eqref{eq:resonantexpansion}.  Here the $\omega$ are real
frequencies while Eq.~\eqref{eq:resonantexpansion} is a discrete sum
over modes with complex frequencies.  We shall return to the resonant
frequencies below when we discuss QNMs.

Using the Fourier modes yields the following two ODEs for the radial parts of
$\widetilde{\Psi}_0$ and $\widetilde{\Psi}_4$~\footnote{Notice that
  the static solution for the Teukolsky
  equations~\eqref{eq:diff-eqs-psi0-psi4-r&v} can be obtained through
  the ansatz $\psi_{lm}^{(0,4)}(v,r)=\psi_{lm}^{(0,4)}(r)$.  The
  resulting differential equations for $\psi_{lm}^{(0,4)}(r)$ can be
  solved analytically in terms of the associated Legendre functions
\begin{subequations}
\begin{align}
    X^{(0)}(r) &= \frac{r K_0}{(r+c)^3} P_l^2\left(1+\frac{2r}{c}\right)\,,\\
    X^{(4)}(r) &= \frac{ K_4}{r(r+c)} P_l^2\left(1+\frac{2r}{c}\right) \, ,
\end{align}
\end{subequations} with $K_0$ and $K_4$ integration constants.} \begin{widetext}
\begin{subequations}\label{eq:teukolsky-extended}
\begin{align}\label{eq:teukolsky-extended:1}
     U\partial_r^2 X^{(0)}_{lm}-[5\rho+2\epsilon+\mu U +i\omega] \partial_r X^{(0)}_{lm} +[4\mu(\rho+\epsilon)+2\Psi_2 -\frac{(l-1)(l+2)}{2(r+c)^2}+i \omega\mu ] X^{(0)}_{lm}=&0\\ \label{eq:teukolsky-extended:2}
     U \partial^2_r X^{(4)}_{lm} +[\partial_r U-5\mu U+4\epsilon-\rho-i\omega]\partial_r X^{(4)}_{lm}+[6\Psi_2-4\mu(5\epsilon-\rho)-\frac{(l+2)(l-1)}{2(r+c)^2} +5i \omega\mu] X^{(4)}_{lm}=&0 \, .
\end{align}
\end{subequations}
\end{widetext} 
The spin coefficients and tetrad functions appearing in Eq.~\eqref{eq:teukolsky-extended}
are the background quantities (see Eqs.~\eqref{eq:Schwarzschild-nonvanishing-data} and~\eqref{eq:tetrad-unperturbed}), which only depend on the radial coordinate. Hence, from this expression, it is manifest that both Teukolsky equations, for $\widetilde{\Psi}_0 $ and $\widetilde{\Psi}_4$, are separable also in our coordinate system.

Now that we have argued that both Teukolsky equations are separable, we can solve them. However, we cannot solve both Teukolsky equations for $\widetilde{\Psi}_0$ and $\widetilde{\Psi}_4$ simultaneously since the system of differential equations at the horizon~\eqref{eq:eqs-horizon} relates the initial values of these two quantities. In fact, a single solution of the radial Teukolsky equation for $\widetilde{\Psi}_0$ is related to a combination of two independent solutions of the radial Teukolsky equation for $\widetilde{\Psi}_4$ and vice versa. As we will discuss more in Sec.~\ref{sec:TSI}, this relation is strongly tied to the Teukolsky-Starobinsky identities and will yield equivalent results. 

Nonetheless, the previous discussion shows that we can find separable solutions for $\widetilde{\Psi}_0$ and $\widetilde{\Psi}_4$, i.e., 
 \begin{subequations}
    \begin{align}
       \widetilde\Psi_0 = & \frac{1}{2\pi} \sum_{l,m} \left(\int \d\omega  e^{-i\omega v} X_{lm}^{(0)} Y^{(1)}_{lm} + \int \d\bar{\omega}  e^{i\bar{\omega} v} \bar{X}_{lm}^{(0)} Y^{(2)}_{lm}\right)\\
       \widetilde\Psi_4 = & \frac{1}{2\pi} \sum_{l,m} \left(\int \d w e^{-i w v} X_{lm}^{(4)}Y^{(3)}_{lm} + \int \d\bar{w} e^{i\bar{w} v} \bar{X}_{lm}^{(4)} Y^{(4)}_{lm}\right) 
    \end{align}
\end{subequations} where $Y^{(i)}_{lm}$ are functions of the angular coordinates $\{z\,,\bar{z}\}$ and $X^{(0,4)}_{lm}$ of the radial coordinate only. Given that the spin weight of the Weyl scalars $\Psi_0$ and $\Psi_4$ is well-defined, we can expand the angular functions in terms of the spin-weight spherical harmonics with spins $s=2$ and $-2$ respectively. However, given the relationship between these two quantities at the horizon we can either choose $Y_{lm}^{(1)}\propto  Y_{lm}^{(2)} \propto {}_2 Y_{lm}$ or $Y_{lm}^{(3)}\propto Y_{lm}^{(4)}\propto {}_{-2} Y_{lm}$ without loss of generality, but not both simultaneously. Both choices yield completely analogous results, but the first one is less cumbersome for our purpose to show that all of the perturbed quantities are sourced by $\widetilde{\Psi}_0$, so we shall choose  $Y_{lm}^{(1)}=a_{lm}^- {}_2 Y_{lm}$ and $Y_{lm}^{(2)}=a_{lm}^+ {}_2 Y_{lm}$, where $a_{lm}^\pm (\omega)$ are functions of the frequency only. 

 Hence, at the horizon $r=0$, we will use an ansatz for $\widetilde \Psi_0$ of the form
\begin{equation}\label{eq:psi0-horizon}
    \widetilde{\Psi}_0 \triangleq  \frac{1}{2\pi} \sum_{l,m} \left(\int \d\omega  a^-_{lm} e^{-i\omega v}  + \int \d\bar{\omega} a^+_{lm} e^{i\bar{\omega} v} \right) \,{}_2Y_{lm}\,,
\end{equation} where the radial function is taken to be normalized to unity at the horizon $X^{(0)}_{lm}(r=0)=1$.  Using Eq.~\eqref{eq:psi0-horizon} and the field equations at the horizon~\eqref{eq:eqs-horizon} we will then derive an expression  for $\widetilde\Psi_4$ at the horizon of the form 
\begin{equation}\label{eq:psi4-horizon}
\begin{split}
     \widetilde{\Psi}_4\triangleq \frac{1}{2\pi} \sum_{l,m} \left(\int \d w [ b^-_{lm}{}_{-2}Y_{lm} + c^-_{lm} {}_{-2}Y_{l\,,-m}] e^{-i w v} \right.\\
     \left.+ \int \d\bar{w} [b^+_{lm}{}_{-2}Y_{lm}+c_{lm}^+ {}_{-2}Y_{l\,,-m} ] e^{i\bar{w} v} \right) \,.  
\end{split}
\end{equation} 
As we will see below, we will find that the coefficients $b^\pm_{lm}$ and $c^\pm_{lm}$ are related to $a^\pm_{lm}$.  Notice that the coefficients $a_{lm}^\pm$ are not independent. This can be deduced by simply counting the degrees of freedom: $\widetilde{\Psi}_0$ has  two independent degrees of freedom (the polarizations of the infalling gravitational wave), while $a_{lm}^\pm$ are two complex constants and have four degrees of freedom. However, how these quantities are related will also depend on properties of the frequencies, so we leave the relationship between these constants unspecified.

Using the ansatz for $\widetilde\Psi_0$ at the horizon in Eq.~\eqref{eq:psi0-horizon}
we can solve explicitly the system of differential equations~\eqref{eq:eqs-horizon} in the order $\widetilde{\sigma}$, $\widetilde{\Psi}_1$, $\widetilde{a}$, $\widetilde{\xi}^{\bar{z}}$, $\widetilde{\Psi}_2$, $\widetilde{\pi}$, $\widetilde{\mu}$, $\widetilde{\lambda}$, $\widetilde{\Psi}_3$ and finally, $\widetilde{\Psi}_4$ 
\begin{widetext}
    \begin{subequations}\label{eq:solution1}
    \begin{align}\label{eq:solution1:1}
        \widetilde{\sigma} \triangleq& -\frac{1}{2\pi} \sum_{l,m} \left(\int_{-\infty}^\infty \d\omega \frac{a_{lm}^-}{\kappa_{(l)}+i\omega} e^{- i \omega v}+\int_{-\infty}^\infty \d\bar{\omega} \frac{a_{lm}^+}{\kappa_{(l)}-i\bar{\omega}} e^{ i \bar{\omega} v} \right)\,{}_2Y_{lm}\,, \\ \label{eq:solution1:2}
        \widetilde{\Psi}_1 \triangleq & \frac{1}{2\pi} \sum_{l,m} \left(\int_{-\infty}^\infty \d\omega\frac{a_{lm}^-\sqrt{(l+2)(l-1)}}{\sqrt{2} c (\kappa_{(l)}+i \omega)} e^{-i \omega v}+\int_{-\infty}^\infty \d\bar{\omega}\frac{a_{lm}^+\sqrt{(l+2)(l-1)}}{\sqrt{2} c (\kappa_{(l)}-i \bar{\omega})} e^{i \bar{\omega} v} \right)\,{}_1 Y_{lm}\,,\\\label{eq:solution1:3}
            \begin{split}
            \widetilde{a} \triangleq &  \frac{1}{2\pi} \left(-\int \d\omega \sum_{l,m} \frac{(-1)^m \bar{a}^+_{lm} e^{-i\omega v}}{i\omega(\kappa_{(l)}+i\omega)} (\bar{a} {}_{-2}Y_{l\,-m} +\frac{\sqrt{(l+2)(l-1)}}{\sqrt{2}c} {}_{-1}Y_{l\,-m})\right.\\
            &\left.+\int \d\bar{\omega} \sum_{l,m} \frac{(-1)^m \bar{a}^-_{lm} e^{i\bar{\omega} v}}{i\bar{\omega}(\kappa_{(l)}-i\bar{\omega})} (\bar{a} {}_{-2}Y_{l\,-m} +\frac{\sqrt{(l+2)(l-1)}}{\sqrt{2}c} {}_{-1}Y_{l\,-m})\right)  \,,
            \end{split}
            \\\label{eq:solution1:4}
     \widetilde{\xi}^{\bar{z}} \triangleq & \frac{\bar{\xi}^{\bar{z}}_0}{2\pi} \sum_{l,m} {}_2Y_{lm} \left(\int \d \omega  \frac{a_{lm}^- e^{-i\omega v}}{i\omega (\kappa_{(l)} +i\omega)}-\int \d\bar{\omega} \frac{a^+_{lm} e^{i\bar{\omega} v}}{i\bar{\omega}(\kappa_{(l)}-i\bar{\omega})}\right) \,,\\\label{eq:solution1:5}
        \widetilde{\Psi}_2 \triangleq& \frac{1}{2\pi} \sum_{l,m} \frac{\sqrt{(l+2)(l+1)l(l-1)}}{2c^2}\left(\int_{-\infty}^\infty \d\omega\frac{a_{lm}^-}{ i \omega (\kappa_{(l)} +i \omega)}e^{- i \omega v}-\int_{-\infty}^\infty \d\bar{\omega}\frac{a_{lm}^+}{i \bar{\omega} (\kappa_{(l)} -i \bar{\omega})}e^{ i \bar{\omega} v} \right)\, Y_{lm} \,,
        \\ \label{eq:solution1:6}
        \widetilde{\pi} \triangleq & \frac{1}{2\pi} \sum_{l,m} (-1)^m\frac{\sqrt{(l+2)(l-1)}}{\sqrt{2}c}\left(\int_{-\infty}^\infty \d\omega \frac{\bar{a}_{lm}^+ }{ i\omega (\kappa_{(l)}+i\omega)} e^{-i\omega v }-\int_{-\infty}^\infty \d\bar{\omega} \frac{\bar{a}_{lm}^- }{ i\bar{\omega} (\kappa_{(l)}-i\bar{\omega})} e^{i\bar{\omega} v }\right) \, {}_{-1} Y_{l\,,-m} \,,
        \\\label{eq:solution1:7}
         \widetilde{\mu} \triangleq & \frac{1}{2\pi}\sum_{l,m}   \frac{\sqrt{(l+2)(l+1)l(l-1)}}{2c^2 } \left( \int_{-\infty}^\infty \d\omega\frac{[a_{lm}^- Y_{lm}+(-1)^m \bar{a}^+_{lm} Y_{l\,,-m}] e^{-i \omega v}}{i \omega (\kappa_{(l)}^2+\omega^2)}-\int_{-\infty}^\infty \d\bar{\omega}  \frac{[a^+_{lm} Y_{lm}+(-1)^m\bar{a}_{lm}^- Y_{l\,,-m}] e^{i \bar{\omega} v}}{i\bar{\omega} (\kappa_{(l)}^2+\bar{\omega}^2)}\right) \,, \\ \label{eq:solution1:8}
         \begin{split}
              \widetilde{\lambda} \triangleq & \frac{1}{2\pi} \sum_{l,m}(-1)^m \left(\int_{-\infty}^\infty \d\omega\frac{\bar{a}_{lm}^+ e^{-i \omega v} }{2ic^2 \omega (\kappa_{(l)}^2+\omega^2)}\left(2ic\omega-(l+2)(l-1)\right)+\right.\\
             & \left.\int_{-\infty}^\infty \d\bar{\omega}\frac{\bar{a}_{lm}^- e^{i \bar{\omega} v}}{2ic^2 \bar{\omega} (\kappa_{(l)}^2+\bar{\omega}^2)}\left(2ic\bar{\omega}+(l+2)(l-1)\right)\right)\,{}_{-2}Y_{l\,,-m}  \,,  
         \end{split} \\ \label{eq:solution1:9}
         \begin{split}
              \widetilde{\Psi}_3 \triangleq & -\frac{1}{2\pi}\sum_{l,m} \frac{\sqrt{(l+2)(l-1)}}{2\sqrt{2}c^3 } \left(\int_{-\infty}^\infty \d\omega \frac{[a_{lm}^- {}_{-1}Y_{lm} (l+1)l+3(-1)^m \bar{a}^+_{lm} {}_{-1}Y_{l\,,-m}] e^{-i \omega v}}{i \omega (\kappa_{(l)}^2+\omega^2)}\right.\\
              &\left.-\int_{-\infty}^\infty \d\bar{\omega} \frac{[a^+_{lm} (l+1)l {}_{-1}Y_{lm}+3(-1)^m\bar{a}_{lm}^- {}_{-1}Y_{l\,,-m}] e^{i\bar{\omega}v}}{i\bar{\omega} (\kappa_{(l)}^2+\bar{\omega}^2) }\right) \,,
         \end{split}
        \\ \label{eq:solution1:10}
         \begin{split}
            \widetilde{\Psi}_4 \triangleq & \frac{1}{8\pi c^4} \sum_{l,m} \left(\int_{-\infty}^\infty \d\omega \frac{[a_{lm}^- (l+2)(l+1)l(l-1) {}_{-2}Y_{lm}+6c(-1)^m i \omega \bar{a}^+_{lm} {}_{-2}Y_{l\,,-m}] e^{-i \omega v}}{i \omega (\kappa_{(l)}^2+\omega^{2}) (2\kappa_{(l)} -i \omega) } \right.\\
            & \left.-\int_{-\infty}^\infty \d\bar{\omega}\frac{[a^+_{lm} (l+2)(l+1)l(l-1){}_{-2}Y_{lm}-6 c (-1)^m i \bar{\omega}\bar{a}_{lm}^- {}_{-2}Y_{l\,,-m} ] e^{i \bar{\omega} v}}{i\bar{\omega}(\kappa_{(l)}^2+\bar{\omega}^{2}) (2\kappa_{(l)} +i\bar{\omega}) } \right)\,. 
        \end{split}      
    \end{align}
\end{subequations}
\end{widetext}
It is important to note that this is \emph{not} the most general
solution to the system of differential equations in
Eq.~\eqref{eq:eqs-horizon}. One reason is that the assumption of a
Fourier decomposition inherently restricts the solution set.
Moreover, if we were to start from a different cross-section of the
horizon and repeat the above procedure, we would find a mode
decomposition on a \emph{different} transverse null surface. We would thus need to be careful when comparing the mode frequencies obtained from the two choices.  Additionally, we have excluded certain
unphysical, unstable particular solutions of the differential
equations~\eqref{eq:eqs-horizon}.
For instance, the differential equation for $\widetilde{\sigma}$~\eqref{eq:eqs-horizon:1} has a second particular solution of the form $c_1 e^{\kappa_{(l)}v}$, where $c_1(z\,,\bar{z})$ is a function of the angular coordinates. This particular solution corresponds to a shear perturbation that grows exponentially with time $v$ (since $\kappa_{(l)}>0$), thus implying that a small flux of gravitational waves causes the horizon to destabilize and have an exponentially growing deformation on the future. This behavior is not physical, so we discard it.  Finally, as we discussed in Sec.~\ref{sec:perturbed-intrinsic-geometry}, in this paper we are only concerned with the perturbations caused by the radiative infalling flux of gravitational waves driven by $\widetilde{\Psi}_0$. The system~\eqref{eq:eqs-horizon} also admits a solution for the perturbed geometry corresponding to a \emph{static} tidal perturbation, which was discussed in great detail in Paper I. Hence, here we shall set the static solution to zero.

Despite these restrictions, the solution in Eq.~\eqref{eq:solution1} is sufficiently general for the
application to the ringdown that we have in mind. In
Sec.~\ref{sec:QNM-frequencies}, we will apply this solution to the
latest phases of the ringdown, where the solution of the spin
coefficients and Weyl scalars~\eqref{eq:solution1} will be expressible
in terms of a sum over an infinite but discrete set of frequencies
(the QNM frequencies). It can be shown that the
frequencies in which the solution of $\widetilde{\Psi}_0$ and
$\widetilde{\Psi}_4$ is spanned is the same for both quantities and in
fact, for all the spin coefficients and Weyl scalars. A simplified
proof of this statement will be provided in
Sec.~\ref{sec:QNM-frequencies}. For now, we will just assume that the
frequencies appearing in the decomposition of $\widetilde\Psi_4$ need
to be the same as those in the decomposition of $\widetilde\Psi_0$.

A very similar solution to Eq.~\eqref{eq:solution1} could
be derived to describe tidal heating during the inspiral, by
considering a periodic solution of Eq.~\eqref{eq:solution1} with the
frequency $\omega$ related to the orbital frequency of the compact
object. Early work studying the geometry of the horizon due to tidal heating is performed in ~\cite{Hartle:1974gy,OSullivan:2014ywd,OSullivan:2015lni}. The
application of the whole perturbative black hole tomography framework to relate the geometry of the horizon to the observed gravitational waves for an EMRI during the inspiral will be discussed in a companion paper.

Comparing Eq.~\eqref{eq:solution1:10} with Eq.~\eqref{eq:psi4-horizon}, we can identify the following constants which appear in Eq.~\eqref{eq:psi4-horizon}
\begin{subequations}\label{eq:b}
    \begin{align}
        b^-_{lm} &=   \frac{ (l+2)(l+1)l(l-1) }{4c^4 i \omega (\kappa_{(l)}^2+\omega^{2}) (2\kappa_{(l)} -i \omega) } a_{lm}^-\,,\\
         c^-_{lm} &=   \frac{3(-1)^m}{2c^3 (\kappa_{(l)}^2+\omega^{2}) (2\kappa_{(l)} -i \omega) } \bar{a}^+_{lm}\,,\\
         b^+_{lm} & =   -\frac{ (l+2)(l+1)l(l-1) }{4c^4 i \bar{\omega}(\kappa_{(l)}^2+\bar{\omega}^{2}) (2\kappa_{(l)} +i \bar{\omega}) } a_{lm}^+\,,\\
          c^+_{lm} & =   \frac{3(-1)^m   }{2c^3(\kappa_{(l)}^2+\bar{\omega}^{2}) (2\kappa_{(l)} +i \bar{\omega}) } \bar{a}^-_{lm}\,,
    \end{align} 
\end{subequations} which are all sourced by $a_{lm}^\pm$ and its complex conjugates.
Notice first that for a decomposition of the form~\eqref{eq:psi0-horizon}, in which we can encode the angular part purely in a spin-weighted spherical harmonic of the appropriate spin, we obtain for $\widetilde{\Psi}_4$ at the horizon an expression that not only depends on ${}_{-2}Y_{lm}$ but also on ${}_{-2} Y_{l\,,-m}$. This dependence could be simplified by rearranging the sum of $m=-l,\,.\,.\,. l$ to $m^\prime=l\,.\,.\,.-l$ for the terms with ${}_{-2}Y_{l\,,-m}$. However, this would give rise to frequencies $\omega_{l\,,-m}$, which have not been defined for the moment. Once we determine that the integral over the frequencies $\omega_{lm}$ reduces to a sum over the QNM frequencies, we will be able to use properties of these modes to simplify these expressions, but for the time being we shall keep the discussion general by maintaining the terms with negative index $m$ in the harmonics.

We could consider two particular cases such that $\widetilde \Psi_0$ and $\widetilde\Psi_4$ only have the $e^{-i\omega v}$ modes. The first case follows by setting $a^+_{lm} =0$ and we obtain a ``single-mode'' excitation for
$\widetilde\Psi_0$. Then, the expression for $\widetilde{\Psi}_4$ at the
horizon shows that a single solution of the radial Teukolsky
equation for $\widetilde{\Psi}_0$ requires a combination of two
linearly independent solutions of the radial Teukolsky equations for
$\widetilde{\Psi}_4$. The converse is also true, as can be seen by requiring a ``single-mode'' perturbation in $\widetilde{\Psi}_4$, which can be attained by choosing the perturbation such that $b_{lm}^+ {}_{-2}Y_{lm}+c_{lm}^+ {}_{-2}Y_{l\,,-m}=0$. This relationship between the boundary conditions at the horizon of $\widetilde{\Psi}_0$ and $\widetilde{\Psi}_4$ reminds us of the boundary conditions of the solution to the Teukolsky equation for $\widetilde{\Psi}_4$ at the horizon and infinity, where having purely ingoing modes at the horizon implies a mixture between ingoing and outgoing modes far away and vice versa.

From the above discussion, it follows that $\widetilde{\Psi}_4$ should be of the form
\begin{equation}\label{eq:psi4-ansatz}
    \begin{split}
        \widetilde{\Psi}_4 = \frac{1}{2\pi} \sum_{l\,,m} \left( \int \d\omega [b^-_{lm} X^{(4)}_{lm} {}_{-2}Y_{lm} +c^-_{lm} Z^{(4)}_{lm} {}_{-2}Y_{l\,,-m}] e^{-i\omega v}\right.\\
        \left.\int \d\bar{\omega} [b^+_{lm} \bar{X}^{(4)}_{lm} {}_{-2}Y_{lm} +c^+_{lm} \bar{Z}^{(4)}_{lm} {}_{-2}Y_{l\,,-m}] e^{i\bar{\omega }v}\right)
    \end{split}
\end{equation} with the constants $b^\pm_{lm}$ and $c^\pm_{lm}$ given by Eq.~\eqref{eq:b} and the purely radial functions $X^{(4)}_{lm}$ and $Z^{(4)}_{lm}$ (a priori independent) normalized such that $X^{(4)}_{lm}(r=0)=\bar{X}^{(4)}_{lm}(r=0)=Z^{(4)}_{lm}(r=0) = \bar{Z}^{(4)}_{lm}(r=0)=1$. Using this ansatz to solve the Teukolsky equation, we need to solve four independent equations, one for each term proportional to the constants $b_{lm}^\pm$ and $c_{lm}^\pm$, i.e.,
\begin{equation}\label{eq:Teukolsky-psi4-split}
    \mathcal{O}_{T_4} \widetilde{\Psi}_4=0 \Rightarrow  
   \left\{ \begin{matrix}
      &   \mathcal{O}_{T_4} \left( X^{(4)}_{lm} e^{-i\omega v} {}_{-2}Y_{lm}\right)=0\,,\\
      &   \mathcal{O}_{T_4} \left( \bar{X}^{(4)}_{lm} e^{i\bar{\omega} v} {}_{-2}Y_{lm}\right)=0\,,\\
      &   \mathcal{O}_{T_4} \left( Z^{(4)}_{lm} e^{-i\omega v} {}_{-2}Y_{l\,,-m}\right)=0\,,\\
      &   \mathcal{O}_{T_4} \left( \bar{Z}^{(4)}_{lm} e^{i\bar{\omega} v} {}_{-2}Y_{l\,,-m}\right)=0\,.
    \end{matrix}\right.
\end{equation}
The first and third expressions yield exactly Eq.~\eqref{eq:teukolsky-extended:2} since the eigenvalue of the angular operator $\bar{\eth} \eth$ does not depend on the index $m$. Hence, it follows that $Z^{(4)}_{lm} = X^{(4)}_{lm}$.
Further, since the spin coefficients and Weyl scalars for the background spacetime are real, the second and fourth expressions in Eq.~\eqref{eq:Teukolsky-psi4-split}, i.e., $ \mathcal{O}_{T_4} \left( \bar{X}^{(4)}_{lm} e^{i\bar{\omega} v} {}_{-2}Y_{lm}\right)=0$ yields the complex conjugate of Eq.~\eqref{eq:teukolsky-extended:2}, so $\bar{X}^{(4)}_{lm}$ is the complex conjugate of $X^{(4)}_{lm}$ and $\bar{Z}^{(4)}_{lm} = \bar{X}^{(4)}_{lm}$. Therefore, we only need to explicitly solve the Teukolsky equation for the first term in Eq.~\eqref{eq:Teukolsky-psi4-split}, i.e., Eq.~\eqref{eq:teukolsky-extended:2}, which yields
\begin{equation}\label{eq:psi4}
\begin{split}
      \widetilde{\Psi}_4 =  \frac{1}{2\pi}\sum_{l,m}  & \left(\int_{-\infty}^\infty \d\omega e^{-i \omega v} [b_{lm}^- {}_{-2} Y_{lm} +c^-_{lm} {}_{-2} Y_{l\,,-m}]   X^{(4)}_{lm}\right.\\
      &\left.+ \int_{-\infty}^\infty \d\bar{\omega}e^{i\bar{\omega}v} [b_{lm}^+ {}_{-2}Y_{lm}+c^+_{lm} {}_{-2}Y_{l\,,-m}]  \bar{X}^{(4)}_{lm}  \right)
\end{split}
\end{equation}
where the radial function $X_{lm}^{(4)}$ is a combination of confluent Heun functions
\begin{equation}\label{eq:radial-psi4}
\begin{split}
     X^{(4)}_{lm}(r)= k_1 H_c[(l-2)(l+3)+5\iota,5\iota,3-\iota,3,\iota,-\frac{r}{c}]\\
     +k_2 \left(-\frac{r}{c}\right)^{\iota-2} H_c[l(l+1) +\iota^2,\iota(3+\iota),\iota-1,3,\iota,-\frac{r}{c}]
\end{split}
\end{equation} with $\iota= 2ci\omega$, and normalized such that 
\begin{equation}\label{eq:confluent-Heun-normalization}
    H_c[\alpha,\beta,\gamma,\delta,\sigma, z] =1-\frac{\alpha}{\gamma} z+\mathrm{O}(z^2)\,.
\end{equation}
The function $\bar{X}_{lm}^{(4)}$ is just the complex conjugate of Eq.~\eqref{eq:radial-psi4}. The integration constants $k_1$ and $k_2$ must be chosen such that 
$\widetilde\Psi_4 (r=0)$ coincides with Eq.~\eqref{eq:solution1:10}. Although the confluent Heun functions are regular at the horizon, the prefactor $(-r/c)^{\iota-2}$ makes the second solution of the radial Teukolsky equation for $\widetilde{\Psi}_4$ singular at the horizon.  Given that we are searching for regular solutions at the horizon we could already discard this second solution by setting $k_2=0$. Nevertheless, we will not fix this constant to keep the discussion general.
In Sec.~\ref{sec:QNM-frequencies},  we will show that $k_2$ should vanish after all, as it describes an unphysical unstable solution in the future.

Eqs.~\eqref{eq:psi4}-\eqref{eq:confluent-Heun-normalization} fully specify the perturbation to the Weyl scalar $\Psi_4$. In Refs.~\cite{PhysRevD.80.124001,fiziev2009classes}, it was already proven that the Teukolsky equation could be solved analytically using confluent Heun functions, but the results here differ given that we use a different coordinate system and different gauge conditions.

Up to this point, we have been using the term ``function'' lightly to
refer to the confluent Heun functions. However, an analytic
``function'' in the mathematical sense should have a representation in
terms of a series with certain convergence properties. This is not
always true for the confluent Heun functions, for which a uniformly
convergent series can only be built from its regular singular point at
the horizon to the irregular singular point at $r=\infty$ for certain
values of $\omega_{lm}$, the so-called eigenfrequencies of the
series~\cite{Leaver:1985ax,Leaver1985}. This is in principle not a
problem, but imposing analyticity in the solution is a very strong
requirement: to define a metric that gives rise to a well-behaved
Riemann curvature tensor, we only need to require solutions that are
finitely differentiable (at least $C^2$ so that the curvature is
continuous \cite{Hawking:1973uf}).
However, in this first application of black hole tomography, we wish to describe the metric of a ringing-down black hole, which is an analytic solution. Requiring the solution for $\Psi_4$ (and the spacetime metric) to be analytic, i.e., that all the relevant functions have a valid series representation between the horizon and infinity, will give rise to the identification of $\omega$ with the frequencies of the QNMs $\omega_n$, as we will see in Sec.~\ref{sec:QNM-frequencies}. 

Further, we need to impose physical boundary conditions to our solution: no radiation can escape from the black hole and no radiation should leak into the spacetime from infinity. To specialize our general solution to this physical situation, we will need to impose boundary conditions for \textit{both} $\widetilde{\Psi}_0$ and $\widetilde{\Psi}_4$ at the horizon and asymptotically. So in order to impose these boundary conditions, we first need to solve the radial differential equation~\eqref{eq:teukolsky-extended:1} for $\widetilde{\Psi}_0$, which we obtain by using the Teukolsky operator for $\widetilde{\Psi}_0$ over the ansatz
\begin{equation}\label{eq:psi0}
    \begin{split}
       \widetilde{\Psi}_0 =   &\frac{1}{2\pi} \sum_{l\,,m} \left(\int \d\omega a_{lm}^-e^{-i\omega v} X^{(0)}_{lm}\right.\\
        &\left.+\int \d\bar{\omega} a_{lm}^+e^{i\bar{\omega} v} \bar{X}^{(0)}_{lm}\right){}_{2}Y_{lm}\,.
    \end{split}    
\end{equation} Analogously to the discussion for $\widetilde{\Psi}_4$, the radial function $X^{(0)}_{lm}$ is obtained by solving $\mathcal{O}_{T_0} \widetilde{\Psi}_0=0$, which yields Eq.~\eqref{eq:teukolsky-extended:1}. The radial function $\bar{X}^{(0)}_{lm}$ satisfies the complex conjugate of Eq.~\eqref{eq:teukolsky-extended:1} and is therefore the complex conjugate of $X^{(0)}_{lm}$.  This function is again assumed to be normalized such that $X^{(0)}_{lm}(r=0)=1$.  However, despite the similarity of the radial equations for $\widetilde{\Psi}_0$ and $\widetilde{\Psi}_4$, the radial part of $\widetilde{\Psi}_0$ does not have a solution in terms of confluent Heun functions. Rather, the solution will be a mixture of polynomial terms, Heun functions, and their derivatives, i.e., 
\begin{widetext}
\begin{equation}\label{eq:psi0-radial-explicit}
    \begin{split}
        X^{(0)}_{lm} (r) = & k_3\left(f^{(1)}_{lm}(r) H_c[(l-2)(l+3)+5\iota,5\iota,3-\iota,3,\iota,-\frac{r}{c}]+g^{(1)}_{lm}(r) \partial_r H_c[(l-2)(l+3)+5\iota,5\iota,3-\iota,3,\iota,-\frac{r}{c}]\right)\\
        +& k_4 \left(-\frac{r}{c}\right)^{\iota} \left( f_{lm}^{(2)}  H_c[l(l+1) +\iota^2,\iota(3+\iota),\iota-1,3,\iota,-\frac{r}{c}] + g_{lm}^{(2)}(r) \partial_r  H_c[l(l+1) +\iota^2,\iota(3+\iota),\iota-1,3,\iota,-\frac{r}{c}]\right)
    \end{split}
\end{equation}  where $f^{(1,2)}_{lm}$ and $g^{(1,2)}_{l,m}$ are complex, rational functions of $r$ and the frequency $\omega_{lm}$. The functions $f_{lm}^{(1)}$ and $g_{lm}^{(1)}$ are regular at the horizon, as expected. The functions $f_{lm}^{(2)}$ and $g_{lm}^{(2)}$ behave as $1/r$ in the $r\to 0$ limit. 
 \end{widetext}
  We give explicit expressions for these function in Eqs.~\eqref{eqs:f-g-fncts}-\eqref{eqs:h2} of Appendix~\ref{sec:complementary}. As before, the integration constants $k_3$ and $k_4$ will be chosen such that the boundary condition at the horizon in Eq.~\eqref{eq:psi0-horizon} is satisfied.

\subsection{The Teukolsky-Starobinsky identities}
\label{sec:TSI}

We showed that by projecting the field equations at the horizon, we can economically solve not only the geometry of the horizon but also directly relate $\widetilde{\Psi}_0$ and $\widetilde{\Psi}_4$ everywhere: Given that the boundary conditions at the horizon already provide the relationship between these two quantities, which can then be used as the boundary conditions to solve their respective Teukolsky equations. 
 Alternatively, we could have used the Teukolsky-Starobinsky identities to show the relationship between the quantities $\widetilde{\Psi}_0$ and $\widetilde{\Psi}_4$~\cite{Teukolsky:1974yv,starobinsky}. 
 These identities are fourth-order partial differential equations that can be derived by differentiating three times the radial and evolution equations~\eqref{eqs:radial-bianchi:0} and ~\eqref{eqs:evolution-bianchi:4} using the $\Delta$ and $D$ derivative operators respectively, and using the perturbative version of the Bianchi identities~\eqref{eqs:radial-bianchi} and~\eqref{eqs:evolution-bianchi} to remove the dependence on $\widetilde{\Psi}_1\,,\widetilde{\Psi}_2$ and $\widetilde{\Psi}_3$. This number of derivatives can be shown to be the minimum needed to derive a differential equation for $\widetilde{\Psi}_0$ sourced by $\widetilde{\Psi}_4$ only (apart from the background quantities), and the other way around. In this sense, the result that we obtained is stronger as it involves fewer derivatives to relate $\widetilde{\Psi}_0$ and $\widetilde{\Psi}_4$.
 
 Here, we present the Teukolsky-Starobinsky identities without proof, as they have been discussed extensively in the literature (see for instance~\cite {Pound_2021}) 
\begin{subequations}\label{eq:starobinsky-identities}
    \begin{align}
        \label{eq:starobinsky-identities:2}
          \text{\textthorn}^{\prime 4}[(r+c)^4 \widetilde{\Psi}_0] &=\eth^4 [(r+c)^4 \widetilde{\Psi}_4] +\frac{3}{2}c\mathcal{L}_{\xi} \widetilde{\bar{\Psi}}_4\,,\\ \label{eq:starobinsky-identities:1}
         \text{\textthorn}^4[(r+c)^4 \widetilde{\Psi}_4] &=\bar{\eth}^4 [(r+c)^4 \widetilde{\Psi}_0] -\frac{3}{2}c\mathcal{L}_{\xi} \widetilde{\bar{\Psi}}_0\,.
    \end{align}
\end{subequations} The thorn operator $\textthorn$ and its primed version are combinations of the directional derivatives. In particular, 
\begin{equation}\label{eq:def-thorn}
     \text{\textthorn} \eta = (D -p \epsilon -q \bar{\epsilon}) \eta\,,\quad  \text{\textthorn}^\prime \eta = (\Delta +p\gamma+q \bar{\gamma}) \eta = \Delta \eta\,,
\end{equation} where the last equality in the second expression follows because we choose a coordinate system and tetrad such that the vectors $\ell^\mu$, $n^\mu$, and $m^\mu$ are parallel propagated along $n^\alpha$ and henceforth $\gamma = 0 $.  In Eq.~\eqref{eq:def-thorn}, $p$ and $q$ are well defined for each spin-coefficient and Weyl scalar (see for instance, Ref.~\cite{Pound_2021}), and determine the spin-weight $s=(p-q)/2$ and the boost-weight $b=(p+q)/2$ of each quantity. The operator $\text{\textthorn}$  raises the boost of the quantity on which it acts. Hence, if $\Psi_4$ has spin-weight -2 and boost-weight $-2$,   $\text{\textthorn} \Psi_4$ has the same spin-weight and boost-weight $-1$. Similarly $\Delta$ acts as a boost-lowering operator. The operator $\mathcal{L}_\xi$ in Eq.~\eqref{eq:starobinsky-identities} is  \begin{equation}
    \mathcal{L}_\xi = -(r+c) [\mu \text{\textthorn} +\rho \Delta +\frac{p}{2} \Psi_2+\frac{q}{2} \bar{\Psi}_2]\,, 
\end{equation} which for a Schwarzschild background simplifies to
\begin{equation}
     \mathcal{L}_\xi = -(r+c) [\mu \partial_v ] = \partial_v\,.
\end{equation}
In the following, we show that our solution (Eq.~\eqref{eq:solution1}, together with the solutions of the Teukolsky equations for $\widetilde{\Psi}_0$ and $\widetilde{\Psi}_4$ in Eqs.~\eqref{eq:psi4} and~\eqref{eq:psi0}) automatically satisfies the Teukolsky-Starobinsky identities.  Furthermore, we can use Eq.~\eqref{eq:starobinsky-identities:1} to rewrite the radial part of $\widetilde{\Psi}_0$ (Eq.~\eqref{eq:psi0-radial-explicit}) more compactly.

We start by showing that the solution we found for $\widetilde{\Psi}_4$ in Eq.~\eqref{eq:psi4} using the data at the horizon is consistent with the solution for  $\widetilde{\Psi}_0$ in Eq.~\eqref{eq:psi0} through the first expression in Eq.~\eqref{eq:starobinsky-identities}.  Recalling the action of the eth operator on the spin-weighted spherical harmonics 
\begin{equation}
    \eth {}_s Y_{lm}=\frac{\sqrt{(l-s)(l+s+1)}}{\sqrt{2} (r+c)} {}_{s+1} Y_{lm}\,,
\end{equation} that the spin-weighted spherical-harmonics satisfy ${}_s\bar{Y}_{l\,,m} = (-1)^{s+m} Y_{l\,,-m}$, 
and collecting terms proportional to $e^{-i\omega v}$ and $e^{i\bar{\omega} v}$, we can rewrite Eq.~\eqref{eq:starobinsky-identities:2} as
\begin{equation}\label{eq:starobinsky-psi0-intermediate}
    \begin{split}
      &  \Delta^4[(r+c)^4\widetilde{\Psi}_0]=\frac{1}{8\pi}\sum_{l,m} 
       \left\{\int \d\omega e^{-i\omega v} X_{lm}^{(4)} \times\right.\\
     &\left( {}_2Y_{lm}[l(l+1)(l+2)(l-1)b_{lm}^- -(-1)^m 6 c i \omega \bar{c}_{lm}^+ ]+ \right.\\
     & \left. {}_2Y_{l\,,-m}[l(l+1)(l+2)(l-1)c_{lm}^- -(-1)^m 6 c i \omega \bar{b}_{lm}^+ ] \right)\\
      &+ \int \d\bar{\omega} e^{i\bar{\omega} v} \bar{X}_{lm}^{(4)} \times \\
      &\left( {}_2Y_{lm}[l(l+1)(l+2)(l-1)b_{lm}^+ +(-1)^m 6 c i \bar{\omega} \bar{c}_{lm}^- ]+ \right.\\
     &\left. \left. {}_2Y_{l\,,-m}[l(l+1)(l+2)(l-1)c_{lm}^+ +(-1)^m 6 c i \bar{\omega} \bar{b}_{lm}^- ] \right)\right\} \, .
    \end{split}
\end{equation} 
Plugging in the expressions for the constants $b_{lm}^\pm$ and $c^\pm_{lm}$~\eqref{eq:b} into the equation above, we see that the terms proportional to ${}_2Y_{l\,,-m}$ vanish, thus yielding
\begin{equation}\label{eq:delta4-psi0}
    \Delta^4 [(r+c)^4 {X}_{lm}^{(0)}] =\frac{ l^2(l-1)^2 (l+1)^2 (l+2)^2 +36 c^2 \omega^2}{16 c^4 i \omega (\kappa_{(l)}^2+\omega^2)(2\kappa_{(l)}-i\omega)} X_{lm}^{(4)}
\end{equation} and its complex conjugate. 
Notice that this expression is independent of the coefficients $a^\pm_{lm}$ and $b^\pm_{lm}/c^\pm_{lm}$, as it should be.
 This identity provides a fourth-order differential equation for the radial part of $\widetilde \Psi_0$, 
which can be solved through the integration of the confluent Heun function. Eq.~\eqref{eq:delta4-psi0} can be shown to hold either by taking the derivatives of Eq.~\eqref{eq:psi0-radial-explicit} or by using the properties of the Heun function~\cite{PhysRevD.80.124001,fiziev2009classes}. Similarly, it can also be shown that the solutions we found for $\widetilde{\Psi}_0$  and $\widetilde{\Psi}_4$ satisfy the second Starobinsky identity~\eqref{eq:starobinsky-identities:1}.  

Given that we have already shown that the Teukolsky-Starobinsky identities are satisfied everywhere, we will use the second expression in Eq.~\eqref{eq:starobinsky-identities} to write the radial part of  $\widetilde{\Psi}_0$ in Eq.~\eqref{eq:psi0-radial-explicit} in a compact form. The idea is to obtain an expression for $X^{(0)}_{lm}$ in terms of a derivative operator acting on $X^{(4)}_{lm}$. 
Using that $\widetilde{\Psi}_0$ is of the form of Eq.~\eqref{eq:psi0} with the radial function for $\widetilde\Psi_0$ in ~\eqref{eq:psi0-radial-explicit} normalized at the horizon $X^{(0)}_{lm} (r=0) =1$, and using Eq.~\eqref{eq:psi4}, we can rewrite Eq.~\eqref{eq:starobinsky-identities:1} as 
\begin{equation}\label{eq:starobinski-1-extended}
    \begin{split}
       & \text{\textthorn}^4[(r+c)^4 \widetilde \Psi_4]=\\
       & \frac{1}{2\pi}\sum_{l,m}   \left(\int \d\omega e^{-i\omega v}X_{lm}^{(0)}\times \right.\\
      & \left[ \frac{a_{lm}^-}{4}l(l+1)(l-1)(l+2){}_{-2}Y_{lm}+\frac{3}{2}c(-1)^m i\omega\bar{a}^+_{lm}{}_{-2}Y_{l\,,-m}\right]\\
       & +\int\d\bar{\omega} e^{i\bar{\omega} v}\bar{X}_{lm}^{(0)} \times\\
     & \left. \left[ \frac{a_{lm}^+}{4}l(l+1)(l-1)(l+2){}_{-2}Y_{lm}-\frac{3}{2}c(-1)^m i\bar{\omega}\bar{a}^-_{lm}{}_{-2}Y_{l\,,-m}\right]\right)\,.
    \end{split}
\end{equation} 
Using Eq.~\eqref{eq:b}, the first term in brackets can be simplified to 
\begin{equation}
    c^4 i\omega (\kappa_{(l)}^2+\omega^2) (2\kappa_{(l)} -i\omega) [b^-_{lm} {}_{-2}Y_{lm} +c^-_{lm} {}_{-2}Y_{l\,,-m}] \, .
\end{equation}
Plugging in Eq.~\eqref{eq:starobinski-1-extended}, the expression for $\widetilde{\Psi}_4$~\eqref{eq:psi4} and taking into account that our solution for $\widetilde{\Psi}_4$ is a linear combination of two solutions of the Teukolsky equation, we can obtain the radial function $X^{(0)}_{lm}$ by separating this equation in two terms, one proportional to $e^{-i\omega v}$, and the other to $e^{i\bar{\omega} v}$. Simplifying, we obtain the following analytic expression for $X_{lm}^{(0)}$ 
\begin{equation}\label{eq:radial-psi0}
    X_{lm}^{(0)} = \frac{e^{i\omega v}\text{ \textthorn}^4[(r+c)^4 e^{-i\omega v} X_{lm}^{(4)}]}{c^4 i\omega (\kappa_{(l)}^2+\omega^2)(2\kappa_{(l)}-i\omega)}\,,
\end{equation} which is independent of $v$. Therefore, the full solution for $\widetilde{\Psi}_0$ is
\begin{equation}\label{eq:psi0-final}
\begin{split}
     \widetilde{\Psi}_0 =\frac{1}{2\pi}\sum_{l,m}  {}_2Y_{lm} \left(\int \d\omega \frac{ a_{lm}^- \text{ \textthorn}^4[(r+c)^4 e^{-i\omega v} X_{lm}^{(4)}]}{c^4 i\omega (\kappa_{(l)}^2+\omega^2)(2\kappa_{(l)}-i\omega)}\right.\\
    \left. - \int \d\bar{\omega} \frac{ a_{lm}^+ \text{ \textthorn}^4[(r+c)^4 e^{i\bar{\omega} v} \bar{X}_{lm}^{(4)}]}{c^4 i\bar{\omega} (\kappa_{(l)}^2+\bar{\omega}^2)(2\kappa_{(l)}+i\bar{\omega})} \right)\,. 
\end{split}
\end{equation} Notice that Eq.~\eqref{eq:psi0-final} is a solution of the Teukolsky equation for $\widetilde{\Psi}_0$ in~\eqref{eq:teukolsky-extended:1}  and satisfies the boundary condition in~\eqref{eq:psi0-horizon}, i.e., 
\begin{equation}
    \widetilde \Psi_0 \triangleq \frac{1}{2\pi} \sum_{l,m} {}_2 Y_{lm} \left(\int \d\omega a_{lm}^- e^{-i\omega v}+\int \d\bar{\omega} a_{lm}^+ e^{i\bar{\omega} v} \right) 
\end{equation} as expected.
It can be shown that the expressions that we found for $\widetilde{\Psi}_0$ and $\widetilde{\Psi}_4$ (Eqs.~\eqref{eq:psi4} and~\eqref{eq:psi0-final})  satisfy the Teukolsky-Starobinsky identities~\eqref{eq:starobinsky-identities}, as well as their decoupled version (see Eq.~(66) in~\cite{Pound_2021})
\begin{subequations}\label{eq:starobinsjy-decoupled}
    \begin{align}
       \begin{split}
       \text{\textthorn}^4[(r+c)^4\Delta^4((r+c)^4 \widetilde{\Psi}_0)] =& \eth^4((r+c)^4 \bar{\eth}^4[(r+c)^4 \widetilde{\Psi}_0])\\
           &-\frac{9}{4}c^2 \mathcal{L}_\xi^2 \widetilde{\Psi}_0
       \end{split} \\
        \begin{split}
       \Delta^4[(r+c)^4\text{\textthorn}^4((r+c)^4 \widetilde{\Psi}_4)] =& \bar{\eth}^4((r+c)^4 \eth^4[(r+c)^4 \widetilde{\Psi}_4])\\
           &-\frac{9}{4}c^2 \mathcal{L}_\xi^2 \widetilde{\Psi}_4
       \end{split}
    \end{align}
\end{subequations}
and are a solution of the Teukolsky equation~\eqref{eq:teukolsky-extended}. As a side remark, a solution of the Teukolsky-Starobinsky identities in Eq.~\eqref{eq:starobinsky-identities} will automatically satisfy their decoupled version (Eq.~\eqref{eq:starobinsjy-decoupled}), but the converse is not true. Since the decoupled system in Eq.~\eqref{eq:starobinsjy-decoupled} is of eighth order and the coupled Starobinsky identities in Eq.~\eqref{eq:starobinsky-identities} are only fourth order, there exist solutions of Eq.~\eqref{eq:starobinsjy-decoupled} which are not solutions of~\eqref{eq:starobinsky-identities}. Hence, that our solution satisfies the decoupled version of these identities in Eq.~\eqref{eq:starobinsjy-decoupled} is a priori trivial and is regarded solely as a consistency check. The same logic applies to the Teukolsky equation and the coupled Teukolsky-Starobinsky identities: A solution of the Teukolsky equations~\eqref{eq:teukolsky-extended}
will automatically satisfy the Teukolsky-Starobinsky identities in Eq.~\eqref{eq:starobinsky-identities}, but the converse is not true. This is why we need to check whether Eq.~\eqref{eq:psi0} satisfies  the Teukolsky equation~\eqref{eq:teukolsky-extended:1}.

In this discussion, we only used the Teukolsky-Starobinsky identities to check the consistency of our solution, as well as to rewrite the solution of the Teukolsky equation for $\widetilde{\Psi}_0$ in a more convenient way. However, we would like to remark that the horizon equations~\eqref{eq:eqs-horizon} already provide the link between $\widetilde{\Psi}_0$ and $\widetilde{\Psi}_4$ on the horizon, and their radial dependence can be found directly by solving the Teukolsky equation for these two quantities~\eqref{eq:teukolsky-extended}. The Teukolsky-Starobinsky identities further imply that the two independent solutions of the radial Teukolsky equation~\eqref{eq:radial-psi4} for $\widetilde{\Psi}_4$ are linked to those of $\widetilde{\Psi}_0$, i.e., if $k_1=0$ in Eq.~\eqref{eq:radial-psi4}, 
then the constant $k_3 =0$ needs to vanish in Eq.~\eqref{eq:psi0-final}.

\section{Boundary conditions and Quasinormal modes}
\label{sec:QNM-frequencies}

So far we have solved the field equations on a null hypersurface,
which represents the isolated horizon, slightly perturbed by an
infalling flux specified by $\widetilde{\Psi}_0$. Given this initial
data at the null hypersurface and the radial dependence of
$\widetilde{\Psi}_0$ and $\widetilde{\Psi}_4$ specified in the bulk
(through the solution of the radial Teukolsky
equations~\eqref{eq:teukolsky-extended}), we can integrate the field
equations everywhere to obtain the full spacetime solution. Additional
restrictions must be imposed to obtain the QNMs.  
There are two natural sets of conditions that select the QNMs.
The usual formulation of QNMs requires purely outgoing waves
from the domain both at the horizon and infinity.
Alternatively, and that is the path we will take here, requiring that the solutions are analytic and stable towards the future (i.e., we discard growing modes) also selects the QNMs. In Sec.~\ref{subsec:Bdy-conditions}, we verify explicitly that the second set of boundary conditions indeed is consistent with only outgoing radiation at future null infinity, no incoming radiation from past null infinity and ingoing modes at the horizon. 

\subsection{QNMs}
\label{subsec:QNM-frequencies}

We start by analyzing the convergence of the two independent solutions~\eqref{eq:radial-psi4} to the radial Teukolsky equation for $\widetilde{\Psi}_4$~\eqref{eq:teukolsky-extended:2}. Using the coordinate shift 
\begin{equation}
    x=r+c
\end{equation} and the transformation
\begin{equation}
    X_{lm}^{(4)} = e^{i\omega x} y(x)   
\end{equation} we can rewrite Eq.~\eqref{eq:teukolsky-extended:2} in the form of Eq.~(1) in Ref.~\cite{LeaverExpansions1986}
\begin{equation}\label{eq:diff-eq-Leaver}
x(x-c) y^{\prime\prime} +[B_1+B_2 x] y^\prime +[\omega^2 x(x-c) -2\eta \omega (x-c) +B_3]y=0    
\end{equation} with
\begin{subequations}\label{eq:coefficients-diff-Eq-Leaver}
  \begin{align}
    &B_1=-3c\,,\quad B_2 = 6-2i c \omega\,,\quad \eta=2i -c\omega\,,\\
    & B_3 = 6-l(l+1) -7 i c \omega +2 c^2 \omega^2\,.
\end{align}  
\end{subequations}
Notice that these coefficients \textit{differ} from the ones in Eq.~(11) of Ref.~\cite{LeaverExpansions1986} for the Schwarzschild geometry\footnote{In  Ref.~\cite{LeaverExpansions1986}, the wave equation for the Schwarzschild background is expressed in the form of the generalized wave equation~\eqref{eq:diff-eq-Leaver} with $c=1$ and the coefficients $B_1= -(2s+1)$, $B_2= 2(s+1-i\omega)$, $\eta=-\omega$, and $B_3=2\omega^2 -l(l+1) +s(s+1) -(2s+1) i\omega$. These coefficients can be extracted by inspection from Eq.~(11) of Ref.~\cite{LeaverExpansions1986}. For a gravitational perturbation $s=2$.}. This difference is to be expected since we are using a different coordinate system from the usual one.
Since a priori, it is not obvious the solution of the differential equation~\eqref{eq:diff-eq-Leaver} with the coefficients~\eqref{eq:coefficients-diff-Eq-Leaver} coincides with the one in Refs.~\cite{Leaver1985,LeaverExpansions1986}, we explicitly construct the series expansion representation of the two independent solutions for the radial part of $\widetilde{\Psi}_4$ ($X^{(4,1)}_{l,m}$ and $X^{(4,2)}_{lm}$) and analyze their eigenfrequencies.   

The first solution of Eq.~\eqref{eq:diff-eq-Leaver} can be written as a regular series expansion around the regular singular point of the differential equation at $x=c$ ($r=0$)~\cite{LeaverExpansions1986}
\begin{equation}\label{eq:Jaffe-series-Psi4-sol1}
    y(x) = e^{i \omega x} x^{-1+2ic\omega} \sum_{k=0}^\infty a_k \left(\frac{x-c}{x}\right)^k\,,
\end{equation} where the coefficients of the series are given by Eqs.~(40) and (41) in Ref.~\cite{LeaverExpansions1986}
\begin{equation}\label{eq:a-series}
    \alpha_0 a_1+\beta_0 a_0=0\,,\quad \alpha_k a_{k+1} +\beta_k a_k +\gamma_k a_{k-1}=0\,,
\end{equation} 
with
\begin{subequations}  \label{eq:alpha-beta-gamma}  \begin{align}\label{eq:alpha-beta-gamma}
    \alpha_k &= (k+1)(3+k-2i c\omega)\\
    \beta_k &= 3-l(l+1) -2k^2+4i c\omega +8 c^2 \omega^2 +k(-2+8ic\omega)\\
    \gamma_k & = (k-2ic \omega) (k-2-2ic \omega)\,.
\end{align}
\end{subequations}
Note that the coefficients $a_k$ also depend on the $l$ index explicitly (through $\beta_k$) and on the $m\,,n$ indices implicitly through the QNM frequencies. To avoid cluttering of notation, we suppress these indices. 
Then, the first independent solution of Eq.~\eqref{eq:teukolsky-extended:2} admits the series expansion 
\begin{equation}\label{eq:Psi4-series1}
    X_{lm}^{(4,1)}=e^{2i \omega r} \left(\frac{r+c}{c}\right)^{-1+2ic\omega} \sum_{k=0}^\infty a_k \left(\frac{r}{r+c}\right)^k\,,
\end{equation} which is regular around $r=0$.  The coefficient $a_0=1$ in Eq.~\eqref{eq:Jaffe-series-Psi4-sol1} is chosen to coincide with the normalization of the confluent Heun function~\eqref{eq:confluent-Heun-normalization}. However, notice that the confluent Heun function only has an analytic representation in the domain $r\in[0,\infty)$  when the convergence of the series~\eqref{eq:Psi4-series1} is uniform~\cite{LeaverExpansions1986}. This occurs when the frequency $\omega$ is a root of the infinite fraction~\cite{LeaverExpansions1986}
\begin{equation}\label{eq:QNM-frequencies}
    0=\beta_0-\frac{\alpha_0\gamma_1}{\beta_1-\frac{\alpha_1\gamma_2}{\beta_2-...}}\,,
\end{equation} which for the differential equation~\eqref{eq:diff-eq-Leaver} with the coefficients~\eqref{eq:coefficients-diff-Eq-Leaver} yields the quasinormal mode (QNMs) frequencies $\omega=\omega_{lmn}^{\text{QNM}}$ (see Tab.~1 in Ref.~\cite{Leaver1985}). It is customary to sort the roots of Eq.~\eqref{eq:QNM-frequencies} according to their ascendent imaginary values using the index $n$, which denotes the overtone number. In other words, the first solution of the Teukolsky equation~\eqref{eq:teukolsky-extended:2} is in reality given by
\begin{equation}\label{eq:psi4-sol1}
    H_c[(l-2)(l+3)+5\iota,5\iota,3-\iota,3,\iota,-\frac{r}{c}]\delta(\omega-\omega_n^{\text{QNM}})\,,
\end{equation} 
where we have once more suppressed the $l\,,m$ harmonic indices from the frequencies.  Notice that it is quite remarkable that we find the QNM frequencies $\omega=\omega_n$ by solving Eq.~\eqref{eq:QNM-frequencies} given that the coefficients (Eq.~\eqref{eq:coefficients-diff-Eq-Leaver}) appearing in the generalized spheroidal  wave equation~\eqref{eq:diff-eq-Leaver} are different from those found by Leaver in~\cite{LeaverExpansions1986,Leaver1985}. 
This happens because 
the coefficients entering the three-term recurrence relation Eq.~\eqref{eq:alpha-beta-gamma}, which are also the ones appearing in the infinite fraction~\eqref{eq:QNM-frequencies}, give rise to the same eigenfrequencies 
as those reported in Refs.~\cite{Leaver1985,LeaverExpansions1986}.

The second independent solution of Eq.~\eqref{eq:teukolsky-extended:2} can be obtained from Eq.~\eqref{eq:diff-eq-Leaver} by further using the transformation
\begin{equation}
    y(x) = (x-c)^{-2+2ic\omega}g(x)\,,
\end{equation} which yields the following differential equation for $g(x)$
\begin{equation}
    x(x-c) g^{\prime\prime} +[\tilde B_1 +\tilde B_2 x] g^\prime +[\omega^2 x(x-c)-2\tilde \eta \omega (x-c) +\tilde B_3]g=0
\end{equation} 
with
\begin{subequations}
\begin{align}\label{eq:coefficients-diff-Eq-Leaver-2}
   & \tilde B_1=-3c \,,\quad \tilde B_2 = 2(1+ic\omega)\,,\quad \tilde \eta= 2i-c\omega\\
   & \tilde B_3 = -l(l+1) -ic\omega +2c^2\omega^2\,.
\end{align}
\end{subequations} The second independent solution of Eq.~\eqref{eq:teukolsky-extended:2}~\footnote{Notice that these two solutions Eqs.~\eqref{eq:psi4-sol1} and~\eqref{eq:psi4-sol2} of Eq.~\eqref{eq:teukolsky-extended:2} are independent. This can be seen by computing the Wronskian of the two solutions.} can be expanded in a regular series at the horizon using Eqs.~(49)-(51) of Ref.~\cite{Leaver1985}, which yields
\begin{equation}\label{eq:Psi4-series2}
    g(x) = e^{-i\omega x} x^{-3-2i c \omega} \sum_{k=0}^\infty b_k \left(\frac{x-c}{x}\right)^k 
\end{equation} with
\begin{equation} \label{eq:b-series}
      \tilde\alpha_0 b_1+\tilde\beta_0 b_0=0\,,\quad \tilde\alpha_k b_{k+1} +\tilde\beta_k b_ k+\tilde\gamma_k b_{k-1}=0\,,
\end{equation} and
\begin{subequations}\label{eq:alpha-beta-gamma-tilde}
    \begin{align}
    \tilde\alpha_k &= (k+1)(k-1+2i c\omega)\\
    \tilde\beta_k &= 3-l(l+1) -2k(k+1)-4i c\omega(1+2k) +8 c^2 \omega^2 \\
   \tilde \gamma_k & = (k+2ic \omega) (k+2+2ic \omega)\,.
\end{align}
\end{subequations} The convergence of the series~\eqref{eq:Psi4-series2} is uniform when the sum $\sum_k b_k$ is finite~\cite{LeaverExpansions1986}. This occurs for the roots of equation~\eqref{eq:QNM-frequencies} (replacing the coefficients $\alpha_k$, $\beta_k$ and $\gamma_k$ by their counterparts in Eq.~\eqref{eq:alpha-beta-gamma-tilde}). The eigenfrequencies turn out to be the complex conjugate of the QNM frequencies. This can be seen by comparing Eqs.~\eqref{eq:alpha-beta-gamma-tilde} and~\eqref{eq:alpha-beta-gamma}. The following relationship holds between these coefficients
\begin{equation}
    \tilde \beta_k = \bar{\beta}_k\,,\quad \tilde \alpha_k\tilde\gamma_{k+1} = \bar{\alpha}_k\bar{\gamma}_{k+1} \, ,
\end{equation} 
so the equation for the eigenfrequencies of the second solution is the complex conjugate of Eq.~\eqref{eq:QNM-frequencies}. Naturally, the solutions of this equation are the complex conjugates of the QNM frequencies. 
Then, the second solution of Eq.~\eqref{eq:teukolsky-extended:2} is 
\begin{equation}
  \delta(\omega-\bar{\omega}_n^{\text{QNM}})  \left(-\frac{r}{c}\right)^{-2+2ic\omega} \left(\frac{r+c}{c}\right)^{-3-2ic\omega} \sum_{k=0} b_k \left(\frac{r}{r+c}\right)^k
\end{equation} 
or using the confluent Heun function expression
\begin{equation}\label{eq:psi4-sol2}
    \delta(\omega-\bar{\omega}_n^{\text{QNM}})  \left(-\frac{r}{c}\right)^{-2+\iota}   H_c[l(l+1) +\iota^2,\iota(3+\iota),\iota-1,3,\iota,-\frac{r}{c}]\,.
\end{equation}

Therefore, the radial function $X^{(4)}_{lm}$ reads
\begin{widetext}
    \begin{equation}\label{eq:radial-psi4-jaffe}
    \begin{split}
         X^{(4)}_{lm}(r) =& \sum_n k_{lmn}^{(1)}\delta(\omega_{lm}-\omega_{lmn}^{\text{QNM}}) e^{2i \omega_{lm} r} \left(\frac{r+c}{c}\right)^{-1+2ic \omega_{lm}} \sum_{k=0}^\infty a_k \left(\frac{r}{r+c}\right)^k\\
        +& \sum_n k_{lmn}^{(2)} \delta(\omega_{lm}-\bar{\omega}_{lmn}^{\text{QNM}}) \left(-\frac{r}{c}\right)^{-2+2ic\omega_{lm}} \left(\frac{r+c}{c}\right)^{-3-2ic\omega_{lm}} \sum_{k=0}^\infty b_k \left(\frac{r}{r+c}\right)^k\\
        =& \sum_n k_{lmn}^{(1)}\delta(\omega_{lm}-\omega_{lmn}^{\text{QNM}}) H_c[(l-2)(l+3)+10 i c \omega_{lm},10 i c \omega_{lm},3-2ic\omega_{lm},3,2ic\omega_{lm},-\frac{r}{c}]\\
         +&\sum_n k_{lmn}^{(2)}\delta(\omega_{lm}-\bar{\omega}_{lmn}^{\text{QNM}})\left(-\frac{r}{c}\right)^{-2+2ic\omega_{lm}}  H_c[l(l+1) -4c^2 \omega_{lm}^2,2ic\omega_{lm}(3+2ic\omega_{lm}),2ic\omega_{lm}-1,3,2ic\omega_{lm},-\frac{r}{c}]\,,
    \end{split}  
\end{equation}
\end{widetext}where $k_{lmn}^{(1,2)}$ are complex, dimensionless integration constants. 
Recall that we fixed the normalization of the radial solution of $\widetilde{\Psi}_4$ to $X_{lm}^{(4)}(r=0) =1$ in order to highlight the role of the boundary conditions at the back hole horizon. Hence, using the second expression in Eq.~\eqref{eq:radial-psi4-jaffe} and the fact that the confluent Heun functions are normalized as in Eq.~\eqref{eq:confluent-Heun-normalization}, it follows that the integration constants $k_{lmn}^{(1,2)}$ need to satisfy
\begin{equation}\label{eq:bdy-conditions-horizon-psi4}
    \lim_{r\to 0} \sum_n \left(k_{lmn}^{(1)} +k_{lmn}^{(2)} \left(-\frac{r}{c}\right)^{-2+2ic\omega_{lmn}^{\text{QNM}}}\right)=1\,.
\end{equation}
This expression emphasizes that the second solution of the radial Teukolsky equation $X_{lm}^{(4,2)}$ is divergent at the horizon in these coordinates. Hence,  for the boundary condition in Eq.~\eqref{eq:bdy-conditions-horizon-psi4} to be satisfied $k_{lmn}^{(2)} =0\, \forall\, n$ and $\re[\sum_n k_{lmn}^{(1)}] =1\,, \im[\sum_n k_{lmn}^{(1)}] =0$. Keeping only the regular solution at the horizon is reasonable in our case because we have explicitly constructed a coordinate system that is horizon penetrating, and therefore should be regular at the horizon. However, to keep the discussion general, we shall not set $k_{lmn}^{(2)}=0$ yet. We will keep these modes in our solution of $\widetilde{\Psi}_4$ to show that these modes correspond to unstable perturbations and are thus not physical.      

The series expansions for the two independent solutions of $\bar{X}_{lm}^{(4)}$ can be found analogously for $\bar{\omega}$. Combining the above discussion with Eq.~\eqref{eq:psi4} and integrating over $\omega$ and $\bar{\omega}$ yields
\begin{widetext}
    \begin{equation}\label{eq:psi4-series}
        \begin{split}
            \widetilde{\Psi}_4 =\frac{1}{2\pi} \sum_{lm}\sum_n & \left[ (b_{lmn}^{-(1)}{}_{-2}Y_{lm} + c_{lmn}^{-(1)} {}_{-2}Y_{l\,,-m}) e^{-i\omega_{lmn}^{\text{QNM}} v} e^{2i \omega_{lmn}^{\text{QNM}} r} \left(\frac{r+c}{c}\right)^{-1+2ic\omega_{lmn}^{\text{QNM}}} \sum_{k=0}^\infty a_k \left(\frac{r}{r+c}\right)^k \right.\\
            + &  (b_{lmn}^{-(2)}{}_{-2}Y_{lm} + c_{lmn}^{-(2)} {}_{-2}Y_{l\,,-m}) e^{-i\bar{\omega}_{lmn}^{\text{QNM}} v} \left(-\frac{r}{c}\right)^{2ic \bar{\omega}_{lmn}^{\text{QNM}}-2 }  \left(\frac{r+c}{c}\right)^{-3-2ic\bar{\omega}_{lmn}^{\text{QNM}}} \sum_{k=0}^\infty b_k \left(\frac{r}{r+c}\right)^k \\
            + & (b_{lmn}^{+(1)}{}_{-2}Y_{lm} + c_{lmn}^{+(1)} {}_{-2}Y_{l\,,-m}) e^{i\bar{\omega}_{lmn}^{\text{QNM}} v} e^{-2i \bar{\omega}_{lmn}^{\text{QNM}} r} \left(\frac{r+c}{c}\right)^{-1-2ic\bar{\omega}_{lmn}^{\text{QNM}}} \sum_{k=0}^\infty \bar{a}_k \left(\frac{r}{r+c}\right)^k\\
            +&\left.(b_{lmn}^{+(2)}{}_{-2}Y_{lm} + c_{lmn}^{+(2)} {}_{-2}Y_{l\,,-m}) e^{i\omega_{lmn}^{\text{QNM}} v} \left(-\frac{r}{c}\right)^{-2ic \omega_{lmn}^{\text{QNM}}-2 }  \left(\frac{r+c}{c}\right)^{-3+2ic\omega_{lmn}^{\text{QNM}}} \sum_{k=0}^\infty \bar{b}_k \left(\frac{r}{r+c}\right)^k \right]\,,
        \end{split}
    \end{equation} or in terms of the confluent Heun functions
        \begin{equation}
        \begin{split}
            &\widetilde{\Psi}_4 = \frac{1}{2\pi} \sum_{lmn} \left[ (b_{lmn}^{-(1)}{}_{-2}Y_{lm} + c_{lmn}^{-(1)} {}_{-2}Y_{l\,,-m}) e^{-i\omega_{lmn}^{\text{QNM}} v} H_c[(l-2)(l+3)+10 i c \omega_{lmn}^{\text{QNM}},10 i c \omega_{lmn}^{\text{QNM}},3-2ic\omega_{lmn}^{\text{QNM}},3,2ic\omega_{lmn}^{\text{QNM}},-\frac{r}{c}] \right.\\
            + &  (b_{lmn}^{-(2)}{}_{-2}Y_{lm} + c_{lmn}^{-(2)} {}_{-2}Y_{l\,,-m}) e^{-i\bar{\omega}_{lmn}^{\text{QNM}} v} \left(-\frac{r}{c}\right)^{2ic \bar{\omega}_{lmn}^{\text{QNM}}-2 }  H_c[l(l+1) +4c^2 \omega_{lm}^2,2ic\omega_{lm}(3+2ic\omega_{lm}),2ic\omega_{lm}-1,3,2ic\omega_{lm},-\frac{r}{c}]  \\
            + & (b_{lmn}^{+(1)}{}_{-2}Y_{lm} + c_{lmn}^{+(1)} {}_{-2}Y_{l\,,-m}) e^{i\bar{\omega}_{lmn}^{\text{QNM}} v} H_c[(l-2)(l+3)-10 i c \bar{\omega}_{lmn}^{\text{QNM}},-10 i c \bar{\omega}_{lmn}^{\text{QNM}},3+2ic\bar{\omega}_{lmn}^{\text{QNM}},3,-2ic\bar{\omega}_{lmn}^{\text{QNM}},-\frac{r}{c}]\\
            +&(b_{lmn}^{+(2)}{}_{-2}Y_{lm} + c_{lmn}^{+(2)} {}_{-2}Y_{l\,,-m}) e^{i\omega_{lmn}^{\text{QNM}} v} \left(-\frac{r}{c}\right)^{-2ic \omega_{lmn}^{\text{QNM}}-2 }  \\
            &\left. H_c[l(l+1) -4c^2 \bar{\omega}_{lm}^2,-2ic\bar{\omega}_{lm}(3-2ic\bar{\omega}_{lm}),-2ic\bar{\omega}_{lm}-1,3,-2ic\bar{\omega}_{lm},-\frac{r}{c}] \right]
        \end{split}
    \end{equation}
\end{widetext}  where we have defined 
\begin{subequations}
\begin{align}\label{eq:blmn&clmn}
    b_{lmn}^{- (1,2)}=k_{lmn}^{(1,2)} b_{lm}^-\,,\quad  c_{lmn}^{- (1,2)}=k_{lmn}^{(1,2)} c_{lm}^-\,.\\
    b_{lmn}^{+ (1,2)}=\bar{k}_{lmn}^{(1,2)} b_{lm}^+\,,\quad  c_{lmn}^{+ (1,2)}=\bar{k}_{lmn}^{(1,2)} c_{lm}^+\,.
\end{align}
\end{subequations}

From  Eq.~\eqref{eq:psi4-series}, it is straightforward to analyze the asymptotic behavior of $\widetilde \Psi_4$. Let us first consider the solution at future null infinity $\mathcal{I}^+$ by taking the limit $v\to\infty$ (with $r=\text{cnt}$). In this limit, we need to study the behavior of 
\begin{equation}\label{eq:psi4-asymptotic}
\begin{split}
    \lim_{\underset{r=\text{cnt}}{v\to\infty}} \widetilde{\Psi}_4 &\sim b_{lmn}^{-(1)} {}_{-2}Y_{lm} e^{-i\omega_{lmn}^{\text{QNM}} v}  +c_{lmn}^{-(1)} {}_{-2}Y_{l\,-m} e^{-i\omega_{lmn}^{\text{QNM}} v}\\
    &+ b_{lmn}^{-(2)} {}_{-2}Y_{lm} e^{-i\bar{\omega}_{lmn}^{\text{QNM}} v}  +c_{lmn}^{-(2)} {}_{-2}Y_{l\,-m} e^{-i\bar{\omega}_{lmn}^{\text{QNM}} v}\\
    &+b_{lmn}^{+(1)} {}_{-2}Y_{lm} e^{i\bar{\omega}_{lmn}^{\text{QNM}} v}  +c_{lmn}^{+(1)} {}_{-2}Y_{l\,-m} e^{i\bar{\omega}_{lmn}^{\text{QNM}} v}\\
    &+ b_{lmn}^{+(2)} {}_{-2}Y_{lm} e^{i\omega_{lmn}^{\text{QNM}} v}  +c_{lmn}^{+(2)} {}_{-2}Y_{l\,-m} e^{i\omega_{lmn}^{\text{QNM}} v}\,,
\end{split}
\end{equation} where the constant factors coming from the evaluation of the radial functions at $r=\text{cnt}\neq0$ have been omitted. Recalling that the sum over the harmonic index $m$ runs from $-l\,,\,.\,.\,. l$, we can split this sum in two contributions $m=0$ and $m\neq0$. For $m=0$ we can factor out the spin-weighted spherical harmonic contribution trivially. For $m\neq0$, we relabel the sum over $m\neq0$ in the even terms of Eq.~\eqref{eq:psi4-asymptotic}, so that we can factor out the spin-weighted spherical harmonic ${}_{-2}Y_{lm}$. To impose physical boundary conditions for this solution, it is important to recall some properties of the QNM frequencies. Given a quasinormal mode frequency 
\begin{equation}
    \omega_{lmn}^{\text{QNM}}=\re[w]+i \im[w]\,,
\end{equation}
\begin{itemize}
    \item Its imaginary part is negative $\im[w]<0$.
    \item We will follow the convention that $m\geq0$, $\re[w]>0$ and for $m<0$, $\re[w]<0$ for the regular modes (notice that for the mirror modes the opposite holds). 
     \item For $m\neq 0$, $\bar{\omega}_{lmn}^{\text{QNM}}=-\omega_{l\,-m n}^{\text{QNM}}$.
\end{itemize} With these properties, we can impose
the stability of the modes towards the future
in Eq.~\eqref{eq:psi4-asymptotic}. 
For $m=0$, Eq.~\eqref{eq:psi4-asymptotic} simplifies to
\begin{equation}\label{eq:psi4-asymptotic-m0}
\begin{split}
    \lim_{\underset{r=\text{cnt}\,, m=0}{v\to\infty}} \widetilde{\Psi}_4 &\propto(b_{l0n}^{-(1)}+c_{l0n}^{-(1)}) e^{-i\omega_{l0n}^{\text{QNM}} v}   \\
    &+ (b_{l0n}^{-(2)}+c_{l0n}^{-(2)})  e^{-i\bar{\omega}_{l0n}^{\text{QNM}} v} \\
    &+(b_{l0n}^{+(1)}+c_{l0n}^{+(1)} ) e^{i\bar{\omega}_{l0n}^{\text{QNM}} v}\\
    &+ (b_{l0n}^{+(2)}+c_{l0n}^{+(2)})  e^{i\omega_{l0n}^{\text{QNM}} v}\,.
\end{split}
\end{equation} Separating the frequency $\omega_{l0n}^{\text{QNM}}$ in its real and imaginary parts, we see that the first and third terms in Eq.~\eqref{eq:psi4-asymptotic-m0} are stable at $\mathcal{I}^+$ since the real part is a negative exponential, i.e., 
\begin{equation}
     \lim_{\underset{r=\text{cnt}\,, m=0}{v\to\infty}} e^{-i \omega_{l0n}^{\text{QNM}} v} =  \lim_{\underset{r=\text{cnt}\,, m=0}{v\to\infty}} e^{\im[\omega_{l0n}^{\text{QNM}}] v}e^{-i \re[ \omega_{l0n}^{\text{QNM}}]v}\to 0\,,
\end{equation} and these modes decay exponentially as $v\to\infty$. 
On the contrary, the second and fourth terms in Eq.~\eqref{eq:psi4-asymptotic-m0} are unstable given that they grow exponentially in the limit $v\to\infty$
\begin{equation}
     \lim_{\underset{r=\text{cnt}\,, m=0}{v\to\infty}} e^{i \omega_{l0n}^{\text{QNM}}v} =  \lim_{\underset{r=\text{cnt}\,, m=0}{v\to\infty}} e^{-\im[\omega_{l0n}^{\text{QNM}}] v}e^{i \re[ \omega_{l0n}^{\text{QNM}}]v}\to \infty\,.
\end{equation} Hence, to impose physical boundary conditions we need to set 
\begin{equation}
    b_{l0n}^{\pm(2)} + c_{l0n}^{\pm(2)}=k_{l0m}^{(2)}(b_{l0}^\pm + c_{l0}^\pm)=0\,.
\end{equation} This can be attained by either restricting the constants $a_{l0}^\pm$ (see Eq.~\eqref{eq:b}) or by setting
$k^{(2)}_{l0n}=0$ in Eq.~\eqref{eq:radial-psi4}. Since we wish to keep the perturbation as general as possible we choose the last option, which yields
\begin{equation}
    b_{l0n}^{-(2)}=c_{l0n}^{-(2)}=b_{l0n}^{+(2)}=c_{l0n}^{+(2)}=0\,.
\end{equation}
For the $m\neq0$ case, Eq.~\eqref{eq:psi4-series} simplifies to
\begin{equation}\label{eq:psi4-asymptotic-mnot0}
\begin{split}
    \lim_{\underset{r=\text{cnt}\,,m\neq0}{v\to\infty}} \widetilde{\Psi}_4 &\propto (b_{lmn}^{-(1)}+c_{l-mn}^{+(1)}) e^{-i\omega_{lmn}^{\text{QNM}} v}  \\
    &+ (b_{lmn}^{-(2)} +c_{l-mn}^{+(2)}) e^{-i\bar{\omega}_{lmn}^{\text{QNM}} v} \\
    &+(b_{lmn}^{+(1)}  + c_{l-mn}^{-(1)}) e^{i\bar{\omega}_{lmn}^{\text{QNM}} v} \\
    &+ (b_{lmn}^{+(2)} +c_{l-mn}^{-(2)}) e^{i\omega_{lmn}^{\text{QNM}} v} \,,
\end{split}
\end{equation} where we have used  $\bar{\omega}_{lmn}^{\text{QNM}}=-\omega_{l\,-m n}^{\text{QNM}}$ to obtain the same exponential terms that we had for the case $m=0$.  Using the same reasoning as before, we only keep stable modes on $\mathcal{I}^+$, so we need to set
\begin{equation}
   b_{lmn}^{\mp(2)} +c_{l-mn}^{\pm(2)} =k_{lmn}^{(2)} b_{lm}^\mp + k_{l-mn}^{(2)}c_{l-m}^\pm=0\,.
\end{equation} In order not to restrict the form of the perturbation, we choose again $k_{lmn}^{(2)}=0$, which yields
\begin{equation}
    b_{lmn}^{-(2)}=c_{l0n}^{-(2)}=b_{l0n}^{+(2)}=c_{l0n}^{+(2)}=0\,. 
\end{equation}
 
Further, at spatial infinity ($v\to\infty$ and $r\to \infty$),   $\widetilde{\Psi}_4 $ scales as
\begin{equation}
\begin{split}
    \lim_{v\,,r\to\infty} \widetilde{\Psi}_4 &\propto \frac{ 1}{r}b_{lmn}^{-(1)}{}_{-2}Y_{lm}e^{i \omega_{lmn}^{\text{QNM}} (-v+2r+2c \log r/c)}\\
    &+ \frac{1}{r} c_{lmn}^{-(1)} {}_{-2}Y_{l\,,-m} e^{i \omega_{lmn}^{\text{QNM}} (-v+2r+2c \log r/c)}  \\
   &+ \frac{ 1}{r}b_{lmn}^{+(1)}{}_{-2}Y_{lm}e^{-i \bar{\omega}_{lmn}^{\text{QNM}} (-v+2r+2c \log r/c)}\\
   &+ \frac{1}{r}c_{lmn}^{+(1)} {}_{-2}Y_{l\,,-m}e^{-i \bar{\omega}_{lmn}^{\text{QNM}} (-v+2r+2c \log r/c)} \,.
\end{split}
\end{equation} 
These direct ($\propto e^{-i \omega_{lmn}^{\text{QNM}}v} $) and mirror ($\propto e^{i \bar{\omega}_{lmn}^{\text{QNM}}v} $)
terms correspond to outgoing modes at spatial infinity. Given the relationship between the gravitational wave strain and $\widetilde\Psi_4$, we find outgoing gravitational radiation at $\mathcal{I}^+$.

The same discussion goes through for $\widetilde \Psi_0$. Starting from the radial Teukolsky equation~\eqref{eq:teukolsky-extended:1} we can again use the coordinate $x=r+c$ and perform the transformation
\begin{equation}
    X_{lm}^{(0)}=e^{i \omega x} (x-c)^a x^b y(x)\,,\quad a=0,2+2i\omega\,,\quad b=-2,-4
\end{equation} to rewrite Eq.~\eqref{eq:teukolsky-extended:1} in the form of Eq.~\eqref{eq:diff-eq-Leaver}. Recall that the Teukolsky equation is a second-order differential equation, so it only has two independent solutions. How is then possible that we have these four transformations that leave the differential equation in the form of a generalized wave equation? The transformation with $a=0$ and $b=-2\,,-4$ selects the first independent solution of the Teukolsky equation for $\widetilde{\Psi}_0$, while  $a=2+2i c \omega$ and $b=-2, -4$ selects the second solution. The existence of two transformations for a single solution of the Teukolsky equation shows a certain degeneracy of this method that was already hinted at in the discussion for $\widetilde{\Psi}_4$: We can have generalized wave equations with different coefficients $B_i$, $\eta$ that nonetheless are solved by the same three-term recurrence relation. This is the case for the transformations $a=0$ and $b=-2,-4$. The series representation for $X_{lm}^{(0,1)}$ that we obtain (using either value of $b$) is unique. For the sake of brevity, we only discuss the case $a=0$, $b=-4$ for the first solution of Eq.~\eqref{eq:teukolsky-extended:1}.
The above transformation yields Eq.~\eqref{eq:diff-eq-Leaver} with 
\begin{align} \label{eq:coefficients-diff-Eq-Leaver-3}
    & B_1^\prime= c\,,\quad B_2^\prime = -2(1+ic \omega)\,,\quad \eta^\prime= -2i - c \omega\\
    & B_3^\prime = 2-l-l^2+ 5i c \omega +2 c^2 \omega^2
\end{align} Using Eqs.(39)-(42) of Ref.~\cite{LeaverExpansions1986}, we can write the expansion of the first solution of $X_{l,m}^{(0)}$ as
\begin{equation}\label{eq:psi0-sol1-series}
    X_{lm}^{(0,1)}= e^{2i\omega r} \left(\frac{r+c}{c}\right)^{-5+2ic \omega} \sum_{k=0}^\infty a_k^\prime \left(\frac{r}{r+c}\right)^k \, ,
\end{equation} where the coefficients $a_k^\prime$ satisfy Eq.~\eqref{eq:a-series} and the coefficients $\alpha_k^\prime$, $\beta_k^\prime$ and $\gamma_k^\prime$ are
\begin{subequations}
\begin{align}\label{eq:alphabetagamma-prime}
     \alpha_k^\prime &= (k+1)(k-1-2i c \omega)\\
     \beta_k^\prime &= 3-l-l^2-2k^2+4i c \omega+8c^2 \omega^2-2k+8ick\omega\\
     \gamma_k^\prime &= (k-2i c \omega) (k+2-2ic\omega) \, .
\end{align}  
\end{subequations} 
The series $\sum_k a_k^\prime$ is finite when the frequencies are the QNM frequencies $\omega=\omega_n$. We again obtain this set of eigenfrequencies because the algebraic equation given by the infinite fraction~\eqref{eq:QNM-frequencies} is the same that we obtained for the first solution of $\widetilde{\Psi}_4$. This occurs because the coefficients $\alpha_k^\prime$, $\beta_k^\prime$ and $\gamma^\prime_k$ are related with $\alpha_k$, $\beta_k$ and $\gamma_k$ as follows
\begin{equation}\label{eq:relationship-parameters}
\beta_k^\prime=\beta_k\,,\quad \alpha_k^\prime \gamma_{k+1}^\prime = \alpha_k \gamma_{k+1}\,.
\end{equation} 
Notice that this discussion constitutes proof that both $\widetilde{\Psi}_0$ and $\widetilde{\Psi}_4$ can be decomposed using the same set of eigenfrequencies, namely the QNM frequencies. Further, this is independent of the relationship between these two quantities at the horizon and the Teukolsky-Starobinsky identities. 

The coefficient $a_0^\prime=1$ is chosen so that Eq.~\eqref{eq:psi0-sol1-series} coincides with the first term in the small $r$ limit expansion of  Eq.~\eqref{eq:confluent-Heun-normalization}
, i.e, 
\begin{equation}
    \widetilde \Psi_0^{(1)} =  1-\frac{2+l(l+1) +2ic\omega}{1+2ic\omega} +\mathrm{O}(r^2)\,.
\end{equation}

Given that $\widetilde{\Psi}_0$ and $\widetilde{\Psi}_4$ are related through the Teukolsky-Starobinsky identities and that $k_{lmn}^{(2)}$ was set to zero by requiring the stability of the solution, the second independent radial solution $X_{lm}^{(0,2)}$ will also vanish.

Defining 
\begin{equation}\label{eq:almn}
    a_{lmn}^{- (1)} = k_{lmn}^{(1)}a_{lm}^{-}\,,\quad a_{lmn}^{+ (1)} = \bar{k}_{lmn}^{(1)}a_{lm}^{+}
\end{equation} and taking into account that we have already set $k^{(2)}_{lmn}=0$,  $\widetilde \Psi_0$ has the series expansion
\begin{widetext}
\begin{equation}
    \begin{split}
         \widetilde{\Psi}_0 =\frac{1}{2\pi} \sum_{l\,,m\,,n} &\left(a_{lmn}^{-(1)} e^{-i \omega_{lmn}^{\text{QNM}} v} e^{2i \omega_{lmn}^{\text{QNM}} r} \left(\frac{r+c}{c}\right)^{-5+2ic \omega_{lmn}^{\text{QNM}}} \sum_{k=0}^\infty a_k^\prime \left(\frac{r}{r+c}\right)^k\right.\\
         &\left.+a_{lmn}^{+(1)} e^{i \bar{\omega}_{lmn}^{\text{QNM}} v} e^{-2i \bar{\omega}_{lmn}^{\text{QNM}} r} \left(\frac{r+c}{c}\right)^{-5-2ic \bar{\omega}_{lmn}^{\text{QNM}}} \sum_{k=0}^\infty \bar{a}_k^\prime \left(\frac{r}{r+c}\right)^k\right) {}_2Y_{lm} \; .
    \end{split}
    \end{equation}
\end{widetext}
From this expression, we can evaluate the asymptotic behavior of $\widetilde\Psi_0$ at spatial infinity 
\begin{equation}\label{eq:psi0-asymptotic-i}
\begin{split}
     \lim_{v\,,r\to \infty} \widetilde\Psi_0 \propto & \frac{1}{r^5}a_{lmn}^{-(1)}  e^{i\omega_{lmn}^{\text{QNM}} (-v+2r+2c \ln r/c) } \\
     &+\frac{1}{r^5}a_{lmn}^{+(1)}  e^{-i\bar{\omega}_{lmn}^{\text{QNM}} (-v+2r+2c \ln r/c) }
\end{split}
\end{equation} 
{{Notice that in Eq.~\eqref{eq:psi0-asymptotic-i} the modes decay towards the future $v\to \infty$.  While the power-law terms are $\mathcal{O}(r^{-5})$ and thus decay faster than $\widetilde{\Psi}_4$, we see that the exponential terms in fact diverge at $i^0$ and at past-null infinity since $\omega_{lmn}$ is complex.  This divergence is common to the behavior of QNMs in the standard formulation discussed earlier and in this regard our formalism is no different.  A complete discussion of this behavior is beyond the scope of this paper, and is likely to involve intricate aspects of functional analysis.  As a tentative proposal for this paper, we shall take the polynomial fall-off of $\widetilde{\Psi}_0$ as implementing a no-incoming radiation from $\mathcal{I}^-$.}}

Summarizing, the above discussion shows that for an asymptotically flat spacetime with ingoing radiation at the horizon and outgoing radiation at infinity, we find the following solution for $\widetilde \Psi_0$
\begin{equation}
    \begin{split}
     \widetilde{\Psi}_0 =\frac{1}{2\pi}\sum_{l,m}  \sum_{n=0}^\infty{}_2Y_{lm} \left( \frac{ a_{lmn}^{-(1)} \text{ \textthorn}^4[(r+c)^4 e^{-i\omega_{lmn}^{\text{QNM}} v} H_{lmn}]}{c^4 i\omega_{lmn} (\kappa_{(l)}^2+\omega_{lmn}^2)(2\kappa_{(l)}-i\omega_{lmn})}\right.\\
    \left.- \frac{ a_{lmn}^+ \text{ \textthorn}^4[(r+c)^4 e^{i\bar{\omega}_{lmn}^{\text{QNM}} v} \bar{H}_{lmn}]}{c^4 i\bar{\omega}_{lmn} (\kappa_{(l)}^2+\bar{\omega}_{lmn}^2)(2\kappa_{(l)}+i\bar{\omega}_{lmn})} \right)\,. 
\end{split}
\end{equation}
where the index $n$ denotes the overtone number, $\omega_n=\omega_{lmn}^{\text{QNM}}$ are the QNM frequencies for each $l\,,m$, and $H_{lmn}$ is a shortcut for
 \begin{equation}\label{eq:shortcut-Hc}
 \begin{split}
     H_{lmn} = & H_c[(l-2)(l+3) +10i c \omega_{lmn},\\
     & 10i c \omega_{lmn}, 3-2ic\omega_{lmn},3,2ic\omega_{lmn},-\frac{r}{c}]\,.    
 \end{split}
 \end{equation} 
 At the horizon, $\widetilde{\Psi}_0$ takes the values
\begin{equation}
    \widetilde \Psi_0 \triangleq \frac{1}{2\pi} \sum_{l,m,n} {}_2 Y_{lm} \left( a_{lmn}^- e^{-i\omega_n v}+ a_{lmn}^+ e^{i\bar{\omega}_n v} \right) 
\end{equation}
where the coefficients $a_{lmn}^\pm$ are defined in Eq.~\eqref{eq:almn}  and the inverse relation is given by
 \begin{equation}
     a_{lm}^\pm = \sum_n  a_{lmn}^\pm \delta(\omega-\omega_{lmn}^{\text{QNM}})\,.
 \end{equation}
 Similarly, the perturbation to $\widetilde\Psi_4$ is given by 
\begin{equation}\label{eq:psi4-final}
\begin{split}
      \widetilde\Psi_4 = \frac{1}{2\pi} \sum_{l,m,n}  \left([b_{lmn}^{-(1)}{}_{-2}Y_{lm} +c_{lmn}^{-(1)}{}_{-2}Y_{l-m}  ]  e^{-i\omega_n v} H_{lmn} \right.\\
     \left. + [b_{lmn}^{+(1)}{}_{-2}Y_{lm}+c_{lmn}^{+(1)}{}_{-2}Y_{l-m}]  e^{i\bar{\omega}_n v} \bar{H}_{lmn}\right) \; .
\end{split}
\end{equation}

In this section we have shown that imposing separability, analicity and stability of the radial solution describing a quasi-isolated horizon yields the late stages of the ringdown, given that the solution has an expansion in terms of QNM frequencies. 
In this construction, we did not impose the standard outgoing radiation condition. Nevertheless, as we discuss next, our solutions naturally satisfy these.
\subsection{Boundary conditions}
\label{subsec:Bdy-conditions}

As highlighted in Sec.~\ref{subsec:standard-qnm}, the QNM frequencies are often obtained by mapping the gravitational wave equation (e.g., the Regge-Wheeler/Zerilli or Teukolsky equations) to an eigenvalue problem with the following boundary conditions:
\begin{itemize}
    \item No incoming radiation from $\mathcal{I}^-$;
    \item Only ingoing/outgoing modes at $\mathcal{H}$/$\mathcal{I}^+$.
  \end{itemize} These boundary conditions discard the continuous spectrum of solutions to the wave-type equation, leaving only the discrete spectrum of QNM frequencies as the solution to the eigenvalue problem the differential equation is mapped to. Our solution also satisfies these boundary conditions, although we have not imposed them explicitly to arrive at the ringdown solution we presented. In fact, selecting analytic and stable solutions in the limit $v\to\infty$ already discards the unphysical outgoing/ingoing modes at the horizon/future null infinity.  In the following, we show explicitly that this is indeed the case. 

\begin{figure}
\centering
\includegraphics[width=\linewidth]{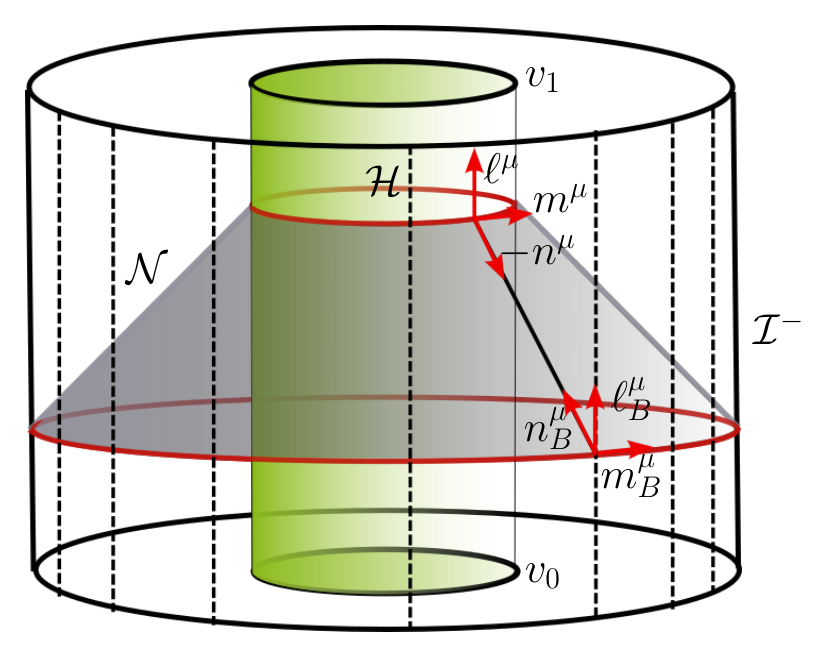}
\caption{Ilustration of a portion of the horizon and $\mathcal{I}^-$ between $v_0$ and $v_1$. The null surface $\mathcal{N}$ generated by past-directed null geodesics emanating from a particular cross-section of the horizon is also depicted. This representation illustrates how the tetrad basis defined at the horizon is parallel propagated along $-n^\mu$ to $\mathcal{I}^-$.  } 
\label{fig:horizon-scrim}
\end{figure}

We start by discussing the absence of incoming radiation from $\mathcal{I}^-$. Given that the perturbed tetrad vectors are parallel propagated along $n$, we can use the perturbed tetrad discussed in Sec.~\ref{sec:perturbed-intrinsic-geometry} to naturally define a Bondi frame at $\mathcal{I}^-$, which has been attached to our spacetime manifold (see Fig.~\ref{fig:horizon-scrim}). The Bondi frame, denoted by a subindex ${}_B$ is related to the perturbed spacetime tetrad through a conformal factor $\Omega$ (that vanishes on $\mathcal{I}^-$)
\begin{equation}
    \ell^\mu_B = l^\mu\,,\quad n^\mu_B = \Omega^2 n^\mu\,,\quad m^\mu_B=\Omega^2 m^\mu\,. 
\end{equation} 
For the moment, we will not need the particular form of the perturbed spacetime tetrad vectors, we have used the fact that they are parallel propagated along $n$ to establish the connection between the tetrad at the horizon and that at $\mathcal{I}^-$, as we represent in Fig.~\ref{fig:horizon-scrim}. In Ref.~\cite{PhysRevD.19.3495}, it was argued that the Weyl scalar $\Psi_0$ in an asymptotically flat spacetime \emph{in the past}~\footnote{The only assumption in Ref.~\cite{PhysRevD.19.3495} is the existence and regularity of the hypersurface $\mathcal{I}^-$, which our solution satisfies.} should decay along a past-directed null geodesics as 
\begin{equation}\label{eq:Psi0-scri-m}
    \Psi_0 = \Psi_0^{(\circ)} \Omega + \Psi_0^{(1)} \Omega^2+ \mathrm{O}(\Omega^3)\,,
\end{equation} where $\Psi_0^{(\circ)}$ is the value of $\Psi_0$ at $\mathcal{I}^-$ and subsequent terms represent the coefficients of the Taylor expansion along the vector $n$ (away from $\mathcal{I^-} $ and into the spacetime). Notice that the discussion in Ref.~\cite{PhysRevD.19.3495} was nonperturbative, so the above argument is general (as long as the structure of $\mathcal{I}^-$ is available) and should also apply to our example of a radiative perturbation. However, to be able to use the results in Ref.~\cite{PhysRevD.19.3495}, we need to establish the connection between the conformal factor $\Omega$ and our coordinate system. Taking into account that we defined the coordinate $r$ as the affine parameter along the past directed null geodesic with tangent vector $-n$, it follows that $\Omega=-1/r$. This means, in particular, that we can identify the coefficients in the Taylor expansion away from $\mathcal{I}^-$ in Eq.~\eqref{eq:Psi0-scri-m} by expanding our perturbative solution for $\widetilde{\Psi}_0$ in Eq.~\eqref{eq:psi0-final} in powers of $1/r$. Using Eq.~\eqref{eq:psi0-sol1-series}, it follows that our ringdown solution has
\begin{equation}\label{eq:psi0-asymptotic-conditions}
    \Psi_0^{(\circ)} =  0\,,\quad \Psi_0^{(I)} =0\,,\quad I=1,...,4\,,
\end{equation} with $\Psi_0^{(5)}$ the first coefficient in the expansion that is nonvanishing. 
  
In  Ref.~\cite{PhysRevD.19.3495}, it was argued that the sufficient conditions for the absence of incoming radiation through $\mathcal{I}^-$ are that i) $\Psi_0$ decays at least as fast as $\Psi_0\sim 1/r^3$ in the $r\to \infty$ limit, i.e., that $\Psi_0^{(\circ)} = \Psi_0^{(1)}=0$, and ii)  $\Psi_1$ satisfies the peeling condition at $\mathcal{I}^-$, i.e., it has the asymptotic behavior $\Psi_1 =\Psi_1^{(\circ)} \Omega^2$. No conditions on the fall-off of the Weyl scalars $\Psi_I$ $I=2,3, 4$ were imposed to guarantee the absence of incoming radiation from $\mathcal{I}^-$. 

From Eq.~~\eqref{eq:psi0-asymptotic-conditions}, we see that in our particular ringdown solution $\widetilde{\Psi}_0$ satisfies an even sharper decay, since $\widetilde{\Psi}_0^{J}$, $J=2,3,4$ also vanish. Hence, the first condition is trivially satisfied for our ringdown solution. The second condition requires more work to assess: It can be explicitly checked by integrating the radial field equations, radial Bianchi identities and radial frame differential equations (Eqs.~\eqref{eqs:radial},~\eqref{eqs:radial-bianchi} and~\eqref{eqs:frame}). This can be done methodologically by rewriting the radial part of $\widetilde{\Psi}_4$, $X_{lmn}^{(4)}$,  as the fourth radial derivative of another function proportional to $X_{lmn}^{(0)}$ (see Sec.~\ref{sec:multipole-moments}). This allows us to obtain analytic, explicit expressions for all the spin coefficients, Weyl scalars and frame functions in terms of  $X_{lmn}^{(0)}$ and its derivatives. This procedure, although straightforward, is quite lengthy, so it will be presented explicitly elsewhere.  However, by following this procedure, we have indeed checked that the second condition is also satisfied. Therefore, there is no incoming radiation from $\mathcal{I}^-$. 

Next, we discuss whether the two remaining boundary conditions, the absence of outgoing radiation at the horizon and incoming radiation from $\mathcal{I}^+$, are satisfied.  Hence, we wish to identify the ingoing and outgoing modes of $\widetilde{\Psi}_0$ and $\widetilde{\Psi}_4$ at the horizon and infinity. The ingoing and outgoing modes can be easily identified in Schwarzschild coordinates $\{t,\mathfrak{r}\}$, which are related to our coordinates through the transformation
\begin{equation}
    t=v-\mathfrak{r}-c\log|\mathfrak{r}-c|\,,\quad \mathfrak{r}=r+c\,.
\end{equation}
Applying this transformation to $\widetilde{\Psi}_0$ and $\widetilde{\Psi}_4$ in their series expansion form Eqs.~\eqref{eq:Psi4-series1} and~\eqref{eq:psi0-sol1-series} and evaluating them at the horizon yields
\begin{align}
\begin{split}
     \lim_{\mathfrak{r}\to c}\widetilde{\Psi}_{4\,,0} \propto\, & e^{-i\omega_{lmn} (t+\mathfrak{r})} |\mathfrak{r}-c|^{-i\omega_{lmn} c}\,, \\
     & e^{i\bar{\omega}_{lmn} (t+\mathfrak{r})} |\mathfrak{r}-c|^{i\bar{\omega}_{lmn} c}
\end{split}
\end{align} which shows that only ingoing modes are present. Similarly, when evaluating these expression in the limit $\mathfrak{r}\to\infty$, we see that the dominant modes are
\begin{equation}
    \begin{split}
     \lim_{\mathfrak{r}\to \infty}\widetilde{\Psi}_{I} \propto\, & e^{-i\omega_{lmn} (t-\mathfrak{r})} \left(\frac{\mathfrak{r}}{c}\right)^{-5+I+ic\omega_{lmn}}\,, \\
     & e^{i\bar{\omega}_{lmn} (t-\mathfrak{r})} \left(\frac{\mathfrak{r}}{c}\right)^{-5+I-ic\bar{\omega}_{lmn}}
\end{split}\quad I=0,4\,,
\end{equation} which are outgoing. Therefore, by selecting an analytic solution that is stable for growing values in $v$, the QNM boundary conditions are automatically satisfied.

\section{ Perturbative black hole tomography}
\label{sec:tomography}

The notion of ``gravitational wave tomography'' has been introduced in
\cite{Ashtekar:2021kqj}, that is, the horizon dynamics in the strong field region 
can be determined from gravitational waves in the weak field region. 
This is not surprising as there is growing evidence that there is a close
interplay between properties of horizons and those of null infinity.
Even in the equilibrium case, isolated horizons and null infinity share
several mathematical structures
\cite{Ashtekar:2024mme,Ashtekar:2024bpi,Ashtekar:2024stm}.  More
generally, the notion of inferring horizon dynamics from properties of
the gravitational waveform has been put forward in
\cite{Jaramillo:2012rr,Jaramillo:2011re,Jaramillo:2011rf,Rezzolla:2010df,Gupta:2018znn}.
In the ringdown
regime specifically, several numerical studies have found empirically that the
infalling fluxes at the horizon can be well modeled by a QNM expansion
\cite{Mourier:2020mwa,Chen:2022dxt,Khera:2023oyf,Forteza:2021wfq}.
These previous efforts have been all heuristic in nature or have been based
on numerical observations. Our formalism provides support for gravitational
wave tomography, for the first time,  from analytical considerations in the perturbative regime. 

Through Sec.~\ref{sec:perturbed-ih}-\ref{sec:QNM-frequencies}, we have stressed the relationship between $\widetilde{\Psi}_0$ and $\widetilde{\Psi}_4$ and made this explicit. However, we have not discussed black hole tomography in detail, i.e., how to reconstruct the geometry of the horizon from the observed gravitational wave. To make this relationship explicit, we first need to connect the asymptotic values of $\widetilde{\Psi}_0$ and $\widetilde{\Psi}_4$ with their respective behavior at the horizon. We sliced our spacetime using past-directed null geodesics, so to take the limit to $\mathcal{I}^+$ we would need to vary simultaneously the coordinates $v\,,r$ at the appropriate rate. Instead, the analysis can be made more transparent by expressing $\widetilde{\Psi}_0$ and $\widetilde{\Psi}_4$ in a more convenient coordinate system. In particular, we will use hyperboloidal coordinates, which directly connect the horizon with future null infinity \cite{Zenginoglu:2007jw,Schinkel:2013zm,Ansorg_2016}. Given that in the gauge in which we expressed our solution, the coordinates $v\,,r$ maintain their original meaning as the affine parameter along the generators of the horizon and the past-directed null geodesics, we can use the background coordinate transformation given in~\cite{Ansorg_2016}
\begin{equation}\label{eq:hyperboloidal}
    v=2c \left(\tau+h(\sigma)\right)\,, \quad r=\frac{c}{\sigma} -c
\end{equation} where $h(\sigma)=\frac{1}{\sigma}-\log\sigma$ is the height function, and $\sigma\in [0,1]$ with $\sigma=1$ being the horizon and $\sigma=0$ future null infinity $\mathcal{I}^+$. 

From the expression we gave for $\widetilde{\Psi}_4$ and $\widetilde{\Psi}_0$ in series form (see Eqs.~\eqref{eq:Psi4-series2} and~\eqref{eq:psi0-sol1-series}), it is straightforward to express them in terms of these hyperboloidal coordinates. We find
\begin{widetext}
    \begin{subequations}\label{eqs:psi0-psi4-hyperboloidal}
    \begin{align}
    \label{eq:psi0-hyperboloidal}
         \widetilde{\Psi}_0 =& \sigma^5 \sum_{l,m,n} \left\{a_{lmn}^{-}e^{-2ci\omega_{lmn} (\tau+1)}\sum_{k=0}^\infty a_k^\prime (1-\sigma)^k+a_{lmn}^{+}e^{2c i \bar{\omega}_{lmn} (\tau+1)}\sum_{k=0}^\infty \bar{a}_k^\prime (1-\sigma)^k\right\}{}_2 Y_{lm}\,,\\    \label{eq:psi4-hyperboloidal}
    \begin{split}
         \widetilde{\Psi}_4 =& \sigma \sum_{l,m,n} \left\{(b_{lmn}^{-}{}_{-2}Y_{lm} + c_{lmn}^{-} {}_{-2}Y_{l\,,-m})e^{-2ci\omega_{lmn} (\tau+1)}\sum_{k=0}^\infty a_k (1-\sigma)^k\right.\\
         &\left.+(b_{lmn}^{+}{}_{-2}Y_{lm} + c_{lmn}^{+} {}_{-2}Y_{l\,,-m})e^{2c i \bar{\omega}_{lmn} (\tau+1)}\sum_{k=0}^\infty \bar{a}_k (1-\sigma)^k\right\}
    \end{split}
    \end{align}
    \end{subequations}
\end{widetext}
    where the coefficients $a_k$, $a_k^\prime$ in the series expansions are defined in Eq.~\eqref{eq:a-series}, together with Eqs.~\eqref{eq:alpha-beta-gamma} and~\eqref{eq:alphabetagamma-prime}~\footnote{Notice that we have dropped the super-index $(1)$ from Eq.~\eqref{eq:psi0-hyperboloidal}. It should be understood that we have set $k_2 =0 $. }. 
    At the horizon, $\sigma=1$ and the only non-vanishing coefficient in the series are $a_0=\bar{a}_0=a_0^\prime = \bar{a}_0^\prime =1$, so we recover the boundary conditions at the horizon, which we rewrite as 
\begin{widetext}
    \begin{subequations}\label{eq:psi4-psi0-horizon-tomography}
        \begin{align}\label{eq:psi4-psi0-horizon-tomography:1}
            \widetilde{\Psi}_0 \triangleq &\sum_{l\,,m\,,n} \left\{\Psi_{0\,,lmn}^{H\,,-} e^{-2ci\omega_{lmn} (\tau+1)}+\Psi_{0\,,lmn}^{H\,,+} e^{2ci\bar{\omega}_{lmn} (\tau+1)} \right\}{}_{2}Y_{lm}\,,\\\label{eq:psi4-psi0-horizon-tomography:2}
            \begin{split}
              \widetilde{\Psi}_4 \triangleq & \sum_{l\,,m=0\,,n} \left\{\Psi_{4\,,l0n}^{H\,,-} e^{-2ci\omega_{l0n} (\tau+1)}+\Psi_{4\,,l0n}^{H\,,+} e^{2ci\bar{\omega}_{l0n} (\tau+1)}\right\}{}_{-2} Y_{l0} \\
              &+ \sum_{l\,,m\neq0\,,n}\left\{\Psi_{4\,,lmn}^{H\,,-} e^{-2ci\omega_{lmn} (\tau+1)}+\Psi_{4\,,lmn}^{H\,,+} e^{2ci\bar{\omega}_{lmn} (\tau+1)}\right\}{}_{-2}Y_{lm}\,,
            \end{split}
        \end{align}
    \end{subequations}
\end{widetext}
where we have just renamed the coefficients in the $\widetilde{\Psi}_0$ expansion as $ \Psi_{0\,,lmn}^{H\,,\mp} = a_{lmn}^\mp$, and we have defined
\begin{subequations}
    \begin{align}
        \Psi_{4\,,l0n}^{H\,,\mp} &= b_{l0n}^\mp+ c_{l0n}^\mp\,,\\
        \Psi_{4\,,lmn}^{H\,,\mp} & = b_{lmn}^\mp+c_{l-mn}^\pm\,,\quad m\neq0\,.
    \end{align}
\end{subequations}

    The expressions in Eq.~\eqref{eqs:psi0-psi4-hyperboloidal} make evident that $\widetilde{\Psi}_0$ decays much faster than $\widetilde{\Psi}_4$ in the limit $\sigma\to 0 $, i.e., while approaching $\mathcal{I}^+$. As a consequence, $\widetilde{\Psi}_4$ will be dominant in the radiation zone, and we shall focus only on this quantity. 
    To leading order in $\sigma$, $\widetilde{\Psi}_4$ at $\mathcal{I}^+$  reads
 \begin{widetext}
     \begin{equation}\label{eq:psi4-infinity}
     \begin{split}
       \lim_{\sigma\to0} \sigma^{-1} \widetilde{\Psi}_4 = & \sum_{l\,,m=0\,,n}  \left\{\Psi_{4\,,l0n}^{H\,,-}  e^{-2ci\omega_{l0n} (\tau+1)} \sum_k a_{k}+\Psi_{4\,,l0n}^{H\,,+} e^{2ci\bar{\omega}_{l0n} (\tau+1)}\sum_k \bar{a}_{k}\right\}{}_{-2} Y_{l0} \\
              &+ \sum_{l\,,m\neq0\,,n}\left\{\Psi_{4\,,lmn}^{H\,,-} e^{-2ci\omega_{lmn} (\tau+1)}\sum_k a_{k}+\Psi_{4\,,lmn}^{H\,,+} e^{2ci\bar{\omega}_{lmn} (\tau+1)} \sum_k \bar{a}_{k}\right\}{}_{-2}Y_{lm}\\
              &:=  \sum_{l\,,m=0\,,n}  \left\{\Psi_{4\,,l0n}^{\mathcal{I}^+\,,-}  e^{-2ci\omega_{l0n} (\tau+1)} +\Psi_{4\,,l0n}^{\mathcal{I}^+\,,+} e^{2ci\bar{\omega}_{l0n} (\tau+1)}\right\}{}_{-2} Y_{l0} \\
              &+ \sum_{l\,,m\neq0\,,n}\left\{\Psi_{4\,,lmn}^{\mathcal{I}^+\,,-} e^{-2ci\omega_{lmn} (\tau+1)}+\Psi_{4\,,lmn}^{\mathcal{I}^+\,,+} e^{2ci\bar{\omega}_{lmn} (\tau+1)} \right\}{}_{-2}Y_{lm}\,,
            \end{split}
     \end{equation}
 \end{widetext}
where we have defined the asymptotic angular modes $\Psi_{4\,,lmn}^{\mathcal{I}^+\,,\pm}$ in relation with the horizon angular modes $\Psi_{4\,,lmn}^{H\,,\pm}$. 
The sums $\sum_{k=0}^\infty a_{k}$, and its complex conjugate, are convergent by construction and depend only on the QNM frequencies and the details of the background. Hence, these sums can be expressed as 
 \begin{equation}\label{eq:Flmn}
     \sum_{k=0}^\infty a_k =\mathcal{F}_{lmn} (\omega_{lmn}\,,c)
 \end{equation} and similarly for its complex conjugate. Notice that to rewrite Eq.~\eqref{eqs:psi0-psi4-hyperboloidal} in the form of Eq.~\eqref{eq:psi4-infinity}, we have used that $\mathcal{F}_{l-mn} = \bar{\mathcal{F}}_{lmn}$ for $m\neq0$ and the relationship between direct and mirror modes.  From this discussion, it follows that the direct and mirror modes of $\widetilde{\Psi}_4$ at the horizon and future null infinity are related by a constant. Namely, 
 \begin{subequations}\label{eq:tomography-psi4}
     \begin{align}
        \Psi_{4\,,lmn}^{\mathcal{I}^+\,,-} = \Psi_{4\,,lmn}^{H\,,-} \mathcal{F}_{lmn} (\omega_{lmn}\,,c)\,,\\
         \Psi_{4\,,lmn}^{\mathcal{I}^+\,,+} = \Psi_{4\,,lmn}^{H\,,+} \bar{\mathcal{F}}_{lmn} (\bar{\omega}_{lmn}\,,c)\,.
     \end{align}
 \end{subequations} 
 This means, in particular, that if we can identify the presence of a given mode in the late stages of the ringdown of the detected gravitational wave, say the fundamental mode $n=0$ in the $l=2$, $m=0$ gravitational wave strain (accounting for the relationship between the strain and $\widetilde{\Psi}_4$), we can calculate the value of the amplitude $\Psi_{4\,,200}^{\mathcal{I}^+\,,\pm}$ and through Eq.~\eqref{eq:tomography-psi4}, that of $\Psi_{4\,,200}^{H\,,\pm}$,  once the coefficient $\mathcal{F}_{200}$ has been computed. An analogous discussion follows for any detected mode $l,m,n$. Further, notice that 
 through Eq.~\eqref{eq:b}, we can obtain $\Psi^{H\,,\pm}_{0\,,lmn}$, which will allow us to reconstruct the geometry of the horizon using Eq.~\eqref{eq:solution1}. In particular, we can explicitly relate the modes of $\widetilde{\Psi}_4$ at infinity with those of $\widetilde{\Psi}_0$ at the horizon. For $m=0$, we find
 \begin{subequations}\label{eq:tomography-psi4-psi0-m=0}
     \begin{align}
         \Psi_{4\,,l0n}^{\mathcal{I}^+\,,-}=&\frac{K_l^2 \Psi_{0\,,l0n}^{H\,,-}+6ic\omega_{l0n} \bar{\Psi}_{0\,,l0n}^{H\,,+}}{4c^4 i\omega_{l0n} (\kappa_{(\ell)}^2+\omega_{l0n}^2)(2\kappa_{(\ell)} -i\omega_{l0n})}
         \mathcal{F}_{l0n}\\
           \Psi_{4\,,l0n}^{\mathcal{I}^+\,,+}=&-\frac{K_l^2 \Psi_{0\,,l0n}^{H\,,+}-6ic\bar{\omega}_{l0n} \bar{\Psi}_{0\,,l0n}^{H\,,-}}{4c^4 i\bar{\omega}_{l0n} (\kappa_{(\ell)}^2+\bar{\omega}_{l0n}^2)(2\kappa_{(\ell)} +i\bar{\omega}_{l0n})}
         \bar{\mathcal{F}}_{l0n}
     \end{align}
 \end{subequations} and for $m\neq 0 $
  \begin{subequations}\label{eq:tomography-psi4-psi0-mneq0}
     \begin{align}
         \Psi_{4\,,lmn}^{\mathcal{I}^+\,,-}=&\frac{K_l^2 \Psi_{0\,,lmn}^{H\,,-}+(-1)^m 6ic\omega_{lmn} \bar{\Psi}_{0\,,lmn}^{H\,,-}}{4c^4 i\omega_{l0n} (\kappa_{(\ell)}^2+\omega_{l0n}^2)(2\kappa_{(\ell)} -i\omega_{l0n})}
         \mathcal{F}_{lmn}\\
           \Psi_{4\,,lmn}^{\mathcal{I}^+\,,+}=&-\frac{K_l^2 \Psi_{0\,,lmn}^{H\,,+}-(-1)^m 6ic\bar{\omega}_{lmn} \bar{\Psi}_{0\,,lmn}^{H\,,+}}{4c^4 i\bar{\omega}_{lmn} (\kappa_{(\ell)}^2+\bar{\omega}_{lmn}^2)(2\kappa_{(\ell)} +i\bar{\omega}_{lmn})}
         \bar{\mathcal{F}}_{lmn}\,,
     \end{align} where we have introduced the shorthand notation $K_l =\sqrt{(l+2)(l+1)l(l-1)}$. These expressions can be inverted, and yield the modes of radiation at the horizon $\Psi_{0\,,lmn}^{H\,,\mp}$ as a function of the modes of the gravitational wave strain observed far away $\Psi_{4\,,lmn}^{\mathcal{I}^+\,,\mp}$ (for the explicit expressions, see App.~\ref{sec:complementary}). 
     \end{subequations}

\begin{figure}
     \centering
     \includegraphics[width=\linewidth]{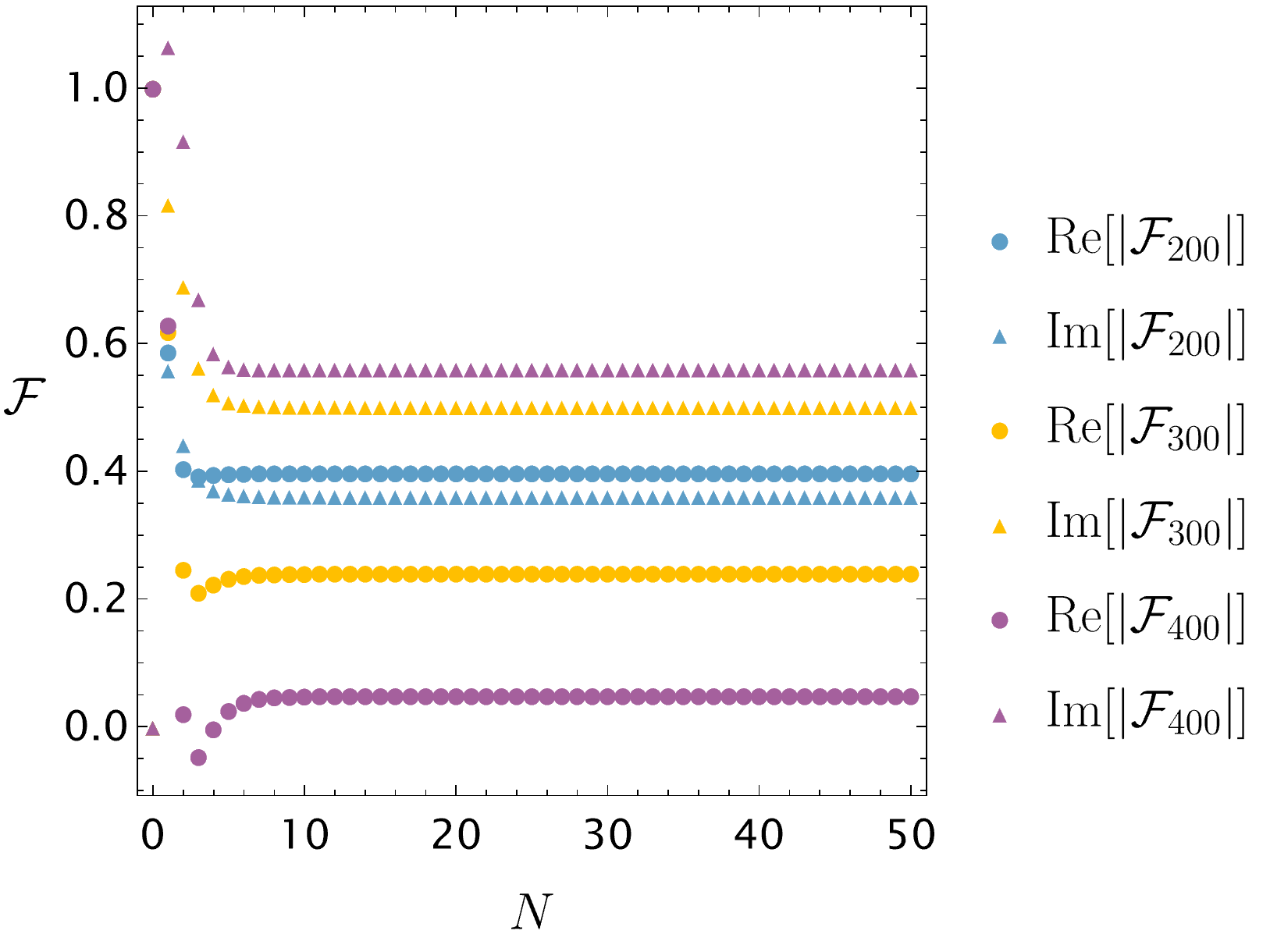}
     \caption{Convergence of $\mathcal{F}_{l00}^{\text{app}}$ for $l=2,3,4$ as a function of the number of elements in the sum $N$. Convergence is reached after $N>10$ and around $N\gtrsim 30$ the deviation with respect to the mean is around $10^{-7}$. 
     }
     \label{fig:Fap-vs-N}
 \end{figure}

 The only thing left to complete this perturbative construction of black hole tomography is to estimate the values of the constants $\mathcal{F}_{lmn}$. 
 Given that the sum $\mathcal{F}_{lmn}$ is convergent when the coefficients $a_k$ are computed using the eigenfrequencies, we can approximate it by a finite number of terms, i.e., we can approximate $\mathcal{F}_{lmn}$ by $\mathcal{F}_{lmn}^\text{app}$, that we define as
 \begin{equation}\label{eq:Fapp}
       \mathcal{F}_{lmn}^\text{app} (N) = \sum_{k=0}^N a_k \,,
 \end{equation} where $N$ will be chosen such that the sum has already converged. In Fig.~\ref{fig:Fap-vs-N}, we show $\mathcal{F}_{lmn}^{\text{app}}$ as a function of $N$ for the fundamental mode of $l=2,3,4$ (for non-rotating black holes like the ones considered here, $\mathcal{F}_{l-mn} =\bar{\mathcal{F}}_{lmn}$ for $m\neq0$, and  $\mathcal{F}_{l0n}=\mathcal{F}_{l1n}=...=\mathcal{F}_{lln}$). The approximate value $ \mathcal{F}_{lmn}^{\text{app}}$ converges quickly in all cases (for $N\gtrsim 30$ deviation of individual points with respect to the average value is at maximum of the order $10^{-7}$).  The value of $\mathcal{F}_{lmn}^{\text{app}}$ also converges quickly when we evaluate the coefficients $a_k$ using higher overtones.  In table~\ref{tab:Fl0n}, we present approximate values of $\mathcal{F}_{lmn}^{\text{app}}$ for $l=2,3,4$ and $n=0,1,2$. Its real and imaginary parts are obtained as the averaged value of Eq.~\eqref{eq:Fapp} for $N\in[30,200]$. However, a word of caution is in order: the numerical techniques we use to evaluate the value of the QNM frequencies do not solve the infinite fraction~\eqref{eq:QNM-frequencies}  exactly. In practice, the QNM frequencies are obtained using some approximation techniques, for instance, by truncating Eq.~\eqref{eq:QNM-frequencies} by a large, but finite number of nested fractions. As a consequence, the QNM frequencies we use are also approximate. Given that Eq.~\eqref{eq:Flmn} is only guaranteed to converge for the eigenfrequencies of Eq.~\eqref{eq:QNM-frequencies}, if we choose a sufficiently large $N\gg 100$, we would obtain numerical instabilities in Fig.~\ref{fig:Fap-vs-N}~\footnote{Heuristically, higher $l$ narrows the stability region, i.e., for higher $l$, $\mathcal{F}_{lmn}^{\text{app}}$ destabilizes for smaller $N$. 
 }. Hence, the values presented in Tab.~\ref{tab:Fl0n} are only meant to give an estimate of the order of magnitude of the constants $\mathcal{F}_{lmn}$, a more careful analysis would be needed to accurately evaluate them and assess their error.  However, we can already see from Eq.~\eqref{eq:tomography-psi4} that if we were to detect the fundamental modes of the $l=2,3,4$ mode in the gravitational strain, the magnitude of the amplitude of that mode at the horizon would be larger than observed at $\mathcal{I}^+$ given that $|\mathcal{F}_{l00}|<1$.  

\begin{table}
    \centering
    \begin{tabular}{|c|c|c|c|}
    \hline
         &  $n=0$ & $n=1$ & $n=2$   \\
        \hline
       $l=2$  & $0.398-0.361 i$ & $0.823 + 0.753 i$ & $-1.695+3.470 i $\\
        $l=3$ & $0.241 -0.502 i$& $0.867 + 0.966 i$ & $-3.602+1.584 i$\\
        $l=4$ & $0.049 -0.561 i$& $1.016 + 0.936 i$ & $-3.885+0.838 i$\\
        \hline
    \end{tabular}
    \caption{Average values of $\mathcal{F}_{l0n}^{\text{app}}$ for $l=2,3,4$ and $n=0,1,2$ and $N\in[30,200]$.   }
    \label{tab:Fl0n}
\end{table}

\subsection{Dependence on the slicing}
Finally, notice that the values reported in Tab.~\ref{tab:Fl0n} are slicing dependent.  In the following, we show this explicitly.

\begin{figure}
    \centering
    \includegraphics[width=\linewidth]{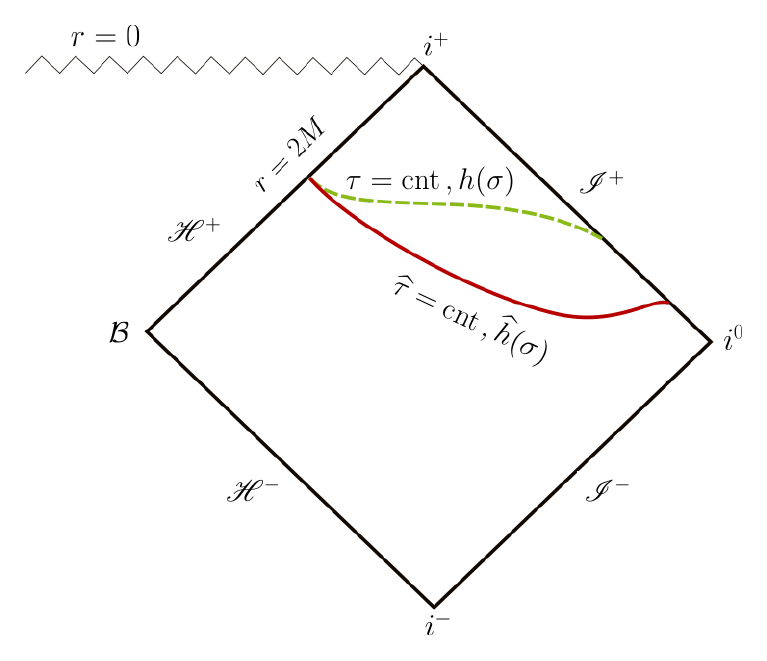}
    \caption{Illustration of a single $\tau=\text{cnt}$ and $\widehat{\tau}=\text{cnt}$ slice with $f(1)=0$. These two slices connect the same cross-section of the horizon $S_v$ with two different points at $\mathcal{I}^+$.}
    \label{fig:hyperboloidal-slicing}
\end{figure}

The idea behind using an hyperboloidal slicing is to create a coordinate system that is regular both at the horizon and at $\mathcal{I}^+$. To accomplish this, we introduced the height function $h(\sigma)=-2c(1/\sigma-\log\sigma)$ in Eq.~\eqref{eq:hyperboloidal} that removes the singularities of the coordinate system both at the horizon and infinity. 
The transformation in Eq.~\eqref{eq:hyperboloidal} expresses the Weyl scalar $\widetilde{\Psi}_4$ in the so-called minimal gauge~\cite{PanossoMacedo:2018hab}. However, other gauge transformations that are regular for $\sigma\in[0,1]$ are also possible and would give rise to a different slicing of the spacetime.  As an example, we could consider the family of transformations
\begin{equation}\label{eq:hyperboloidal-2}
    v=2c\left(\widehat{\tau}+\frac{1}{\sigma}-\log\sigma +f(\sigma)\right)\,,\quad r=\frac{c}{\sigma}-c
\end{equation} where we are using the height function $\widehat{h}(\sigma) =-2c(1/\sigma -\log(\sigma) +f(\sigma))$, with $f(\sigma)$  a real, regular function, and $f(0)$ and $f(1)$ finite.  
Applying this coordinate transformation to the Weyl scalar $\widetilde{\Psi}_4$ and evaluating it at the horizon yields Eq.~\eqref{eq:psi4-psi0-horizon-tomography:1} with $\tau\to\widetilde{\tau}$ and the functions
\begin{subequations}
    \begin{align}
\widehat{\Psi}_{4,lmn}^{H,-} =& (b_{lmn}^- {}_{-2} Y_{lm} +c_{lmn}^- {}_{-2}Y_{l-m}) e^{-2ic\omega_{lmn} f(1)}\\
\widehat{\Psi}_{4,lmn}^{H,+} =& (b_{lmn}^+ {}_{-2} Y_{lm} +c_{lmn}^+ {}_{-2}Y_{l-m}) e^{2ic\bar{\omega}_{lmn} f(1)}\,.
\end{align}
\end{subequations} Evaluating  $\widetilde{\Psi}_4$ at $\mathcal{I}^+$ (around $ \sigma=0$) in the coordinate system~\eqref{eq:hyperboloidal-2} yields the second expression in Eq.~\eqref{eq:psi4-infinity} with the functions 
\begin{subequations}
\begin{align}
    \widehat{\Psi}_4^{\infty,-}= &e^{2ic\omega_{lmn}(f(1)-f(0))} \sum_{k} a_k \widehat{\Psi}_{4,lmn}^{H,-} \\
    \widehat{\Psi}_4^{\infty,+}= &e^{-2ic\bar{\omega}_{lmn}(f(1)-f(0))} \sum_{k} \bar{a}_k \widehat{\Psi}_{4,lmn}^{H,+} \,
\end{align} to first order in $\sigma$. 
\end{subequations} This expression explicitly evidences that changing the slicing by a function $f(\sigma)$ changes also the relationship between the value of $\widetilde{\Psi}_4$ at the horizon and $\mathcal{I}^+$, given that
\begin{equation}\label{eq:F-hat}
    \widehat{\mathcal{F}}_{lmn} = e^{2ic\omega_{lmn} (f(1)-f(0))} \mathcal{F}_{lmn}
\end{equation} with $\mathcal{F}_{lmn}$ defined in Eq.~\eqref{eq:Flmn}. This highlights the fact that the mapping between the horizon and $\mathcal{I}^+$ is slicing dependent. The physical meaning of this dependence is that by choosing different slicing we connect different cross-sections  of the horizon with different points of $\mathcal{I}^+$. To see this, consider a function $f(\sigma=1) = 0$. Then, the minimal gauge in Eq.~\eqref{eq:hyperboloidal} and the transformation in Eq.~\eqref{eq:hyperboloidal-2} map the cross-section of the horizon $S_v$ to two distinct points in $\mathcal{I}^+$ as we show in Fig.~\ref{fig:hyperboloidal-slicing}.  

As long as we use a ``time coordinate'' that is linearly related to a ``well-defined'' notion of time of the background, changing the slicing will yield the same QNM frequencies, although the relationship between the modes at the horizon and far away will be different, as shown by Eq.~\eqref{eq:F-hat}. Still, notice that the discussion above relating the modes of $\widetilde{\Psi}_0$ at the horizon, and those of $\widetilde{\Psi}_4$ at $\mathcal{I}^+$ goes trough by replacing $\mathcal{F}\to\widehat{\mathcal{F}}$ using Eq.~\eqref{eq:F-hat}. Hence, the above discussion about black hole tomography still applies once we fix the gauge.

\section{$\widetilde{\Psi}_2$ at the horizon and null infinity}
\label{sec:multipole-moments}
In this section, we also provide the explicit expressions for the scalar component of the Weyl tensor $\widetilde{\Psi}_2$. In paper I, we found that the geometric information about the deformation of the horizon geometry was encoded in  $\widetilde{\Psi}_2$.
In this dynamic setting, unfortunately, an interpretation in terms of horizon and field multipole moments, and related surficial and field Love numbers, is not straightforward. We will defer such an interpretation to later work.

To obtain an expression for $\widetilde{\Psi}_2$ everywhere, we need to integrate its radial equation. 
For this, it comes in handy to derive a differential equation for the radial part of $\widetilde{\Psi}_2$ as a function of $\widetilde{\Psi}_4$. We start by taking the $\Delta$ derivative of Eq.~\eqref{eqs:radial-bianchi:2}
\begin{equation}\label{eq:delta2psi2}
    \Delta^2 \Psi_2 -\Delta\eth\Psi_3= -3 (\Delta\mu \Psi_2 + \mu\Delta\Psi_2) +\Delta (\bar{\pi} \Psi_3) + \Delta(\sigma \Psi_4)
\end{equation} where we have \emph{not} yet expanded this equation to first order in the radiative perturbation. The second term on the left-hand-side can be rewritten as 
\begin{equation}\label{eq:deltaethPsi3}
\begin{split}
     \Delta\eth\Psi_3  &= \eth^2\Psi_4 -4\eth\mu \Psi_3 -4\mu \eth\Psi_3 +2\eth\bar{\pi} \Psi_4 +2\bar{\pi}\eth \Psi_4  \\
     &+\bar{\pi} \Delta\Psi_3-\mu \eth\Psi_3 -\bar{a} \mu \Psi_3 -\bar{\lambda} \bar{\eth} \Psi_3 +\bar{\lambda} a\Psi_3 {- \Delta \bar{a} \Psi_3} \, ,
\end{split}
\end{equation} 
where we used the commutation relation between $\delta$ and $\Delta$ in Eq.~\eqref{eqs:commutation} and Eq.~\eqref{eqs:radial-bianchi:3}. Plugging Eq.~\eqref{eq:deltaethPsi3} in Eq.~\eqref{eq:delta2psi2}, expanding the $\eth, \bar \eth$ operators in terms of the angular directional derivatives $\delta\,,\bar{\delta}$, using Eqs.~\eqref{eqs:radial-bianchi:3},~\eqref{eqs:radial:2},~\eqref{eqs:radial:8}, and $\pi= \alpha + \bar \beta$ we obtain 
\begin{equation}
\begin{split}
    \Delta^2\Psi_2 = & \delta^2\Psi_4 -4\delta\mu\Psi_3 -5\mu\delta\Psi_3+4\delta(\beta \Psi_4) +\bar{\pi} \Delta\Psi_3 \\
    &-\bar{\lambda}\bar{\delta} \Psi_3 -3\mu\Delta\Psi_2 +3(\mu^2+|\lambda|^2) \Psi_2 \\
    &+2\beta \Delta\Psi_3 -2(\mu\beta+\alpha \bar{\lambda}) \Psi_3 +\Delta(\sigma\Psi_4)\,.
\end{split}
\end{equation} Collecting the terms with $\Psi_3$ and using Eq.~\eqref{eqs:radial-bianchi:2} to eliminate the terms $\delta\Psi_3$, we obtain
\begin{equation}
    \begin{split}
        \Delta^2 \Psi_2 = &\delta^2\Psi_4 +4\delta(\beta\Psi_4) +(\bar{\alpha}+3\beta) \delta\Psi_4 -8\mu \Delta\Psi_2\\
        &+4\beta (\bar{\alpha}+3\beta)\Psi_4  +(3|\lambda|^2 -12\mu^2) \Psi_2\\
        &+[\Delta(\sigma \Psi_4) -4\delta\mu\Psi_3 -\bar{\lambda}\bar{\delta} \Psi_3 -2\alpha \bar{\lambda} \Psi_3 \\
        &+5\mu\sigma \Psi_4 -4\mu\bar{\pi} \Psi_3]\,.
    \end{split}
\end{equation} Now expanding to first order in the radiative perturbation around the Schwarzschild background, we see that the expression between squared parenthesis vanishes to first order in the perturbation. Rearranging the terms, we obtain that at first order
\begin{equation}
\begin{split}
    \Delta^2 \widetilde{\Psi}_2+&8\mu\Delta \widetilde{\Psi}_2 +12\mu^2\widetilde{\Psi}_2 = \delta^2\widetilde{\Psi}_4 +4\delta(\beta \widetilde{\Psi}_4) \\
    &+(\bar{\alpha}+3\beta) \delta \widetilde{\Psi}_4 +4\beta (\bar{\alpha} +3\beta) \widetilde{\Psi}_4
    \end{split}
\end{equation} Finally, rewriting this expression in terms of the $\eth$ operator and taking into account that $\pi=0$ for the background spacetime, we obtain the concise expression
\begin{equation}
    \Delta^2 \widetilde{\Psi}_2 +8\mu \Delta\widetilde{\Psi}_2 +12\mu^2 \widetilde{\Psi}_2 = \eth^2 \widetilde{\Psi}_4\,,
\end{equation} relating the perturbation of $\Psi_2$ with the gravitational wave. Notice that this is a second-order differential equation, and as such, we must check that its solution is a solution of the radial first-order differential equation for $\Psi_2$ (see Eq.~\eqref{eqs:radial-bianchi:2} in App.~\ref{sec:field-equations}). The general solution of this differential equation is of the form
\begin{equation}\label{eq:psi2-tilde}
    \begin{split}
         \widetilde{\Psi}_2 &= \frac{q_1}{(r+c)^4} + \frac{q_2}{(r+c)^3} \\
         &-\frac{1}{(r+c)^4} \int_{0}^r (r^\prime+c)^5 \eth^2\widetilde{\Psi}_4 \d r^\prime\\
         &+\frac{1}{(r^\prime+c)^3} \int_0^r (r^\prime +c)^4 \eth^2 \widetilde{\Psi}_4 \d r^\prime 
    \end{split}
\end{equation}
where the integration constants  $q_i=q_i(v\,,z\,,\bar{z})$ are functions on the horizon hyperfurface. For the ringdown solution we discussed above, we can find an explicit solution in terms of a function $g$ (and its complex conjugate $\bar{g}$), which we define as being proportional to the radial part of $\widetilde{\Psi}_0$, $X^{(0)}_{lmn}$, i.e., 
\begin{equation}
     g_{lmn}(r) = \frac{16 (r+c)^4 c^4 i\omega_{lmn} (\kappa_{(l)}^2+\omega_{lmn}^2)(2\kappa_{(l)}-i\omega_{lmn})}{l^2(l-1)^2(l+1)^2(l+2)^2+6^2 c^2\omega_{lmn}^2 } X_{lmn}^{(0)}\,.
\end{equation}
Defining this function is useful because through the Teukolsky-Starobinsky identities, the radial part of $\widetilde{\Psi}_4$ can be written as the fourth radial derivative of the function $g$
\begin{equation}
     X_{lmn}^{(4)} = \Delta^4[g_{lmn}(r)]\,. 
\end{equation} This allows to integrate Eq.~\eqref{eq:psi2-tilde} explicitly, yielding
\begin{widetext}
\begin{equation}\label{eq:psi2-full}
\begin{split}
     \widetilde{\Psi}_2 = \frac{q_1}{(r+c)^4} + \frac{q_2}{(r+c)^3} +\frac{1}{2\pi}\sum_{l\,,m\,,n} \frac{\sqrt{l(l+1)(l-1)(l+2)}}{2} & \left\{e^{-i\omega_{lmn}v}(b_{lmn}^- Y_{lm} + c_{lmn}^- Y_{l-m}) \Delta^2\left[\frac{g}{(r+c)^2}\right]\right. \\
     &\left.+e^{i\bar{\omega}_{lmn}v}(b_{lmn}^+ Y_{lm} + c_{lmn}^+ Y_{l-m}) \Delta^2\left[\frac{\bar{g}}{(r+c)^2}\right] \right\} \,.
\end{split}  
\end{equation}The functions $q_1$ and $q_2$ are fixed by the boundary condition at the horizon in Eq.~\eqref{eq:solution1}, and the condition that it satisfies the first-order radial differential equation for $\widetilde{\Psi}_2$. Explicitly, we find
\begin{subequations}\label{eq:q1&2}
      \begin{align}
    q_1 =& \frac{1}{2\pi}\sum_{lmn}  \frac{3c^2}{2} K_l (l^2+l-3)\left\{\frac{(a_{lmn}^- Y_{lm} +(-1)^m \bar{a}_{lmn}^+ Y_{l\,,-m}) e^{-i\omega_{lmn}v}}{(K_l^2-6i c \omega_{lmn})(\kappa_{(\ell)}^2+\omega_{lmn}^2)}+\frac{(a_{lmn}^+ Y_{lm} +(-1)^m \bar{a}_{lmn}^- Y_{l\,,-m}) e^{i\bar{\omega}_{lmn}v}}{(K_l^2+6i c \bar{\omega}_{lmn})(\kappa_{(\ell)}^2+\bar{\omega}_{lmn}^2)} \right\} \\
    \begin{split}
         q_2 = &- \frac{1}{2\pi}\sum_{lmn}  \frac{3c}{2} K_l \left\{(l^2+l-2-2ic\omega_{lmn})\frac{(a_{lmn}^- Y_{lm} +(-1)^m \bar{a}_{lmn}^+ Y_{l\,,-m}) e^{-i\omega_{lmn}v}}{(K_l^2-6i c \omega_{lmn})(\kappa_{(\ell)}^2+\omega_{lmn}^2)}\right.\\
    &\left.+(l^2+l-2+2ic\bar{\omega}_{lmn})\frac{(a_{lmn}^+ Y_{lm} +(-1)^m \bar{a}_{lmn}^- Y_{l\,,-m}) e^{i\bar{\omega}_{lmn}v}}{(K_l^2+6i c \bar{\omega}_{lmn})(\kappa_{(\ell)}^2+\bar{\omega}_{lmn}^2)} \right\}
    \end{split}
\end{align} where we have defined $K_l=\sqrt{(l+2)(l+1)l(l-1)}$. 
\end{subequations}
\end{widetext}

This explicit expression shows that the perturbation of $\Psi_2$ at the horizon does not vanish when $\widetilde{\Psi}_4$ (or alternatively $\widetilde{\Psi}_0$) is nonvanishing at the horizon. Additionally, Eq.~\eqref{eq:q1&2} shows that the functions $q_1$ and $q_2$ do not vanish in general. However, if the constants $a^\pm_{lmn}$ satisfy some relationship such that, e.g., $q_1=0$, then $q_2=0$. The converse is also true. 

In paper I, we identified the term $r^{-3}$ in $\widetilde{\Psi}_2$ with the field mass monopole and the term $r^{-4}$ with its field spin dipole.  Furthermore, the Weyl scalar $\Psi_2$ at the horizon was related to the horizon multipole moments. 
Such an identification in the dynamic case is not obvious, as the scaling of the different $l$ modes of the spherical harmonic decomposition is no longer directly associated with a given power of $r$.  This is reminiscent of the structure of the field multipole moments in the post-Newtonian setting \cite{Blanchet:2013haa}. 

\section{Conclusions}
\label{sec:conclusions}

In this paper, we have applied the characteristic initial value
formulation to study the near horizon geometry of perturbed horizon
including the effects of small amounts of infalling radiation. In the context of a binary system, this effect of infalling radiation in the horizon geometry is called tidal heating.  We have
focused our attention on the ringdown phase, and we have reformulated
the black hole QNM problem in a fully 4-dimensional setting.  We
recover the usual Schwarzschild QNM frequencies. The reformulation also sheds light on the minimal conditions required for QNM solutions to appear. Specifically, we did not need to impose the absence of incoming modes, but found that demanding analyticity and stable solutions towards the future naturally selected the QNM frequencies. Furthermore, we showed that, in the presence of a small flux of infalling gravitational radiation, the unique solution of the perturbative equations when we demand a mode decomposition, separability of the horizon equations, analyticity and stability towards the future is this QNM solution. This may have ramifications for data analysis, as it makes transparent when the QNM description is valid.

Our analysis also addresses the ``mystery'' of why the infalling
radiation has the same features as the ringdown waveform which is
a superposition of damped sinusoidal signals.  This is an example of
black hole ``tomography'' in a perturbative setting: The present work
can be seen as providing analytical support for the existence of
correlations between gravitational wave observations and horizon
dynamics.

In forthcoming work, we will complete the final remaining step,
i.e., reconstruct the near horizon metric. For the QNM case, this
will yield an explicit spacetime metric near the remnant black hole
resulting from a binary black hole merger when
the final black hole is slowly spinning. We will also study physical
aspects of this spacetime, such as properties of the light-ring and
the effective potential near the black hole. The dynamical multipole moments of the horizon and asymptotic field, as well as the Love numbers will be further investigated along with
implications for gravitational wave astronomy.  
 
This
analysis will also be particularly suited to the case of extreme-mass-ratio
systems wherein linear perturbation theory provides an excellent
approximation.  This would eventually provide a different route to the
spacetime mapping problem where one used the motion of a small object
around a supermassive black hole to map the spacetime around it with
great precision \cite{Ryan:1995wh}.  Here, one would be using the EMRI
to measure the source multipole moments of the large black hole.  

Our work highlights several open issues.  There is the natural
question of extending this work to arbitrary spins, i.e. for a general
Kerr black hole. While technically challenging, there are no questions
of principle in that direction. 
Furthermore, several works have
found instabilities in the QNM frequencies as the effective potential
is varied
\cite{Jaramillo:2020tuu,Destounis:2021lum,Jaramillo:2022kuv,Boyanov:2022ark,Cardoso:2024mrw}. It would be interesting to investigate whether such instabilities
occur also in our formulation of the QNM problem.

We have been working in the setting for which the lowest multipole
moments, namely the mass and angular momentum, do not change at linear order. This is a consequence of the assumption that the flux at the horizon is small. This approximation ensures that the horizon inherits many structures from the isolated horizon framework, simplifying its treatment. However, this assumption can, and will be, relaxed.
This requires the horizon to be modeled as a dynamical horizon transitioning to the perturbed isolated horizon regime described in this study \cite{Booth:2012xm,Booth:2003ji}. Such a transition could be captured using second-order perturbation theory to account for the infalling flux, as this suffices to describe the spacelike nature of the dynamical horizon. A particularly exciting direction is to develop a second-order black hole tomography for the ringdown. This would enable us to investigate whether the nonlinearities observed at infinity correspond directly to those experienced by the horizon.

 However, already at linear order we gained invaluable insight.

While we have not analyzed the multipole moments explicitly, it is clear that the higher multipole moments depend on the infalling radiation already at linear order. This is generically not accounted for in the self-force program, but might
have potentially observable consequences for EMRI observations and the spacetime mapping problem.

\section*{Acknowledgments} 
We are grateful to Abhay Ashtekar, Patrick Bourg, Scott Hughes, Neev Khera, Jos\'e Luis Jaramillo, and
 Bangalore Sathyaprakash  for valuable discussions.

\bibliographystyle{apsrev4-1}
\bibliography{main}{}

\appendix
\section{The Newman-Penrose formalism and our coordinate system}
\label{sec:field-equations}
\subsection{The Newman-Penrose formalism in a nutshell}
\label{subsec:np-nutshell}

The results discussed in the main text are expressed in the language of the Newman-Penrose formalism, which we summarize in the following. Given that we want to exhaust the null structures of our spacetime (given by the quasi-isolated horizon and null infinity), it is natural to introduce a tetrad basis of four \emph{null} vectors $(\ell^a\,, n^a\,,m^a\,,\bar{m}^a)$ at each point of spacetime. 
These vectors satisfy the following inner-product relations
\begin{equation}
  \ell\cdot n = -1\,,\quad m\cdot\bar{m} = 1\,, \label{eq:tetrad-orthonormality}
\end{equation}
with all other inner-products vanishing. The vector $n$ is chosen to be a past-directed, null geodesic, affinely parametrized by the coordinate $r$, and emanating from a cross-section of the horizon at $r=0$. Hence, $n$ takes the form 
 \begin{equation}
  \label{eq:2}
n^a\nabla_a
  := -\frac{\partial}{\partial r}\,.
\end{equation}  The motivation to use this coordinate system and tetrad was explained in great detail in Paper I, here we shall assume this form for the tetrad component $n$ without proof. To satisfy the inner product relations in Eq.~\eqref{eq:tetrad-orthonormality}, the other basis vectors must be of the form:
\begin{subequations}\label{eq:tetrad-general}
  \begin{align}
  \label{eq:3}
  \ell^a\nabla_a &:= D = \frac{\partial}{\partial v} + U\frac{\partial}{\partial r}
                   + X\frac{\partial}{\partial z} + \bar{X}\frac{\partial}{\partial \bar{z}}\,,\\
    m^a\nabla_a&:= \delta =
  \Omega\frac{\partial}{\partial r} + \xi_1\frac{\partial}{\partial z} + \xi_2\frac{\partial}{\partial \bar{z}}\,.
  \end{align}
\end{subequations}
The frame function $U$ is real while $X,\Omega,\xi_i$ are complex. Furthermore, as explained in Paper I, the frame functions can be chosen such that at the horizon the frame functions
\begin{equation}
U\triangleq \Omega\triangleq X \triangleq 0\,, 
\end{equation} vanish. The symbol $\triangleq$ denotes equality only at the horizon.

Associated with this vector basis at each point of spacetime we can also define directional covariant derivatives, which we denote
\begin{equation}\label{eq:directional-derivatives}
  D:= \ell^a\nabla_a\,,\quad \Delta := n^a\nabla_a\,,\quad \delta := m^a\nabla_a\,,\quad \bar{\delta} := \bar{m}^a\nabla_a\,.
\end{equation} The connection is represented as a set of 12 complex scalars, the spin coefficients, which are defined using the directional derivatives of the basis vectors:
\begin{subequations}
  \label{eq:spincoeffs}
  \begin{align}
    D\ell& = (\epsilon + \bar{\epsilon})\ell - \bar{\kappa}m -
    \kappa\bar{m} \,,\\
    Dn& = -(\epsilon + \bar{\epsilon})n + \pi m + \bar{\pi}m \,,\\
    Dm& = \bar{\pi}\ell - \kappa n + (\epsilon - \bar{\epsilon})m \,,\\
    \Delta \ell& = (\gamma + \bar{\gamma})\ell - \bar{\tau}m -
    \tau\bar{m}\,,\label{eq:Deltal}\\
    \Delta n & = -(\gamma + \bar{\gamma})n + \nu m + \bar{\nu}\bar{m}\,,\label{eq:Deltan}\\
    \Delta m & = \bar{\nu}\ell - \tau n + (\gamma-\bar{\gamma})m\,,\label{eq:Deltam}\\
    \delta\ell & = (\bar{\alpha} + \beta)\ell -\bar{\rho}m -
    \sigma\bar{m}\,,\\
    \delta n & = -(\bar{\alpha} + \beta)n + \mu m +
    \bar{\lambda}\bar{m}\,,\\
    \delta m & = \bar{\lambda}\ell - \sigma n + (\beta-\bar{\alpha})m\,,\label{eq:deltam}\\
    \bar{\delta}m & = \bar{\mu}\ell - \rho n +
    (\alpha-\bar{\beta})m\,. \label{eq:deltabarm}
  \end{align}
\end{subequations}
Eq.~\eqref{eq:spincoeffs} is useful to understand the geometric meaning of the spin coefficients. Some important quantities for us are the optical scalars of $\ell$ and $n$: the expansion of $\ell$
 and $n$ are given by the real parts of $\rho$ and $\mu$ respectively, while their twist is encoded in the imaginary part of these spin coefficients. The shears of $\ell$ and $n$ are $\sigma$ and $\lambda$. Apart from the optical scalars, the spin coefficients characterizing the tetrad basis are:  $\kappa$ and $\nu$ indicate that $\ell$  and $n$ are geodetic, while the $\epsilon+\bar{\epsilon}$ and $\gamma+\bar{\gamma}$ are their respective accelerations. The quantity $a=\alpha-\bar{\beta}$ codifies the connection in the $m-\bar{m}$ plane, and is therefore related to the curvature of the 2-manifold spanned by $m\,,\bar{m}$.  
 
Since the null tetrad is typically not a coordinate basis, the above definitions
of the spin coefficients lead to non-trivial commutation relations:
\begin{subequations}\label{eqs:commutation}
    \begin{align}
    \Delta D-D \Delta &= (\epsilon+\bar\epsilon) \Delta-\pi \delta-\bar\pi\bar\delta \\
        \delta D-D\delta& = \kappa \Delta-(\bar\rho+\epsilon-\bar\epsilon) \delta-\sigma\bar\delta \\
        \delta\Delta-\Delta\delta&=-\bar\pi \Delta+\mu\delta+\bar\lambda \bar\delta \\
        \bar\delta\delta-\delta\bar\delta&= (\bar\rho-\rho)\Delta+a\delta-\bar a\bar\delta \, .
    \end{align}
\end{subequations}
The Weyl tensor $C_{abcd}$ breaks down into five complex scalars
\begin{subequations}
  \begin{align}
  \Psi_0 &= C_{abcd}\ell^am^b\ell^cm^d\,,\quad  \Psi_1 = C_{abcd}\ell^am^b\ell^cn^d\,,\\
  \Psi_2 &=  C_{abcd}\ell^am^b\bar{m}^cn^d\,,\quad  \Psi_3 = C_{abcd}\ell^an^b\bar{m}^cn^d\,,\\
  \Psi_4 &= C_{abcd}\bar{m}^an^b\bar{m}^cn^d\,.
  \end{align}
\end{subequations}

Projecting the Einstein field equations in the tetrad basis yields a system of 16 complex first-order differential equations relating the spin coefficients with the curvature scalars. These are the so-called Newman-Penrose field equations. The
Bianchi identities, $\nabla_{[a}R_{bc]de} = 0$, are written explicitly
as eight complex equations involving both the Weyl and Ricci tensor
components, and three real equations involving only Ricci tensor
components.  See
\cite{Penrose:1985jw,Chandrasekhar:1985kt,Stewart:1991} for the full
set of field equations and Bianchi identities (but beware that they
use slightly different conventions such as the sign for the metric
signature and normalization of the null tetrad, leading to some
minus sign changes).

Contrary to the usual formulation of the Newman-Penrose field equations, it will be useful to introduce the notion of spin-weights and to work with the $\eth$ operator for derivatives in the $m-\bar{m}$ plane. 
  A tensor $X$ projected on the
$m$-$\bar{m}$ plane is said to have spin weight $s$ if under a spin
rotation $m\rightarrow e^{i\psi}m$, it transforms as $X\rightarrow
e^{i s\psi}X$.  Thus, $m^a$ itself has spin weight $+1$ while
$\bar{m}^a$ has weight $-1$.  For  instance, the scalar $X= m^{a_1}\cdots
m^{a_p}\bar{m}^{b_1}\cdots \bar{m}^{b_q}X_{{a_1}\cdots {b_q}}$ 
has spin weight $s=p-q$ and the Weyl tensor component
$\Psi_k$ has spin weight $2-k$.  

The $\eth$ and $\bar{\eth}$ operators are defined as
\begin{eqnarray}
  \label{eq:19}
  \eth X = m^{a_1}\cdots m^{a_p}\bar{m}^{b_1}\cdots
  \bar{m}^{b_q}\delta X_{{a_1}\cdots {b_q}}\,,\\
  \bar{\eth} X = m^{a_1}\cdots m^{a_p}\bar{m}^{b_1}\cdots
  \bar{m}^{b_q}\bar{\delta} X_{{a_1}\cdots {b_q}}\,.
\end{eqnarray}
From Eqs.~(\ref{eq:deltam}) and (\ref{eq:deltabarm}), after projecting
onto the $m$-$\bar{m}$ plane, we get
\begin{equation}
  \delta m^a = (\beta-\bar{\alpha})m^a\,,\qquad \bar{\delta}{m}^a =
  (\alpha-\bar{\beta})m^a\,. 
\end{equation}
A short calculation shows that
\begin{equation}
  \label{eq:55}
  \eth X = \delta X + s(\bar{\alpha}-\beta)X\,,\qquad \bar{\eth}X =
  \bar{\delta}X - s(\alpha-\bar{\beta})X\,.
\end{equation}
The $\eth$ and $\bar{\eth}$ act as spin raising and
lowering operators. This means in particular,  that when acting over the spin-weighted spherical harmonic basis defined in the 2-manifold spanned by $m\,,\bar{m}$, they satisfy the following useful relationships
\begin{subequations}
\begin{align}
    \eth \, {}_s Y_{lm} & = \frac{1}{\sqrt{2} (r+c)}\sqrt{(l-s)(l+s+1)} {}_{s+1} Y_{lm}\,,\\
    \bar{\eth}\, {}_s Y_{lm} & =-\frac{1}{\sqrt{2} (r+c)} \sqrt{(l+s)(l-s+1)} {}_{s-1} Y_{lm}\,,\\
    \bar{\eth}\eth \, {}_sY_{lm}&=-\frac{(l-s)(l+s+1)}{2(r+c)^2} {}_sY_{lm}\,.
\end{align}
\end{subequations}
 See \cite{Goldberg:1966uu} for further properties of the $\eth$ operator and its connection to representations of the rotation group.

Rewriting the field equations and Bianchi identities using the $\eth$ and $\bar{\eth}$ operators (for those spin coefficients with a well-defined spin-weight) and splitting them into angular, evolution, and radial equations, we obtain the following five sets of differential equations, encompassing the system of angular equations for the spin coefficients  
\begin{subequations}
  \label{eqs:angular}
  \begin{align}\label{eqs:angular:1}
    \eth\rho - \bar{\eth}\sigma &= \bar{\pi}\rho -
                                    \pi \sigma -\Psi_1\,, \, \\ \label{eqs:angular:2}
    \delta\alpha - \bar{\delta}\beta &= \mu\rho -\lambda\sigma +
                                       |\alpha|^2 + |\beta|^2 -2\alpha\beta -\Psi_2\,, \, \\ \label{eqs:angular:3}
    \eth\lambda -\bar{\eth}\mu &= \pi\mu -
                                     \bar{\pi}\lambda -\Psi_3\,,  \end{align}
\end{subequations}
their evolution equations
\begin{subequations}
  \label{eqs:timeevolution}
  \begin{align} 
    D\rho -\bar{\eth}\kappa &= \rho^2 + |\sigma|^2 +
                                (\epsilon+\bar{\epsilon})\rho -\pi\kappa\,, \, \label{eq:drho}\\ 
    D\sigma-\eth\kappa &= (\rho+\bar{\rho} +3\epsilon-\bar{\epsilon})\sigma
                           -\bar{\pi}\kappa + \Psi_0\,,\, \label{eq:dsigma}\\ \label{eqs:timeevolution:1}
    D\alpha-\bar{\delta}\epsilon &= (\rho+\bar{\epsilon}-2\epsilon)\alpha +
                                   \beta\bar{\sigma} - \bar{\beta}\epsilon -\kappa\lambda +
                                   (\epsilon+\rho)\pi\,,\, \\ \label{eqs:timeevolution:2}
    D\beta -\delta\epsilon &= (\alpha+\pi)\sigma +
                             (\bar{\rho}-\bar{\epsilon})\beta -\mu\kappa
                             -(\bar{\alpha}-\bar{\pi})\epsilon + \Psi_1\,,\, \\ \label{eq:timeevolution:lambda}
    D\lambda -\bar{\eth}\pi &= (\rho-2\epsilon)\lambda +
                                \bar{\sigma}\mu + \pi^2\,,\, \\ \label{eqs:timeevolution:mu}
    D\mu - \eth\pi &= (\bar\rho - \epsilon-\bar{\epsilon})\mu + \sigma\lambda +
                       \bar{\pi}\pi + \Psi_2\,,
  \end{align}
\end{subequations} and their system of radial equations
\begin{subequations}
  \label{eqs:radial}
  \begin{align}\label{eqs:radial:1}
    \Delta\lambda &= -2\lambda\mu -\Psi_4\,,\\\label{eqs:radial:2}
    \Delta\mu &= -\mu^2 - |\lambda|^2\,,\\\label{eqs:radial:3}
    \Delta\rho &= -\mu\rho -\sigma\lambda -\Psi_2\,,\\\label{eqs:radial:4}
    \Delta\sigma &= -\mu\sigma -\bar{\lambda}\rho\,.\\\label{eqs:radial:5}
    \Delta\kappa &= -\bar{\pi}\rho - \pi\sigma -\Psi_1\,,\\\label{eqs:radial:6}
    \Delta\epsilon &= -\bar{\pi}\alpha - \pi\beta - \Psi_2\,,\\\label{eqs:radial:7}
    \Delta\pi &= -\pi\mu-\bar{\pi}\lambda -\Psi_3\,,\\\label{eqs:radial:8}
    \Delta\beta &= -\mu\beta -\alpha\bar{\lambda}\,,\\\label{eqs:radial:9}
    \Delta\alpha &= -\beta\lambda -\mu\alpha-\Psi_3\,.
  \end{align}  
\end{subequations}
Similarly, we can also split the eight complex Bianchi equations into a system describing the time-evolution of the Weyl scalars  

\begin{subequations}\label{eqs:evolution-bianchi}
\begin{align}\label{eqs:evolution-bianchi:1}
  D\Psi_1 -\bar{\eth}\Psi_0 &= -\pi \Psi_0 +
  2(2\rho+\epsilon)\Psi_1 -3\kappa\Psi_2\,,\, \\\label{eqs:evolution-bianchi:2}
  D\Psi_2-\bar{\eth}\Psi_1 &= -\lambda\Psi_0 +\pi \Psi_1 +
  3\rho\Psi_2 -2\kappa\Psi_3\,,\, \\  \label{eqs:evolution-bianchi:3}
  D\Psi_3 -\bar{\eth}\Psi_2 &= -2\lambda\Psi_1 + 3\pi\Psi_2
  +2(\rho-\epsilon)\Psi_3 -\kappa\Psi_4 \,,\, \\\label{eqs:evolution-bianchi:4}
  D\Psi_4-\bar{\eth}\Psi_3 &= -3\lambda\Psi_2 +5\pi \Psi_3 +
  (\rho-4\epsilon)\Psi_4 \,,
\end{align}
\end{subequations}
and their radial differential equations
\begin{subequations}
  \label{eqs:radial-bianchi}
  \begin{align} \label{eqs:radial-bianchi:0}
    \Delta\Psi_0 - \eth\Psi_1 &= -\mu\Psi_0 -\bar{\pi}\Psi_1 +
                                  3\sigma\Psi_2\,,\\\label{eqs:radial-bianchi:1}
    \Delta\Psi_1 -\eth\Psi_2 &= -2\mu\Psi_1 + 2\sigma\Psi_3\,,\\\label{eqs:radial-bianchi:2}
    \Delta\Psi_2 -\eth\Psi_3 &= -3\mu\Psi_2 + \bar{\pi}\Psi_3 +
                                 \sigma\Psi_4\,,\\ \label{eqs:radial-bianchi:3}
    \Delta\Psi_3 - \eth\Psi_4 &= -4\mu\Psi_3 + 2\bar{\pi}\Psi_4\,.
  \end{align}
\end{subequations}

Notice that Eqs.~\eqref{eqs:angular:2}, ~\eqref{eqs:timeevolution:1} and~\eqref{eqs:timeevolution:2} can be rewritten in a more convenient way in terms of the 2-manifold connection $a$ and the spin coefficient $\pi$. From the real and imaginary parts of Eq.~\eqref{eqs:angular:2} we can extract the following two equations
\begin{subequations}
    \begin{align}
       -2\re \Psi_2 &=\delta a +\bar{\delta}\bar{a} -2 a\bar{a} -\mu(\rho+\bar{\rho}) +\lambda \sigma + \bar{\lambda} \bar{\sigma}\,,\\
       -2i\im\Psi_2 & = \eth \pi-\bar{\eth}\bar{\pi} -\mu(\rho-\bar{\rho}) +\lambda\sigma -\bar{\lambda} \bar{\sigma}\,.
    \end{align}
\end{subequations} Furthermore, combining Eq.~\eqref{eqs:timeevolution:1} with the complex conjugate of~\eqref{eqs:timeevolution:2} we obtain
\begin{subequations}
    \begin{align}
        D\pi-\bar{\delta}(\epsilon+\bar\epsilon) &= \pi (2\rho+\bar\epsilon-\epsilon)+2\bar\sigma\bar\pi-\kappa\lambda -\bar\kappa\bar\mu+\bar\Psi_1 \\
        Da-\bar\delta(\epsilon-\bar\epsilon) &= a(\rho+\bar\epsilon-\epsilon)+\rho\pi-\bar\sigma(\bar\pi+\bar a)-\kappa\lambda +\bar\kappa\bar\mu \notag \\
        & \quad -\bar\Psi_1
    \end{align}
\end{subequations} This form of Eqs.~\eqref{eqs:angular:2}, \eqref{eqs:timeevolution:1}, and~\eqref{eqs:timeevolution:2} is the one that we use in the derivations in Sec.~\ref{sec:perturbed-intrinsic-geometry}.

Finally, the frame functions $U$, $\Omega$, $X$, and $\xi^A$ defined in Eq.~\eqref{eq:tetrad-general} also satisfy time-evolution
\begin{subequations}
  \label{eqs:alongv}
  \begin{align}
    D\Omega - \delta U&= \kappa + \rho\Omega + \sigma\bar{\Omega}\,,\\ 
    D\xi^i - \delta X^i &= (\bar{\rho}+\epsilon-\bar{\epsilon})\xi^i + \sigma\bar{\xi}^i\,.
  \end{align}
\end{subequations}
 and radial equations
\begin{subequations}
  \label{eqs:frame}
  \begin{align}
    \Delta U &= -(\epsilon+\bar{\epsilon}) - \pi\Omega -
               \bar{\pi}\bar{\Omega}\,,\\ 
    \Delta X^i &= -\pi\xi^i - \bar{\pi}\bar{\xi}^i\,,\\
    \Delta\Omega &= -\bar{\pi} - \mu\Omega -\bar{\lambda}\bar{\Omega}
                   \,,\\ 
    \Delta\xi^i &= -\mu\xi^i -\bar{\lambda}\bar{\xi}^i\,, 
  \end{align}
\end{subequations} which can be obtained from the commutation relations in Eq.~\eqref{eqs:commutation}.

\subsection{Summary of our gauge choices}
\label{subsec:gauge-choices}
In this work, we restrict ourselves to perturbations of a Schwarzschild isolated horizon. However, the framework we detail in the main text can in principle be applied to spacetimes that are \emph{not} of type D, as long as they contain an isolated horizon (e.g., the Robinson-Trautman spacetime). This is because we do not impose that the two tetrad vectors $\ell$ and $n$ be aligned along the two principal null directions of a type D spacetime. Instead, we require the two tetrad vectors $\ell$ and $n$ to be geodesic, and one of them, $\ell$,  to be a null generator of the horizon. The tetrad vector $n$ is chosen such that the tetrad basis is parallel propagated along it. Notice that this choice of tetrad coincides with aligning the two vectors  $\ell$ and $n$ along the principal null directions only for the Schwarzschild spacetime (so for instance, $\Psi_3\neq 0$ in general, even for type D spacetimes). This particular choice of coordinate system and tetrad was detailed in Paper I. For completeness, here we recap the gauge choices we made in the main text both for the background spacetime and the perturbation. 

For a general spacetime containing an isolated horizon, we choose the vector $n^a$ to be an affinely parameterized geodesic, along which $\ell$, $m$ and itself are parallel propagated. Then,  we have
$\Delta n = \Delta \ell = \Delta m = 0$.  From Eqs.~(\ref{eq:Deltal}),
(\ref{eq:Deltan}) and (\ref{eq:Deltam}), this leads to
\begin{equation}
  \label{eq:gamma-tau-nu}
  \gamma = \tau = \nu = 0\,.
\end{equation}
We first impose these conditions in the commutation relations in
Eqs.~(\ref{eqs:commutation}).  Then, setting $f=v$ in those equations
leads to
\begin{equation}
  \label{eq:pi-mu}
  \pi = \alpha + \bar{\beta}\,,\quad \mu = \bar{\mu}\,.
\end{equation}
These must hold throughout the region where the coordinate system is
valid.

When specifying the Schwarzschild isolated horizon as our unperturbed background (see Paper I for an explicit construction), the background spacetime can be shown to be the usual Schwarzschild spacetime in a special set of horizon-penetrating coordinates (analogous to the Eddington-Finkelstein ones), which we denote by $(v,r,z,\bar{z})$. The nonvanishing spin coefficients and Weyl scalars in these coordinates are   
\begin{subequations} \label{eq:Schwarzschild-nonvanishing-data}
    \begin{align} 
        \mu_\circ&= -\frac{1}{c+r}\,,\quad  a_\circ  = \frac{z}{\sqrt{2}(r+c)} \,,
        \quad\epsilon_\circ = \frac{c}{4(c+r)^2}\,,\\
        \rho_\circ &= -\frac{ r}{2(c+r)^2 }\,, \quad\Psi_2^\circ =-\frac{c}{2(c+r)^3}\,, 
    \end{align} and the unperturbed tetrad
\end{subequations} 
\begin{subequations}\label{eq:tetrad-unperturbed}
    \begin{align}
        l_\circ^a &= \partial_v +\frac{ r}{2(c+r)} \partial_r\,,\\
        n_\circ^a &= -\partial_r\,,\\
        m_\circ^a & = \frac{P_0}{ (c+r)} \partial_z\,.
    \end{align}
\end{subequations}We remove the subindex ${}_\circ$
 to denote the unperturbed quantities whenever possible.

In the main text, we perturb the basis vectors as follows
\begin{subequations}
    \begin{align}
        \ell^a = \ell^a_\circ+ \widetilde{\ell}^a\\
        n^a = n^a_\circ+ \widetilde{n}^a\\
        m^a = m^a_\circ+ \widetilde{m}^a
    \end{align}
\end{subequations} and analogously for $\bar{m}$. The perturbation to the basis vectors is taken to be of the form
\begin{subequations}
    \begin{align}
        \widetilde{l}^a\partial_a & = \widetilde{U}\partial_r + \widetilde{X}^A\partial_A\\
        \widetilde{m}^a\partial_a & = \widetilde{\Omega} \partial_r + \widetilde{\zeta}^A\partial_A
    \end{align} where $A=\{z\,,\bar{z}\}$. The outgoing transverse vector $n$ is unperturbed everywhere, so 
\end{subequations}
 \begin{equation}
    \widetilde{n}=0\,.
\end{equation} At the horizon, the perturbing functions have been chosen such that
\begin{equation}
    \widetilde{\Omega} \triangleq\widetilde{U}\triangleq\widetilde{X}^i \triangleq0\,, \widetilde{\xi}^z\triangleq0\,,
\end{equation} with 
\begin{equation}
    \widetilde{\xi}^{\bar{z}} \neq 0\,, \quad \widetilde{\bar{\xi}}^z \neq 0
\end{equation} the only nonzero quantities. This tetrad choice can be shown to still satisfy the orthonormality conditions
\begin{equation}
    (\ell_\circ+\widetilde{\ell})\cdot n_\circ=-1\,,\quad (m_\circ+\widetilde{m})\cdot (\bar{m}_\circ + \widetilde{\bar{m}}) =1\,,
\end{equation}
and to be compatible with the gauge choices we made at the horizon.

For the background and the perturbed spacetime we chose
\begin{equation}
    \gamma_\circ=\nu_\circ=\tau_\circ=0\,,\quad \widetilde{\gamma}=\widetilde{\nu}=\widetilde{\tau}=0\,,
\end{equation} and we work in a gauge in which 
\begin{equation}
    \pi_\circ = \alpha_\circ+\bar{\pi}_\circ\,,\quad a_\circ=\alpha_\circ-\bar{\beta}_\circ\,.
\end{equation} The perturbation to these spin coefficients is expressed in a gauge where analogous expressions hold for these spin coefficients, so
\begin{equation}
    \widetilde{\pi} = \widetilde{\alpha}+\widetilde{\bar{\beta}}\,, \quad \widetilde{a}=\widetilde{\alpha}-\widetilde{\bar{\beta}} \, .
\end{equation} Furthermore, 
\begin{equation}
    \mu=\bar{\mu}\,,\quad \rho=\bar{\rho}
\end{equation} and the same applies to the perturbed quantities. Regarding the conditions at the horizon, for the background
\begin{equation}
    \Psi_0^\circ\triangleq\Psi_1^\circ\triangleq0
\end{equation} and the surface gravity $\kappa_{(\ell)}$ is related to the following spin coefficients 
\begin{equation}
    \epsilon_\circ+\bar{\epsilon}_\circ\triangleq \kappa_{(\ell)}\,,\quad \epsilon_\circ-\bar{\epsilon}_\circ\triangleq 0\,.
\end{equation} The perturbation to these quantities satisfy
\begin{equation}
   \widetilde{\Psi}_0\neq 0\,,\quad \widetilde{\Psi}_1\neq0\,,\quad \widetilde{\epsilon} \triangleq0\,.
\end{equation} This gauge choice is not unique.  We could also choose instead $\widetilde{\epsilon} \neq 0$ and $\widetilde{\mu}\triangleq 0 $, this choice is explained in App.~\ref{sec:alternative-gauge}. 

Finally, we can choose the expansion to vanish at the horizon to first order in the perturbation, i.e.,
\begin{equation}
   \rho_\circ\triangleq0\,,\quad  \widetilde{\rho}\triangleq 0 \, .
\end{equation} The vector $\ell$ is chosen to be geodesic at the horizon
\begin{equation}
    \kappa_\circ\triangleq0\,,\quad \widetilde{\kappa}_\circ\triangleq 0 
\end{equation} and the shear also vanishes for the background spacetime, but not for the perturbation
\begin{equation}
    \sigma_\circ\triangleq0\,,\quad\sigma\neq0 \, .
\end{equation} These choices are general as long as we work with an isolated horizon background (these would also be valid for Kerr if we work in the coordinate system given by the initial value formulation). Because we are restricting ourselves to Schwarzschild, we also have
\begin{equation}
    \pi_\circ=\lambda_\circ=0
\end{equation} and 
\begin{equation}
    \Psi_0^\circ=\Psi_1^\circ=\Psi_3^\circ=\Psi_4^\circ=0 \, .
\end{equation}
In the main text, we consider perturbations of a Schwarzschild isolated horizon, and therefore, these simplifications have been used.

\section{Alternative gauge choice}
\label{sec:alternative-gauge}
Rather than considering $\widetilde{\epsilon}$ to vanish at the horizon, we can choose instead a coordinate system such that the past directed lightcones coincide with those of the unperturbed spacetime. This choice corresponds to $\widetilde{\mu}\triangleq 0$ and implies $\widetilde{\mu}=0$ given the radial equation for $\mu$~\eqref{eqs:radial:2}. Notice that given the remaining gauge freedom that we have, we cannot set both $\widetilde{\epsilon}\triangleq0$ and $\widetilde{\mu}\triangleq 0 $ simultaneously. In the following, we derive the system of ten differential equations that determine the initial data at the horizon and the respective constraint equations when $\widetilde{\mu}=0$. The procedure is the same as the one we used in the main text, so we shall be brief. Setting
\begin{equation}
    \widetilde{\mu}\triangleq 0\,,\quad \widetilde{\epsilon}-\widetilde{\bar\epsilon}\triangleq 0
\end{equation} we obtain the following system of ten differential equations
\begin{subequations}\label{eq:eqs-horizon-alternative}
    \begin{align}
        D\widetilde{\sigma} -\kappa_{(l)} \widetilde{\sigma} \triangleq& \widetilde{\Psi}_0\\
        D\widetilde{\Psi}_1-\kappa_{(l)} \widetilde{\Psi}_1 \triangleq & \bar{\eth}\widetilde{\Psi}_0\\
        D\widetilde{\Psi}_2 \triangleq &\bar{\eth}\widetilde{\Psi}_1\\
        D\widetilde{\pi} \triangleq & \bar{\eth} (\epsilon+\bar{\epsilon})+\widetilde{\bar{\Psi}}_1\\
        \mu D(\widetilde{\epsilon}+\widetilde{\bar\epsilon}) -\eth\bar\eth (\widetilde{\eth}+\widetilde{\bar\epsilon})\triangleq&\eth \widetilde{\bar{\Psi}_1}+\bar{\eth} \widetilde{\Psi}_1\\
        D\widetilde{\lambda} +\kappa_{(l)} \widetilde{\lambda} \triangleq & \bar{\eth} \widetilde{\pi} +\mu \widetilde{\bar{\sigma}}\\
        D\widetilde{\Psi}_3+\kappa_{(l)} \widetilde{\Psi}_3 \triangleq &  \bar{\eth}\widetilde{\Psi}_2+3\widetilde{\pi}\Psi_2 \\   
        D\widetilde{\Psi}_4+2\kappa_{(l)}\widetilde{\Psi}_4 \triangleq & \bar{\eth} \widetilde{\Psi}_3-3\widetilde{\lambda} \Psi_2 \\
       D\widetilde{a}\triangleq & -\bar{a}\widetilde{\bar{\sigma}}-\widetilde{\bar{\Psi}}_1\,,\\
       D\widetilde{\xi}^{\bar{z}}& \triangleq\widetilde{\sigma}  \bar{\xi}^{\bar{z}}
    \end{align}
\end{subequations} and a group of five constraint equations
\begin{subequations}\label{eq:eqs-constraint-alternative}
    \begin{align}
        \eth \widetilde{\lambda}\triangleq& \mu\widetilde{\pi}-\widetilde{\Psi}_3\\
        -2\re \widetilde{\Psi}_2 \triangleq& \eth\widetilde{a}+\bar{\eth}\widetilde{\bar{a}}+\widetilde{\eth}a+\widetilde{\bar\eth}\bar{a}\\
        -2i\im \Psi_2\triangleq&\eth\widetilde{\pi}-\bar{\eth}\widetilde{\bar{\pi}}\\
        \bar{\eth}\widetilde{\sigma}\triangleq & \widetilde{\Psi}_1\\
        \eth\widetilde{\pi} \triangleq& \mu (\widetilde{\epsilon}+\widetilde{\bar\epsilon}) -\widetilde{\Psi}_2\,.
    \end{align}
\end{subequations}
Notice that the expressions for $\widetilde{\sigma}$, $\widetilde{\Psi}_0$, $\widetilde{\Psi}_2$ and $\widetilde{\Psi}_4$ at the horizon do not change. 

\section{Complementary material}
\label{sec:complementary}
In this appendix, we compile some lengthy expressions that have been omitted from the main text for the sake of conciseness. In Eq.~\eqref{eq:psi0-radial-explicit}, we presented the solution to the radial Teukolsky equation for $\widetilde{\Psi}_0$ as a function of Heun functions, their derivatives and the polynomial functions $f_{lm}^{(1\,,2)}$ and $g_{lm}^{(1,2)}$. These polynomial functions are 
\begin{widetext}
\begin{subequations}\label{eqs:f-g-fncts}
    \begin{align}
        f^{(1)}_{lm} =& \frac{h_0^{(1)}(r) +h_1^{(1)}(r) \omega +h_2^{(1)}(r) \omega^2+h_3^{(1)}(r) \omega^3 + h_4^{(1)}(r) \omega^4}{16c^4 (r+c)^2 \omega(2\kappa_{(\ell)} -i\omega)(\kappa_{(\ell)}^2+\omega^2)}\,,\\
        g^{(1)}_{lm} = & \frac{r}{r+c} \frac{2r(r+c)(l^2+l-1)-(2r^2+c^2) -4(r+c)^4 \omega^2}{8c^5 (2\kappa_{(\ell)} -i\omega)(\kappa_{(\ell)}^2+\omega^2)}\,, \\
        f^{(2)}_{lm} = & \frac{h_0^{(2)}(r) +h_1^{(2)}(r) \omega +h_2^{(2)}(r) \omega^2+h_3^{(2)}(r) \omega^3 + h_4^{(2)}(r) \omega^4}{16c^4 (r+c)^2 \omega(2\kappa_{(\ell)} -i\omega)(\kappa_{(\ell)}^2+\omega^2)}\,,\\
        g^{(2)}_{lm} = & \frac{c^2}{r^2} g_{lm}^{(1)}\,,
    \end{align}
\end{subequations}
 \end{widetext}
where we introduced the following auxiliary functions $h^{(1\,,2)}_i(r)$ for $i=0\,, ..., 4$
\begin{subequations}\label{eqs:h1}
    \begin{align}
        h^{(1)}_0 =& -i (-1 + l) l (1 + l) (2 + l) r^2\,,\\
        \begin{split}
            h^{(1)}_1 = & 4 c^3 - 2 r [3 c^2 (-2 + l + l^2) \\
            &+ c (-9 + 7 l (1 + l)) r + 
     4 (-2 + l + l^2) r^2]\,,
        \end{split}\\
        \begin{split}
             h_2^{(1)} =& -4 i (c + r)^2 [c^2 + c (1 - 3 l (1 + l)) r \\
             &- 3 (-2 + l + l^2) r^2]\,,
        \end{split}\\
     h_3^{(1)} =& 8 (c + r)^4 (2 c + r)\,,\\
     h_4^{(1)} =& -16 i (r+c)^6\,,
     \end{align}
\end{subequations}and 

\begin{subequations}\label{eqs:h2}
    \begin{align}
         h^{(2)}_0 =& -i (-1 + l) l (1 + l) (2 + l) r\,,\\
        h^{(2)}_1 = &2 c (-3 r + l (1 + l) (c + r))\,,\\
        \begin{split}
             h^{(2)}_2 =& 4 i (c + r) [c^2 l (1 + l) + c (-3 + 4 l (1 + l)) r\\
             &+ 
   3 (-2 + l + l^2) r^2]\,,
        \end{split} \\
   h^{(2)}_3 = &-8(r+c)^4\,,\\
   h^{(2)}_4 = &-16i (r+c)^5\,.
    \end{align}
\end{subequations}
In Sec.~\ref{sec:tomography}, we discussed black hole tomography and showed that we could obtain explicit expressions relating the mode decomposition of $\widetilde{\Psi}_4$ at future null infinity as a function of the modes of $\widetilde{\Psi}_0$ at the horizon in Eqs.~\eqref{eq:tomography-psi4-psi0-m=0} and~\eqref{eq:tomography-psi4-psi0-mneq0}. These expressions can be inverted to describe the infalling gravitational radiation modes at the horizon as a function of the outgoing gravitational wave strain. Explicitly, we obtain  for $m=0$
\begin{widetext}
\begin{subequations}
    \begin{align}
        \Psi_{0\,,l0n}^{H\,,-}=&\frac{4c^4 i\omega_{l0n} (\kappa_{(\ell)}^2+\omega_{l0n}^2)(2\kappa_{(\ell) }-i\omega_{l0n})}{\mathcal{F}_{l0n} (K_l^4+36c^2 \omega_{l0n}^2)} (K_l^2\Psi_{4\,,l0n}^{\mathcal{I}^+\,,-} +6ic \omega_{l0n} \bar{\Psi}_{4\,,l0n}^{\mathcal{I}^+\,,+}  )\\
         \Psi_{0\,,l0n}^{H\,,+}=&-\frac{4c^4 i\bar{\omega}_{l0n} (\kappa_{(\ell)}^2+\bar{\omega}_{l0n}^2)(2\kappa_{(\ell) }+i\bar{\omega}_{l0n})}{\bar{\mathcal{F}}_{l0n} (K_l^4+36c^2 \bar{\omega}_{l0n}^2)} (K_l^2\Psi_{4\,,l0n}^{\mathcal{I}^+\,,+} -6ic \bar{\omega}_{l0n} \bar{\Psi}_{4\,,l0n}^{\mathcal{I}^+\,,-}  )\,,
    \end{align}
\end{subequations} and for $m\neq0$
\begin{subequations}
    \begin{align}
        \Psi_{0\,,lmn}^{H\,,-}=&\frac{4c^4 i\omega_{lmn} (\kappa_{(\ell)}^2+\omega_{lmn}^2)(2\kappa_{(\ell) }-i\omega_{lmn})}{\mathcal{F}_{lmn} (K_l^4+36c^2 \omega_{lmn}^2)} (K_l^2\Psi_{4\,,lmn}^{\mathcal{I}^+\,,-} +(-1)^m 6ic \omega_{lmn} \bar{\Psi}_{4\,,lmn}^{\mathcal{I}^+\,,-}  )\\
         \Psi_{0\,,lmn}^{H\,,+}=&-\frac{4c^4 i\bar{\omega}_{lmn} (\kappa_{(\ell)}^2+\bar{\omega}_{lmn}^2)(2\kappa_{(\ell) }+i\bar{\omega}_{lmn})}{\bar{\mathcal{F}}_{lmn} (K_l^4+36c^2 \bar{\omega}_{lmn}^2)} (K_l^2\Psi_{4\,,lmn}^{\mathcal{I}^+\,,+} -6(-1)^mic \bar{\omega}_{lmn} \bar{\Psi}_{4\,,lmn}^{\mathcal{I}^+\,,+}  )\,.
    \end{align}
\end{subequations} where again we use the notation $K_l=\sqrt{(l+2)(l+1)l(l-1)}$.
\end{widetext}

\end{document}